\documentclass[12pt,preprint]{aastex}
\begin{document}

\title{A Comprehensive Study of Gamma-Ray Burst Optical Emission: II. Afterglow Onset and Late Re-Brightening Components}
\author{En-Wei Liang \altaffilmark{1,2,3}, Liang Li\altaffilmark{1}, He Gao\altaffilmark{3}, Bing Zhang\altaffilmark{3}, Yun-Feng Liang\altaffilmark{1}, Xue-Feng Wu\altaffilmark{4}, Shuang-Xi Yi\altaffilmark{5}, Zi-Gao Dai\altaffilmark{5}, Qing-Wen Tang\altaffilmark{1}, Jie-Min Chen\altaffilmark{1}, Hou-Jun L\"{u}\altaffilmark{3}, Jin Zhang\altaffilmark{2,6}, Rui-Jing Lu\altaffilmark{1}, Lian-Zhong L\"{u}\altaffilmark{1}, and Jian-Yan Wei\altaffilmark{2}}
\altaffiltext{1}{Department of Physics and GXU-NAOC Center for Astrophysics
and Space Sciences, Guangxi University, Nanning 530004, China; lew@gxu.edu.cn}
\altaffiltext{2}{National Astronomical Observatories, Chinese Academy of Sciences, Beijing, 100012, China}
\altaffiltext{3}{Department of Physics and Astronomy, University of
Nevada, Las Vegas, NV 89154; zhang@physics.unlv.edu}
\altaffiltext{4}{Purple Mountain Observatory, Chinese Academy of Sciences, Nanjing 210008, China}
\altaffiltext{5}{School of Astronomy and Space Science, Nanjing University, Nanjing, Jiangsu 210093, China}

\altaffiltext{6}{College of Physics and Electronic Engineering, Guangxi Teachers Education
University, Nanning, 530001, China}

\begin{abstract}
We continue our systematic statistical study of various components in gamma-ray burst (GRB) optical lightcurves. We decompose the early onset bump and the late re-brightening bump with empirical fits and analyze their statistical properties. Among the 146 GRBs that have well-sampled optical lightcurves, the onset and re-brightening bumps are observed in 38 and 26 GRBs, respectively. It is found that the typical rising and decaying slopes for both the onset and re-brightening bumps are $\sim 1.5$ and $\sim -1.15$, respectively. No early onset bumps in the X-ray band are detected to be associated with the optical onset bumps, while an X-ray re-brightening bump is detected for half of the re-brightening optical bumps. The peak luminosity is anti-correlated with the peak time, $L_p\propto t_{p}^{-1.81\pm 0.32}$ for the onset bumps and $L_p\propto t_{p}^{-0.83\pm 0.17}$ for the re-brightening bumps. Both $L_p$ and the isotropic energy release of the onset bumps are correlated with $E_{\gamma, \rm iso}$, whereas no similar correlation is found for the re-brightening bumps. These results suggest that the afterglow onset bumps are likely due to the deceleration of the GRB fireballs. Taking the onset bumps as probes for the properties of the fireballs and their ambient medium, we find that the typical power-law index of the relativistic electrons is 2.5 and the medium density profile behaves as $n\propto r^{-1}$  within the framework of the synchrotron external shock models. With the medium density profile obtained from our analysis, we also confirm the correlation between initial Lorentz factor ($\Gamma_0$) and $E_{\rm iso, \gamma}$ in our previous work. The jet component that produces the re-brightening bump seems to be on-axis and independent of the prompt emission jet component. Its typical kinetic energy budget would be about one order of magnitude larger than the prompt emission component, but with a lower $\Gamma_0$, typically several tens.
\end{abstract}

\section{Introduction\label{sec:intro}}
The understanding of the gamma-ray burst phenomenon has been greatly advanced in recent years. While the prompt gamma-ray emission is generally interpreted as due to internal dissipation processes within the relativistic ejecta due to internal shocks near or far above the photosphere (Rees \& M\'esz\'aros 1994; Kobayashi et al. 1997; Daigne \& Mochkovitch 1998; Rees \& M\'esz\a'ros 2005; Pe'er et al. 2006) or internal magnetic energy dissipation processes (e.g. Usov 1992; Thompson 1994; Drenkhahn \& Spruit 2004; Giannios \& Spruit 2006; Zhang \& Yan 2011), the afterglow is usually interpreted as arising from the external shock formed as the fireball is decelerated by the ambient medium (M\'esz\'aros \& Rees 1997; Sari et al. 1998).

In the pre-{\em Swift} era, afterglow observations were mostly made in the optical bands. The data were well explained by the external shock model (e.g., M\'esz\'aros \& Rees 1997; Sari et al. 1998; Panaitescu et al. 1998; Panaitescu \& Kumar 2001;  Huang et al. 2000; see Zhang \& M\'{e}szaros 2004 for review). However, simultaneous observations with the XRT and UVOT on board {\em Swift} (Gehrels et al. 2004) as well as ground-based optical telescopes in the early afterglow phase have opened a new window to understand GRB afterglow (M\'{e}sz\'{a}ro 2006; Zhang 2007; Liang 2010). In particular, early X-ray afterglow observations revealed erratic flares and early plateaus that are difficult to interpret within the standard theoretical framework (Zhang et al. 2006, Nousek et al. 2006). The flares are believed to be produced by late central engine activity (Burrows et al. 2005; Fan \& Wei 2005; Zhang et al. 2006; Dai et al. 2006; Proga \& Zhang. 2006; Perna et al. 2006), and the shallow decay segment likely signals a long-lasting wind powered by the GRB central engine after the prompt gamma-ray phase (Dai \& Lu 1998a,b; Zhang et al. 2006). These features indicate that the GRB central engine does not die out quickly. The observed afterglow emission is a superposition of the traditional external shock afterglow and an afterglow related to the late central engine activity (Zhang 2011).

The mix of various emission components makes afterglow lightcurves more diverse (e.g., Liang et al. 2008). One may decompose the lightcurves using two different approaches: one through theoretical modeling and the other through empirical fitting. Theoretical modeling prevailing in the pre-{\em Swift} era (e.g., Panaitescu et al. 1998; Panaitescu \& Kumar 2001;  Huang et al. 2000; Wu et al. 2005) is found increasingly difficult in the {\em Swift} era, because of the large amount of data, and more importantly, the chromatic behavior that defeats the simplest external shock model.
We therefore take the more empirical approach to perform the analysis (e.g., Liang \& Zhang 2006; Panaitescu \& Vestrant 2008, 2011; Kann et al. 2010; 2011).

Our results are presented in a series of papers. In the first paper of this series (Li et al. 2012; paper I), we presented the general features of a ``synthetic" optical lightcurve based on our decomposition analysis. The "synthetic'' optical lightcurve shows eight possible components with distinct physical origins (see Figure 2 of Paper I): these include prompt optical flares; an early optical flare of the reverse shock origin; an early shallow decay segment; the standard afterglow component (an onset hump followed by a normal decay segment); the post jet break phase; optical flares; a re-brightening hump; and a late supernova (SN) bump. The optical flares and the shallow decay segment may signal late activity of the GRB central engine, which have been discussed in detail in Paper I. This paper is dedicated to the onset bump (component III) and the late re-brightening bump (component VI) in the synthetic lightcurve. The reason to discuss them together is because physically they share a similar origin of jet-medium interaction, and they are potentially good probes for the fireball and the ambient density profile (e.g., Rykoff et al.2009; Oates et al. 2009; Liang et al. 2010; L\"{u} et al. 2012; Ghirlanda et al. 2012; Yi et al. 2012).

The fireball model predicts a clear early onset hump in the afterglow lightcurve as the fireball is decelerated by the ambient medium (Rees \& M\'esz\'aros 1992; M\'esz\'aros \& Rees 2003; Sari \& Piran 1999 for the thin shell case, and Kobayashi et al. 1999; Kobayashi \& Zhang 2007 for the thick shell case). This prediction was examined by some authors with early optical and X-ray afterglow observations (e.g., Molinari et al. 2007; Xue et al. 2009; Melandri et al. 2010; Liang et al. 2010). The epoch of deceleration depends on the energy of the fireball, density of the medium, and more sensitively the initial Lorentz factor of the ejecta, $\Gamma_0$. The rising slopes of the onset bumps are determined by both the medium density profile (the $k$ value by assuming $n \propto r^{-k}$) and the electron spectral index, $p$. Therefore, the afterglow onset bumps would be an ideal probe to study the properties of the fireball and the circumburst medium. The re-brightening feature is likely related to another emission component, and hence, may probe the jet structure. We present the data and lightcurve fitting results in \S 2 and compare the properties of the onset and late re-brightening bumps in \S 3. We measure $\Gamma_0$, $p$, and $k$ with the onset bumps in \S 4. An analysis for the physical origin of the re-brightening bumps is presented in \S5. A summary and discussion of our results are presented in \S 6. Throughout,a convention $Q_{n} = Q/10^{n}$ is used for the physical parameters in cgs units, and the slope of a lightcurve and the power-law index of a spectrum is defined with a convention $F\propto t^\alpha \nu ^{-\beta}$.

\section{Data and Lightcurve Fitting \label{sec:data}}
All the GRBs that have optical afterglow detections by November 2011 (from Feb. 28 1997 to Nov. 2011) are included in our analysis. A sample of 225 optical lightcurves are compiled from published papers, or from GCN Circulars if no published paper is available. Well-sampled lightcurves are available for 146 GRBs. The sample has been presented in Table 1 of Paper I, in which GRB name, redshift, optical spectral information, observational time interval, as well as $\gamma$-ray spectral parameters are summarized (Li et al. 2012). We collected the optical spectral indices $\beta_O$ \footnote{An optical spectral index $\beta_O=0.75$ is adopted for those GRBs whose $\beta_O$ is not available.} and the host galaxy extinction $A_{\rm V}$ for each burst from the same literature in order to reduce the uncertainties introduced by different authors. Galactic extinction correction is made by using a reddening map presented by Schlegel et al. (1998). Since the $A_{\rm V}$ values are available only for some GRBs and the $A_{\rm V}$ is derived from the spectral fits using different extinction curves, we do not make correction for the GRB host galaxy extinction. The $k$-correction in magnitude is calculated by $k=-2.5(\beta_O-1)\log(1+z)$. The late epoch data ($\sim 10^6$ seconds after the GRB triggers) are heavily contaminated with the flux from the host galaxy. We fit the host galaxy flux with the late time data and subtract it from the entire light curve. The isotropic gamma-ray energy ($E_{\rm \gamma, iso}$) is derived in the rest frame $1-10^4$ keV energy band using the spectral parameters. We fit the lightcurves with a model of multiple power-law components. The basic component of our model is either a power-law function, $F = F_0 t^{\alpha}$ or a smooth broken power-law function, $F =  F_0 [(t/t_{\rm p})^{\alpha_1 s}+(t/t_{\rm p})^{\alpha_2 s}]^{1/s}$, where $\alpha$, $\alpha_1$, $\alpha_2$ are the temporal slopes, $t_{\rm p}$ is the peak time of a bump, and $s$ measures the sharpness of a peak of the lightcurve component. In some cases, we adopt a tripple broken-power-law model. We developed an IDL code to make best fits with a subroutine called MPFIT\footnote{http://www.physics.wisc.edu/$\sim$craigm/idl/fitting.html.}.  For the details of our lightcurve fitting, see Paper I.

\begin{figure*}
\includegraphics[angle=0,scale=0.35]{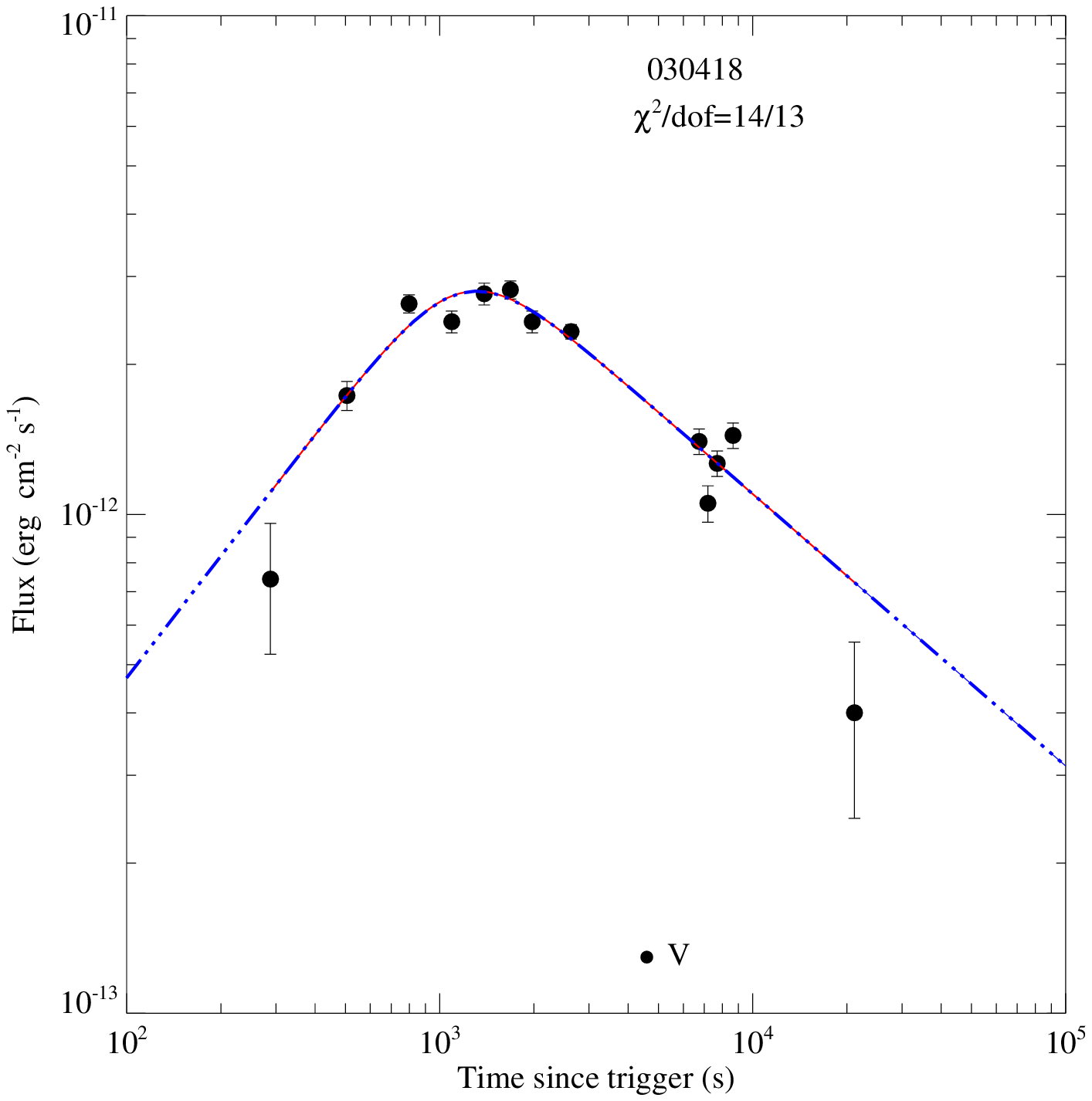}
\includegraphics[angle=0,scale=0.35]{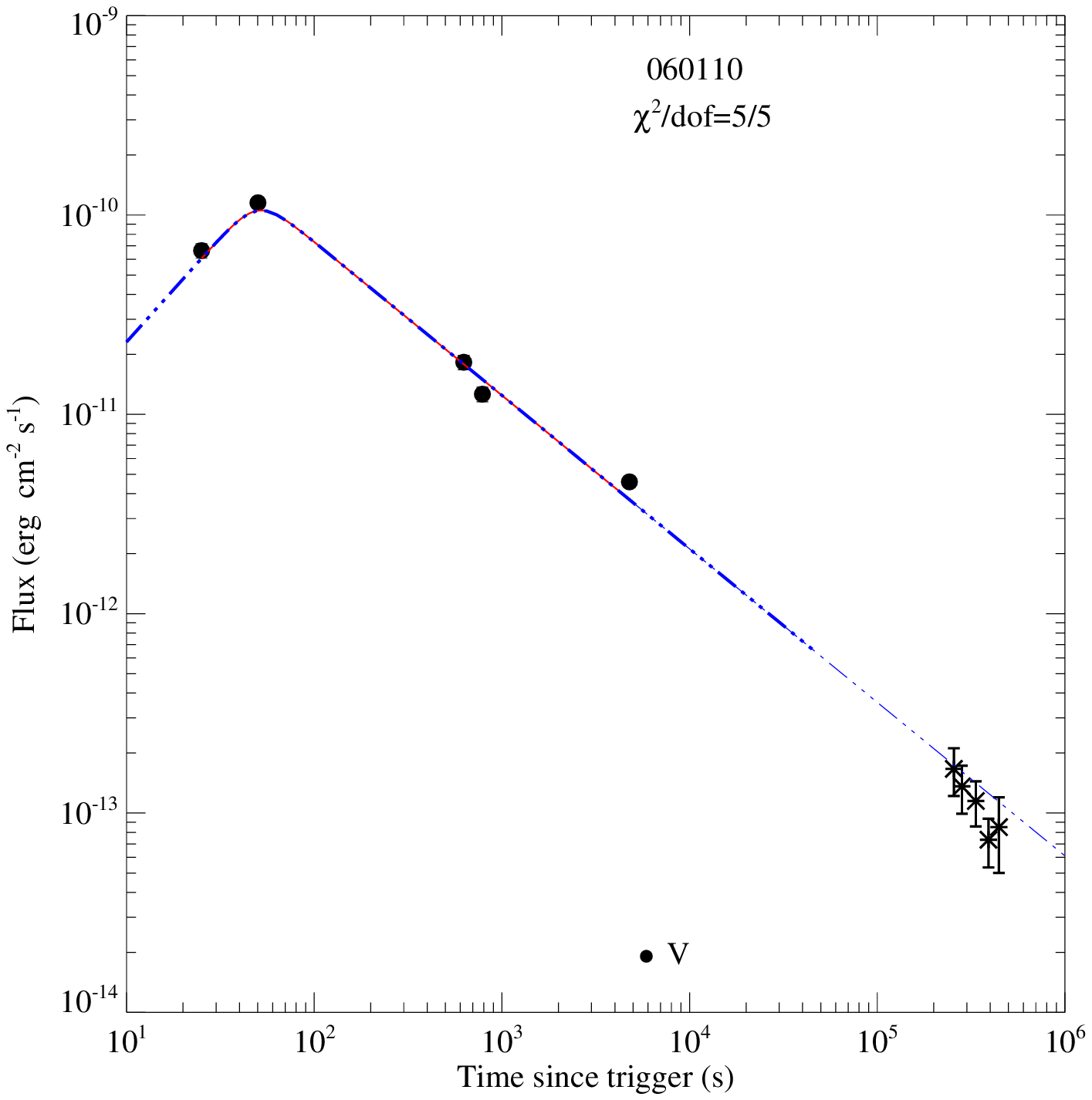}
\includegraphics[angle=0,scale=0.35]{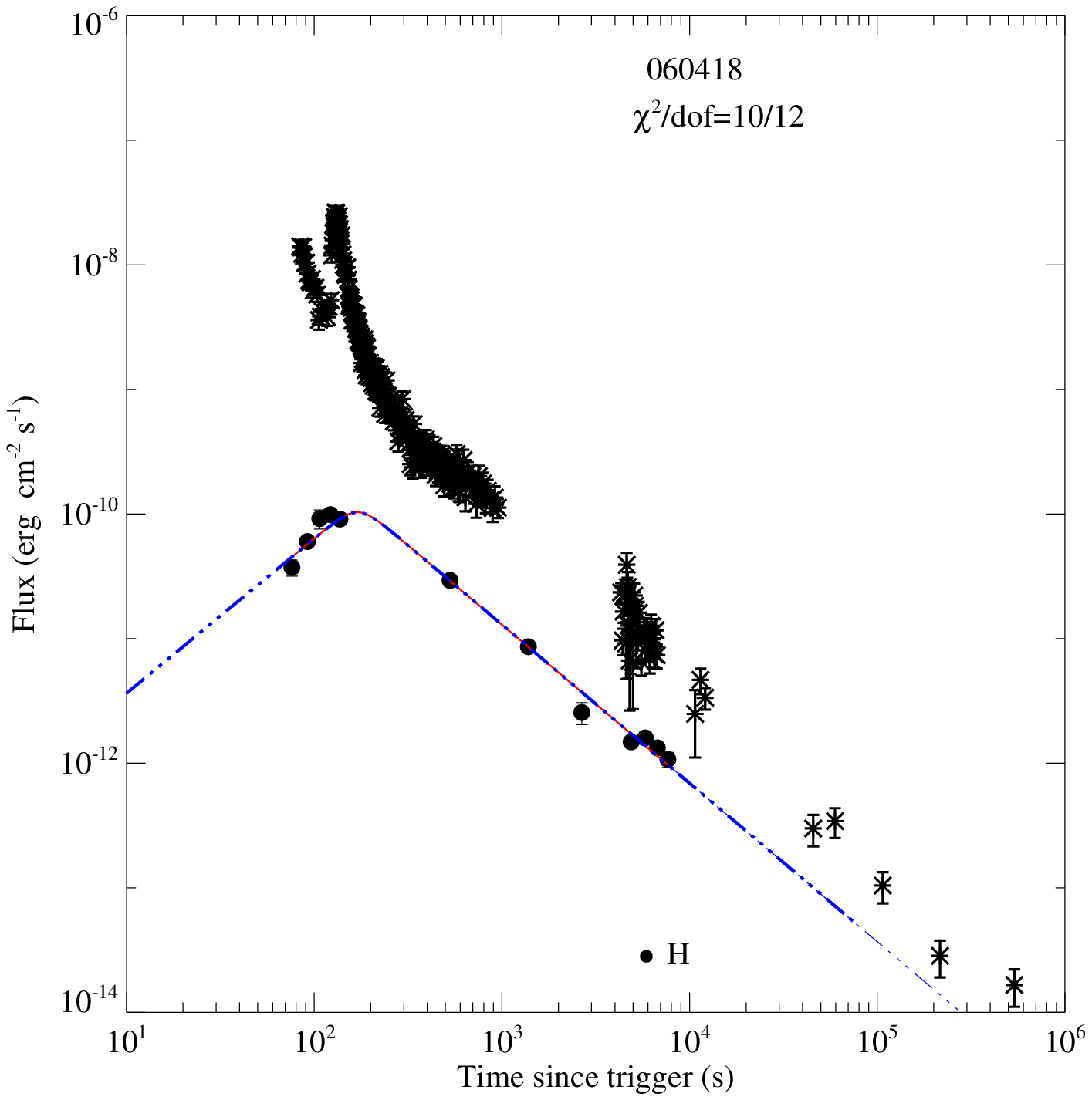}
\includegraphics[angle=0,scale=0.35]{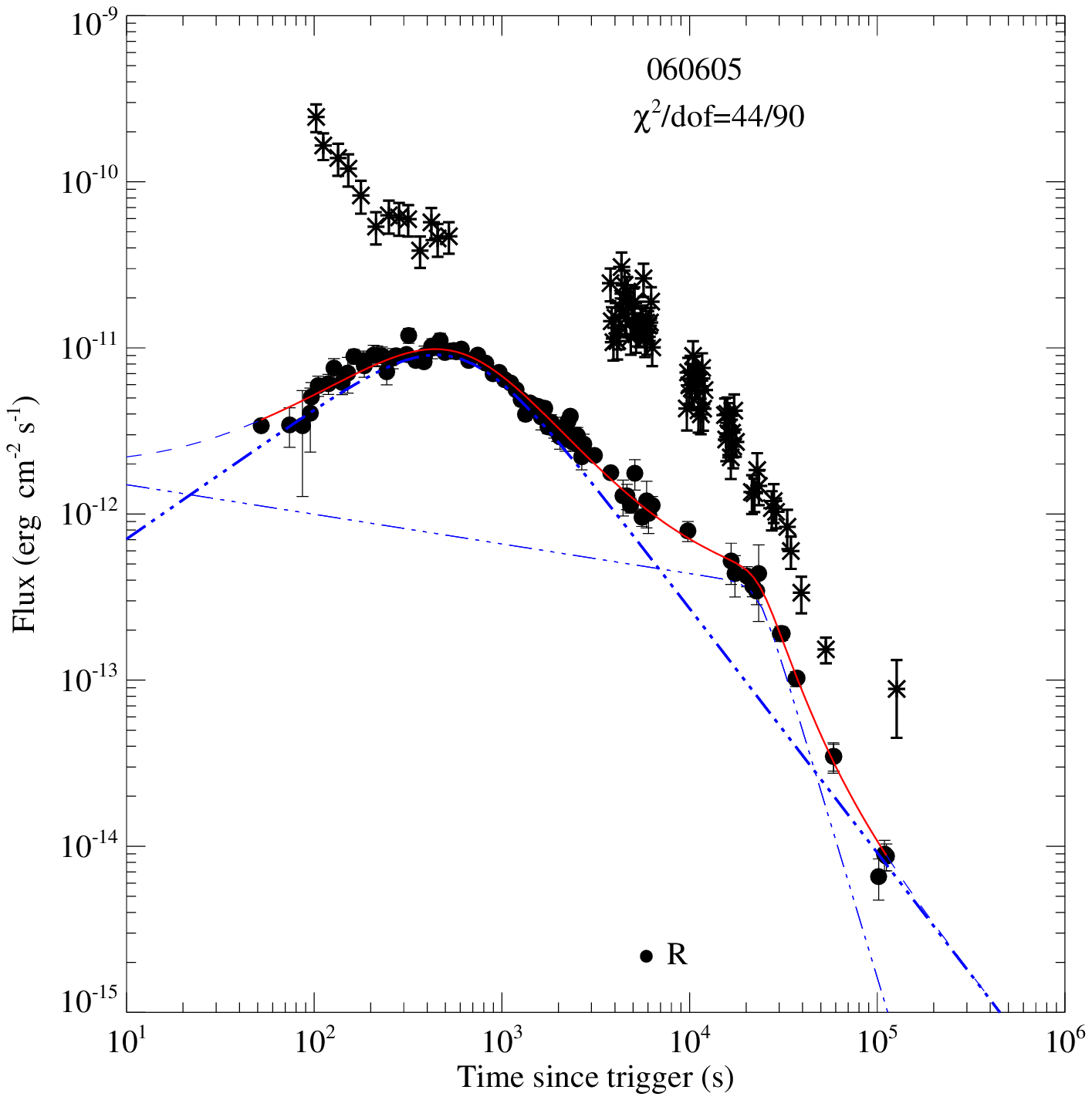}
\includegraphics[angle=0,scale=0.35]{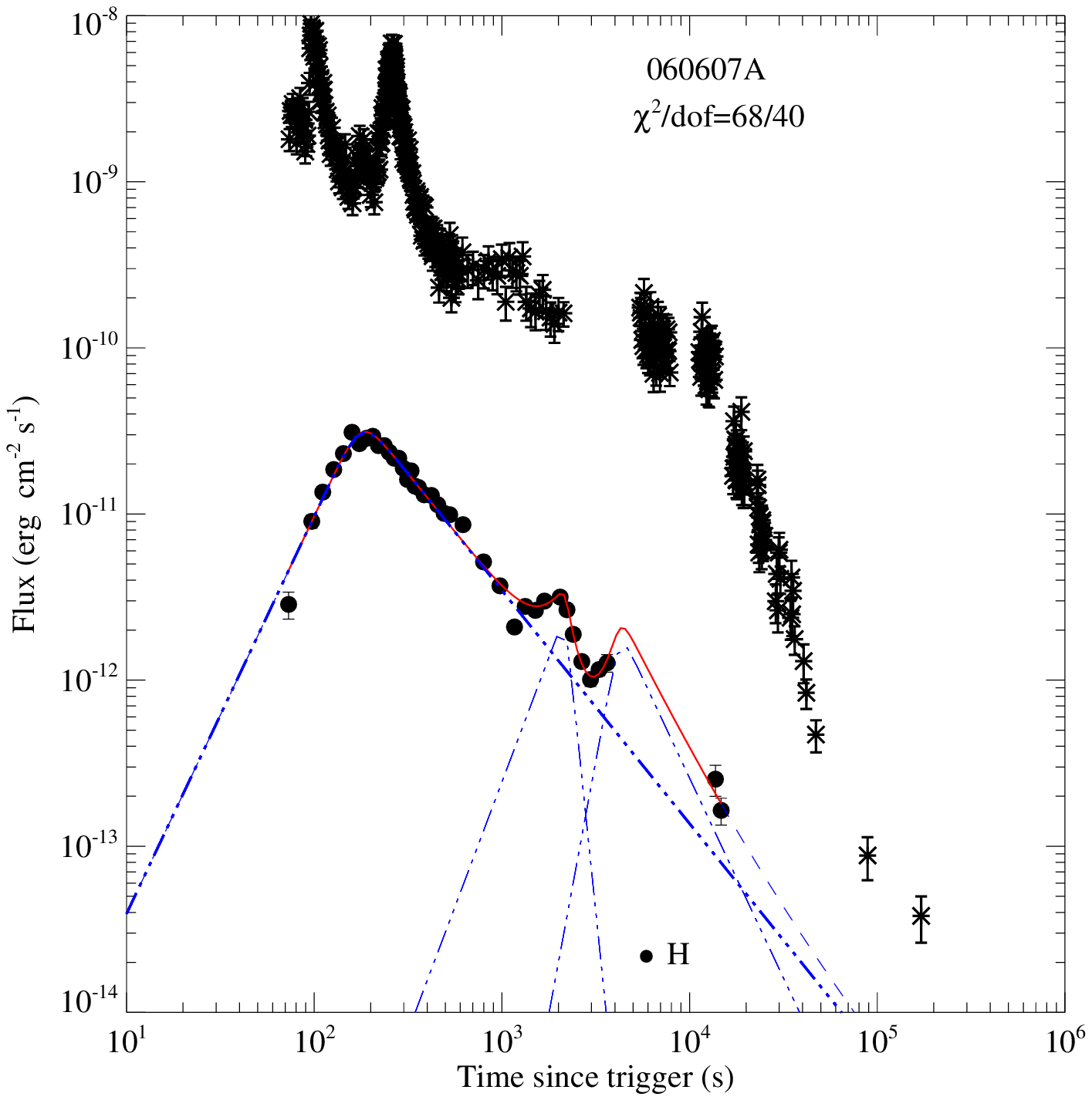}
\includegraphics[angle=0,scale=0.35]{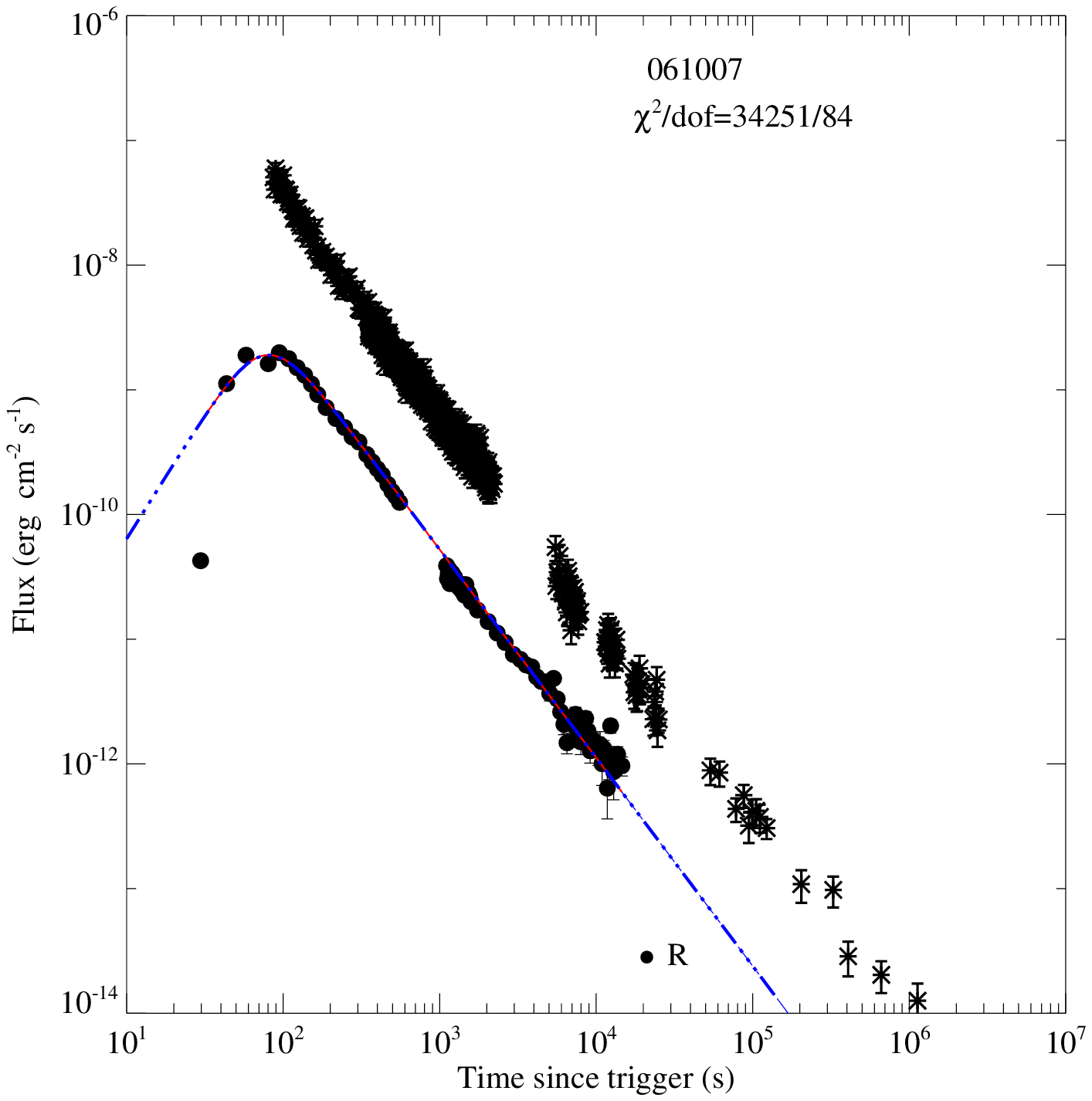}
\includegraphics[angle=0,scale=0.35]{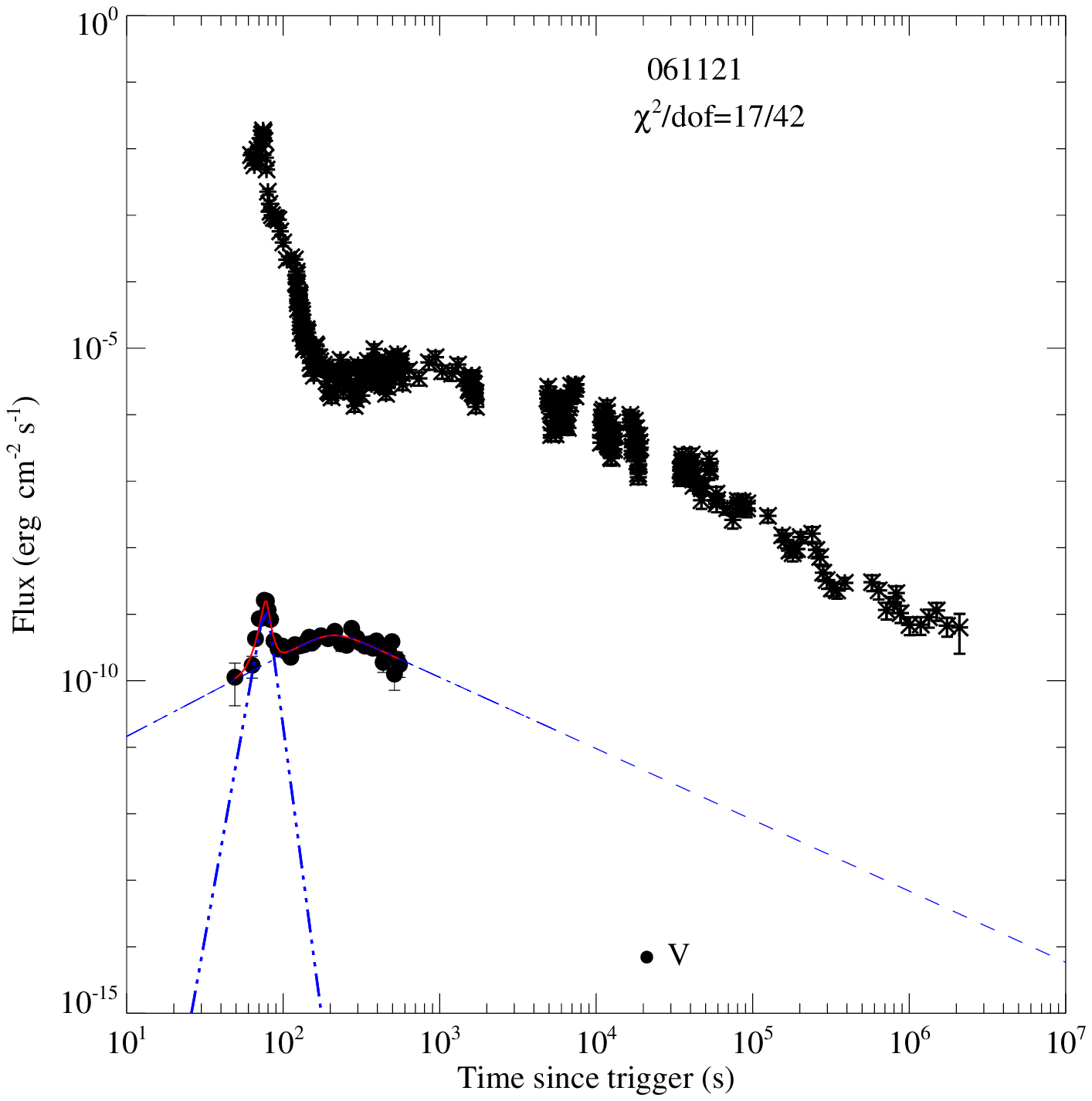}
\includegraphics[angle=0,scale=0.35]{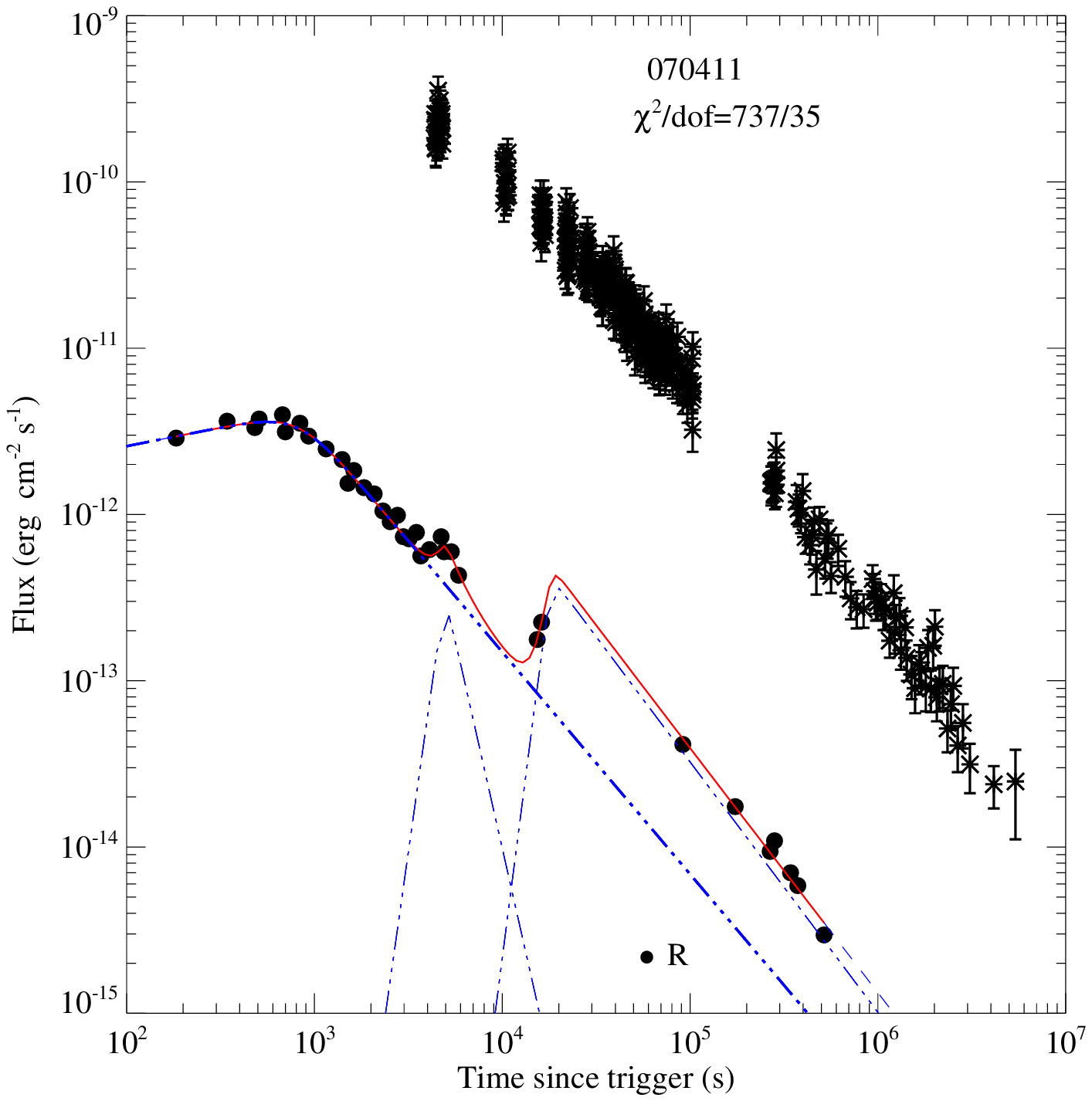}\hfill
\includegraphics[angle=0,scale=0.35]{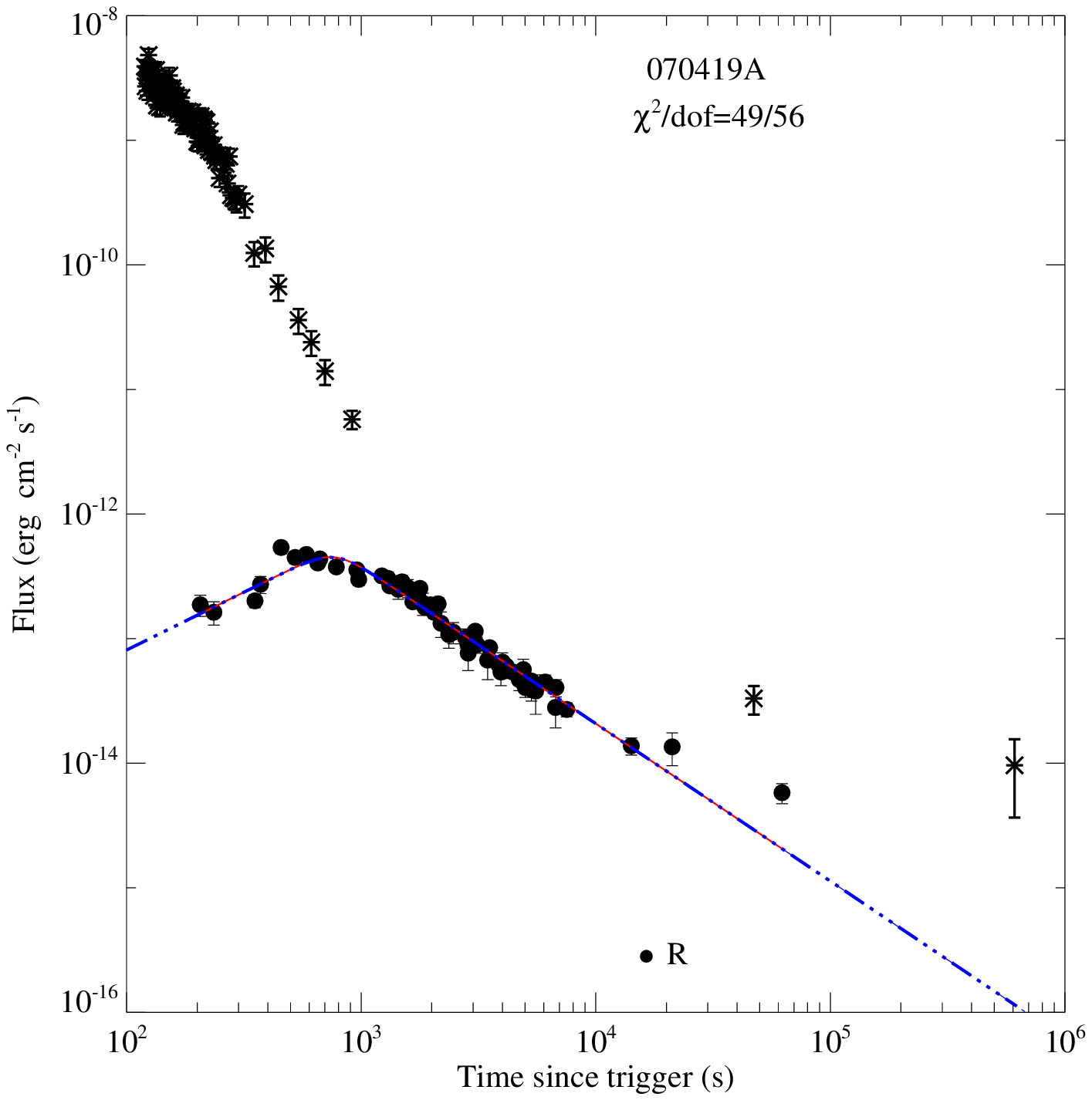}
\caption{Observed optical afterglow lightcurves as predicted by the standard fireball model. Erratic flares are detected in some GRBs. The solid line in each panel is the best fit with our multi-component power-law model in the temporal coverage of the data and the dashed line show the extrapolation of the best fit. The dash-dotted lines mark the components in our model. The simultaneous X-ray data observed with {\em Swift}/XRT (crosses with error bars) are also presented.}\label{Onset}
\end{figure*}

As shown in Paper I, eight components are decomposed from the observed optical lightcurves. This paper focuses on the early afterglow onset and late re-brightening afterglow components. The reason to discuss them together is because they may be both related to the similar physical origin, i.e., multiple fireball components are decelerated by the circum-burst medium with the same or different jet axes so that the line-of-sight may view them with on-axis or off-axis geometries.  An afterglow onset feature is characterized by a smooth bump with a peak less than 1 hour post trigger, which is followed by a normal (decay index between -0.75 and -2) power-law decay component. A rebrightening bump is similar to an onset bump but peaks at a much later time. The supernova bump is a special late rebrightening peaking at around 1-2 weeks after a GRB trigger, which usually shows a red color. We exclude the supernovae bumps in this analysis by restricting the peak time of the bumps to be earlier than 7 days after a GRB trigger. Throughout, we mark the parameters of the early onset bumps and the late re-brightening bumps with the superscripts ``on'' and ``re'', respectively. We classify the optical lightcurves with the detections of an onset bump and/or a late re-brightening bump into five groups as described below.
\begin{itemize}
 \item Group I: This group includes bursts with an onset bump followed by a normal decay segment without a late re-brightening bump, as shown in Figure \ref{Onset}. These are a textbook version of lightcurves as predicted by the standard fireball model. Erratic flares are superimposed on some of the lightcurves, but they may be produced by late internal shocks as presented in paper I. We have 27 cases in the sample.
 \item Group II: This group includes bursts with an initial afterglow onset bump followed by a late re-brightening hump. We have 11 cases in the sample. They are shown in Figure \ref{Onset_RB}.
 \item Group III: This group includes bursts with an initial shallow decay segment followed by a late re-brightening hump. We have 8 cases in the sample (Figure \ref{Shallow_RB}).
 \item Group IV: This group includes bursts with an initial normal decay followed by a late re-brightening bump. We have 7 cases in the sample (Figure \ref{Normal_RB}).
 \item Group V: This group includes bursts with a single bump that peaks at $>10^4$ seconds post the GRB trigger. We have 3 cases (GRBs 060614, 070306 and 100418) in the sample (Figure \ref{Late_bump}). Since we do not detect a decay component before the bump, one cannot decide with confidence whether it is an onset or re-brightening bump. We therefore do not include them in the statistical analysis of the two types of bumps.
\end{itemize}
Altogether we have 38 early afterglow onset bumps\footnote{
An early onset bump may be also embedded in the data of GRBs 080319A and 090716 with our lightcurve fits. However, the bumps are highly contaminated with bright flares or reversed shock emission. The rising slopes are very uncertain. We therefore do not include these bumps in our sample.} and 26 late re-brightening bumps for our statistical analysis. The parameters of the onset and re-brightening bumps in our sample derived from our fits, include the peak flux ($F_{\rm m}$), slopes ($\alpha_1$ and $\alpha_2$), peak time ($t_{\rm p}$), the full-width-at-half-maximum (FWHM, $w$), the rising and decaying timescales measured at FWHM ($t_{r}$ and $t_{d}$), the ratio ($R_{\rm rd}$) of $t_{\rm r}/t_{\rm d}$, and the ratio ($R_{\rm rp}$) of $t_{\rm r}/t_{\rm p}$, are reported in Tables 1 and 2. We derive the isotropic energy release by integrating the emission over the $R$ band (with the central frequency and the full-width-half-maximum as $7000{\AA}$ and $2200{\AA}$, respectively), in units of $10^{48}$ ergs ($E_{\rm R, iso, 48}$) for the afterglow onset and re-brightening bumps in the time interval [$t_{\rm p}/5, 5t_{\rm p}$]. They are also listed in Tables 1 and 2. The redshift ($z$) and the isotropic gamma-ray energy in units of $10^{52}$ ergs ($E_{\gamma, \rm iso, 52}$) of these GRBs are also reported in Tables 1 and 2.

\section{Comparison between the Properties of the Early Onset and Late Re-Brightening bumps}
Figure \ref{Onset_RB_Dis} compares the distributions of $\alpha_1$, $\alpha_2$, $t_{\rm p}$, the R-band luminosity at $t_{\rm p}$ ($L_{R, \rm p}$) and the ratio \textbf{($R_{\rm rd}$)} of rising-to-decaying timescales measured at FWHM. It is found that both $\alpha^{\rm on}_1$ and $\alpha^{\rm re}_1$ distributions are well consistent with each other, falling in the range of $0.3\sim 4$, with a typical value 1.5. Both $\alpha^{\rm on}_2$ and $\alpha^{\rm re}_2$ are also consistent. They are narrowly distributed in the range [-1.8, -0.6] with a typical value $ -1.15$, except for three outliers whose $\alpha^{\rm re}_2$ are steeper than 2. The distribution of the ratio \textbf{$\rm R_{\rm rd}$} is clustered around $\sim 0.4$. The distribution of $t^{\rm on}_{\rm p}$ spans the range 30 to 3000 seconds, while $t^{\rm re}_{\rm p}$ is distributed in a much wider range from several hundreds of seconds to days. The $L^{\rm on}_{R, \rm p}$ distribution shows a bimodal feature that peaks at $10^{46}$ erg s$^{-1}$ and $10^{48}$ erg s$^{-1}$, respectively, but this feature is not statistically significant in the current sample\footnote{Since the visual bimodal distribution feature depends on the bin size selection, we examined this feature with a statistical algorithm proposed by Keith et al. (1994), the so-called KMM algorithm, which is independent of bin size selection effect. The KMM algorithm yields a likelihood ratio test statistics ($r^{\rm KMM}$) and a corresponding chance probability $p^{\rm KMM}$. A bimodal feature is statistically acceptable if $p^{\rm KMM}<10^{-4}$. We get $r^{\rm KMM}=2.2$ and $p^{\rm KMM}=0.33$, which means that the bimodal feature is still not statistically acceptable.}.  A sharp cutoff at the left side of the first peak is likely due to an instrumental detection limit, which selects against faint optical afterglow onset bumps. The log-normal function fit to the $L^{\rm re}_{\rm R, p}$ distribution yields $\log L^{\rm re}_{\rm R, p}=45.76\pm 0.90$ ($1\sigma$).

Figure \ref{Onset_RB_Corr} shows $w$, $L_{\rm p}$, and $E_{\rm R, iso}$ as a function of $t_{\rm p}$ for both the onset and re-brightening bumps. We find that $\log w=(0.17\pm 0.20)+(1.06\pm 0.08) \log t_{p}$ with a Spearman correlation coefficient $r=0.93$ and $p<10^{-4}$ for the onset bumps. The rebrightening bumps share the same $w-t_{\rm p}$ relation, with a slope of $0.94\pm 0.04$. The peak luminosity is anti-correlated with the peak time for both the onset and re-brightening bumps, but the power-law indices of the correlations are different. The best fits give $\log L^{\rm on}_{\rm R, p,48}=(2.83\pm 0.69)-(1.81\pm 0.32) \log [t^{\rm on}_{\rm p}/(1+z)]$ (with $r=-0.74$, $p<10^{-4}$) and $\log L^{\rm re}_{\rm R, p,48}=(0.65\pm 0.61)-(0.83\pm 0.17) \log [t^{\rm re}_{\rm p}/(1+z)]$ (with $r=-0.73$, $p<10^{-4}$). The $E^{\rm on}_{\rm R, iso}$ is tentatively anti-correlated with $t^{\rm on}_{\rm p}$, i.e., $\log E^{\rm on}_{\rm R, iso,48}=(3.38\pm 0.67)-(0.83\pm 0.32) \log [t^{\rm on}_{\rm p}/(1+z)]$ with $r=-0.46$ and $p=0.015$. A correlation analysis between $E^{\rm re}_{\rm R, iso}$ and $t^{\rm re}_{\rm p}$ gives $r=0.04$ and $p=0.83$, confidently indicating that there is no correlation between the two quantities. Both onset and re-brightening bumps are detected for 11 GRBs. The luminosity $L^{\rm re}_{\rm R, p}$ is usually smaller than  $L^{\rm on}_{\rm R, p}$, but the energy $E^{\rm re}_{\rm R, iso}$ is comparable to $E^{\rm on}_{\rm R, iso}$ or even larger than $E^{\rm on}_{\rm R, iso}$,  as shown in Figure \ref{LR_LB_Corr}. We examine the pair correlations for $L^{\rm on}_{\rm R, p}-L^{\rm re}_{\rm R, p}$ and $E^{\rm on}_{\rm R, p}-E^{\rm re}_{\rm R, iso}$, and obtain $r=0.24$ with $p=0.47$ for $L^{\rm on}_{\rm R, p}-L^{\rm re}_{\rm R, p}$ and $r=0.55$ with $p=0.07$ for $E^{\rm on}_{\rm R, p}-E^{\rm re}_{\rm R, iso}$. These indicate that $L^{\rm on}_{\rm R, p}$ is not corrected with $L^{\rm re}_{\rm R, p}$, and there is only a weak correlation between $E^{\rm on}_{\rm R, p}$ and $E^{\rm re}_{\rm R, iso}$.

Figure \ref{Eiso_Onset_RB_Corr} shows $L_{\rm R, p}$ and $E_{\rm R, iso}$ as a function of $E_{\gamma, \rm iso}$ for both the onset and re-brightening bumps. It is found that both $L^{\rm on}_{\rm R, p}$ and $E^{\rm on}_{\rm R, iso}$  are correlated with $E_{\gamma, \rm iso}$. The best fit gives  $\log L^{\rm on}_{\rm R, p,48}=(3.77\pm 0.47)+(1.03\pm 0.16) \log E_{\gamma, \rm iso,50}$ (with $r=0.79$, $p=10^{-4}$) and $\log E^{\rm on}_{\rm R, p, 48}=(-0.30\pm 0.38)+(0.72\pm 0.13) \log E_{\gamma, \rm iso,50}$ (with $r=0.75$ and $p<10^{-4}$). However, we do not find similar correlations for the late re-brightening bumps in the sample.

Both X-ray and optical afterglows were observed for most GRBs in our sample. The X-ray afterglow lightcurves observed with {\em Swift}/XRT are also shown in Figures 1-5. We check whether an onset or re-brightening bump also exist in the X-ray lightcurve at the time of optical onset/rebrightening bump. For the onset bumps, the X-ray lightcurves are usually characterized by a steep decay segment or flares. No associated onset bumps are identified (see Figure 1)\footnote{The decay slope post the onset peak is roughly consistent with that observed in the X-ray band for GRBs 060418, 061007, 061121, 071010B, 071025, 071031, 071112C, 080603A, 080810, 081008, 081109A, 081203A, 090812, 110205A, 070318, 080310, 070411, 080710, 090102, 100901, 060729, 081029, 100219A. The early afterglow onset bump in the X-ray band, if exist, would be highly contaminated by the early steep decay segment and flares. Therefore, the onset bump is difficult to be identified in the X-ray band (e.g. Rykoff et al 2009; Oates et al 2009).}. For the re-brightening bumps, on the other hand, the associations seem common. A simultaneous bump in the X-ray band is detected in about half of the optical re-brightening bumps (Figure 2). These results suggest that the optical lightcurves are less contaminated by the prompt emission tail and late central engine activity emission, and therefore a good probe for the external shock afterglow (e.g., Rykoff et al.2009; Oates et al. 2009; Liang et al. 2010; L\"{u} et al. 2012; Ghirlanda et al. 2012; Yi et al. 2012).

\section{Onset bumps: Probing properties of fireball and circumburst medium density profile}
Within the framework of the synchrotron external shock model, a clear onset bump is expected in the early afterglow lightcurve as the fireball is decelerated by the surrounding medium (e.g., Sari \& Piran 1999; Kobayashi \& Zhang 2007). The tight  $ L^{\rm on}_{\rm R, p}- E_{\gamma, \rm iso}$ correlation and the values of the rising and decay slopes are all consistent with the model that invokes early deceleration of the GRB fireball\footnote{A smooth bump may be also attributed to an off-axis viewing angle effect (Panaitescu \& Vestrant 2008). In this case the rising part of the bump is defined by the off-beam effect, while the asymptotic decaying index would be the post-jet-break deceleration phase (with a decay slope steeper than 1.5). For a constant density model, the transition to this phase is quick enough, so that the typical value of the observed $\alpha_2$ (shallower than 1.5) would disfavor such a possibility. For a more stratified medium, the transition time to a steeper phase can be longer, which may accommodate some of the observations. However, such a model predicts a gradual steepening feature in the lightcurve, which is not observed from the data.}.  In this section we take the deceleration model to interpret the onset bumps, and use the data to probe the properties of fireball and circumburst medium density profile.

\subsection{Circumburst medium density profile and electron spectral index}
Within the deceleration model, the rising and decaying slopes of an afterglow onset bump depend on the circumburst medium density profile parameter $k$ and the radiation spectral regime (e.g., Sari et al. 1999; Xue et al. 2009). As shown in Table 1 of Paper I, the spectral index ($\beta_{\rm O}$) for most optical afterglows are between 0.5-0.7, indicating that the optical emission is usually in the spectral regime $\nu_m<\nu<\nu_c$ and $p>2$, where $\nu_m$ and $\nu_c$ are the typical and cooling frequencies of synchrotron radiation. In this spectral regime, we have $p=2\beta_{\rm O}+1$, $\alpha_1=3$ and $\alpha_2=3(1-p)/4$ for the constant density ISM model, and  $\alpha_1=(1-p)/2$ and $\alpha_2=(1-3p)/4$ for the wind model. With the $p$ values derived from $\beta_{\rm O}$, we find that the predicted rising slopes of both the ISM and wind medium are significantly inconsistent with the observed $\alpha_1$ values of most GRBs in our sample. The observed $\alpha_2$ values, on the other hand, generally agree with the prediction of the ISM model. Therefore, we consider a circumburst medium density profile
\begin{eqnarray}
\label{n} n = \left\{ \begin{array}{ll}
n_{0} \left(\frac{R}{R_{t}}\right)^{-k}, &  R <=R_t, \\
n_{0}, &  R > R_t,
\end{array} \right.
\end{eqnarray}
where $R_{\rm t}$ is the transition radius at which the medium turns into a constant density ISM,  $n_0=1$ cm$^{-3}$. If the condition $R_{\rm t} \geq R_{\rm dec}$ is satisfied, where $R_{\rm dec}$ is the deceleration radius, the thin-shell external shock model gives a rising slope\footnote{If instead $R_{\rm t} < R_{\rm dec}$ is satisfied, in principle one would see a transition from a slow-rising phase to a fast-rising phase. This is not observed, which suggests that usually $R_{\rm t} \geq R_{\rm dec}$ is satisfied.}
\begin{eqnarray}
\label{al1}\alpha_1=3-\frac{k(p+5)}{4}.
\end{eqnarray}
Since the decaying phase after the peak usually lasts a much longer time, its decay slope is essentially defined by the phase of $R\gg R_t$, so that the long-term decay slope is defined by the constant-density ISM scaling\footnote{It is also possible that the medium profile transits to the wind medium since the radius of the wind medium would be $> 10^{19}$ cm in some GRBs (e.g. Dai \& Wu 2003). In this case the decay slope $\alpha_2=(1-3p)/4$. }, i.e.
\begin{eqnarray}
\label{al2}\alpha_2=-\frac{3(p-1)}{4}.
\end{eqnarray}
With the observed $\alpha_1$ and $\alpha_2$ one can derive the $k$ and $p$ values for each burst. We find that 32 out of the 38 bursts in our sample can be explained with this simple model. Six GRBs cannot be explained with this model. The early bumps in GRBs 090102, 110205, and 110906 have a very rapid rising with an index 5.2, 4.0, and 3.4$\pm$0.2, respectively. This may be due to the contamination of the reversed shock emission (e.g., Zhang et al. 2003, Zheng et al. 2012). The decay slope of the bumps in GRBs 030418, 071010B and 081126, on the other hand, are too shallow (-0.55$\pm$ 0.04, -0.60$\pm$0.06, -0.39$\pm$ 0.01, respectively) to be interpreted in the standard fireball model, and requires significant energy injection\footnote{We should note that the lightcurves of GRBs 030418, 071010B and 081126, are poorly sampled or only covered for a short time interval.}. We therefore only include 32 GRBs in our following analysis.

Figure \ref{pkonset} shows the distributions of $p$ and $k$. The $p$ values range from 2 to 3.5, with a typical value of 2.5. Both analytical and numerical studies show that the spectrum of electrons accelerated by ultra-relativistic shocks are a power-law with an index in this range (Kirk et al. 2000, Achterberg et al. 2001, Bednarz \& Ostrowski 1998, Lemonine \& Pelletier 2003). The derived $p$ values from our analysis are generally consistent with this, but they are distributed in a wider range (see also Panaitescu \& Kumar 2001; Yost et al. 2003; Shen et al. 2006; Curran et al. 2010).

The $k$ values vary from 0 to 1.5 among GRBs, and its typical value is 1. Most of them are narrowly distributed in the range $0.75-1.25$. This is an intermediate regime between ISM and stellar wind. It may be formed via episodic energy ejection of matter by the progenitor star some time before the GRB. The exact mechanism to shape such a density profile is subject to further theoretical study.

\subsection{Relation between Fireball Lorentz Factor and Isotropic Gamma-ray Energy}
Liang et al. (2010) derived the intial Lorentz factor $\Gamma_0$ and $R_{\rm dec}$ assuming that the early optical afterglow onset bumps are due to the deceleration of a GRB fireball in an ISM medium, and discovered a tight correlation between $\Gamma_0$ and $E_{\rm \gamma, iso}$.  We revisit this relation in view of the modified density profile as discussed above.
We define a dimensionless parameter
\begin{eqnarray}
\xi\equiv\frac{R_t}{R_{\rm dec,cr}},
\end{eqnarray}
where $R_{\rm dec,cr}$ is the deceleration radius of the GRB fireball in the
ISM model, which can be calculated by
\begin{eqnarray}
R_{\rm dec,cr}=2.0\times\left[\frac{3E_{\rm\gamma, iso}t_{\rm
p}}{32\pi \eta nm_{\rm p}c(1+z)}\right]^{\frac{1}{4}},
\end{eqnarray}
where $n$ is the ambient density and $\eta = E_{\gamma, {\rm iso}} / E_{\rm
K,iso}$ is the ratio between the isotropic gamma-ray energy and the isotropic
blastwave kinetic energy, $m_{\rm p}$ is the mass of proton, $c$ speed of light. In the moderate density profile $n=n_0r^{-k}$ as we derived above, the deceleration radius should be
\begin{eqnarray}
\label{Rd} R^{'}_{\rm dec}=\left[\frac{(3-k)E_{\rm\gamma, iso}}{2\pi
\eta n_0\xi^kR_{\rm
dec,cr}^km_{\rm p}\Gamma_0^2c^2}\right]^{\frac{1}{3-k}}.
\end{eqnarray}
The initial Lorentz factor can be then estimated as\footnote{The constant factor 2 shown in Sari \& Piran (1999) is replaced by 1.4 through rigorous integration, as suggested by L\"{u} et al. 2012.}
\begin{eqnarray}
\label{g0} \Gamma^{'}_0 = \left\{ \begin{array}{ll}
G(k)\times\left[\frac{{\rm (3-k)}E_{\rm\gamma,
iso}(1+z)^{3-k}}{32\pi\eta n_0\xi^kR_{\rm dec,cr}^km_{\rm p}c^{\rm
5-k}t_{\rm p}^{\rm 3-k}}\right]^{\frac{1}{\rm 8-2k}}, &  \xi \gg 1 \\
1.4\times\left[\frac{{\rm 3}E_{\rm\gamma, iso}(1+z)^{3}}{32\pi\eta
n_0m_{\rm p}c^{\rm 5}t_{\rm p}^{\rm 3}}\right]^{\frac{1}{\rm 8}}, & \xi \ll 1,
\end{array} \right.
\end{eqnarray}
where $G(k)$ is a dimensionless parameter as a function of $k$ (Figure \ref{Gk}). We take $n_0=1$ cm$^{-3}$, $\eta=0.2$ and $\xi=3$ to calculate $\Gamma^{'}_0 $ and $R^{'}_{\rm dec}$. The distributions of $\Gamma^{'}_0 $ and $R^{'}_{\rm dec}$ are shown in Figure \ref{Gamma_0}. It is found that $\Gamma_0^{'}$ is distributed in the range [50, 500], and $R_{\rm dec}^{'}$ is around 10$^{17}$ cm. The relation between $\Gamma^{'}_{0}$ and $E_{\rm iso}$ is shown in Figure \ref{Gamma_0-Eiso}. The best fit yields
\begin{equation}
\log \Gamma^{'}_{0}=(1.43\pm 0.10)+(0.26\pm 0.03)\log E_{\rm iso,50}\label{Gamma0_Eiso_Relation},
\end{equation}
or $\Gamma'_0 = (27\pm 6) E_{\rm iso,50}^{0.26\pm 0.03}$.
The slope of this relation is consistent with that reported in Liang et al. (2010), but the coefficient is smaller since here we adopt the density profile as discussed above and replace the constant factor 2 by 1.4 in our calculation of the initial Lorentz factor.

\section{Re-brightening Bumps: possible origins}
As summarized in \S 2, late rebrightenning bumps are observed in 26 GRBs. Several models may produce a rebrightening bump, including a medium density jump (Dai \& Lu 2002; Dai \& Wu 2003; Lazzati et al. 2002), a refreshed shock (Zhang \& M\'esz\'aros 2002a; Bj\"{o}rnsson et al. 2004); a structured jet (Nakar et al. 2003; Berger et al. 2003; Huang et al. 2004; Liu et al. 2006; Jia et al. 2012), and emission from a long-lived reverse shock due to certain stratified ejecta profile (Uhm et al. 2012). However, with a density bump it is difficult to produce very sharp rebrightenings (Nakar \& Granot 2007). The refreshed shock scenario requires a large total energy budget to be at least comparable to the existing energy in the blastwave, even though a reverse shock can be bright enough to power a rapid optical rebrightening feature (Zhang \& M\'esz\'aros 2002a). The long-lasting reverse shock model (Uhm et al. 2012) needs to have the forward shock suppressed in order to have a significant rebrightening feature from the reverse shock.

In the following we focus on the structured jet model. Three scenarios involving structured jets may be employed to explain the late re-brightening bumps. First, the outflow may be ``patchy'', and have at least two components with different jet axes. The on-axis one gives rise to the prompt emission and early afterglow, while the more energetic off-axis one enters the field of view at a later time and gives the re-brightening feature (e.g. Granot et al. 2002). This model interprets the rebrightening bump as the full emergence of the off-axis jet in the field of view (Huang et al. 2004; Panaitescu \& Vestrand 2008; Guidorzi et al. 2009, Margutti et al. 2010). This requires a uniform jet with a sharp edge, with the line of sight initially outside the jet cone. The peak of the bump corresponds to the epoch when the $1/\Gamma$ cone enters the line of sight, i.e., $(\theta_v-\theta_j) =1/\Gamma$, where $\theta_v$ and $\theta_j$ are the viewing angle and the jet opening angle, respectively. This model predicts that the rising index of the lightcurve is steep, i.e. $\alpha^{\rm re}_1 \sim (3-4)$ (Panaitescu \& Vestrand 2008), and the decaying should correspond to the post jet break phase, which should be steeper than 1.5 (e.g., Liang et al. 2008). As shown in Figure \ref{Onset_RB_Dis}, the $\alpha_1$ values are $\leq 3$ and the $\alpha^{\rm re}_2$ values are shallower than 1.5 the late re-brightening bumps in our sample. All these suggest that the off-axis two-component jet model is not favored by the data of the majority GRBs that show a rebrightening bump, even though with proper contrived parameters, some GRBs can be still interpreted by this model.

The second model invokes a structured jet with smoothly varying energy per solid angle, usually as a power law or Gaussian distribution (e.g. M\'esz\'aros et al. 1998; Zhang \& M\'esz\'aros 2002b; Rossi et al. 2002). For the Gaussian model (e.g. Zhang et al. 2004), an off-axis observer at a large viewing angle would see a smooth rebrightening feature (Kumar \& Granot 2003). However, this model predicts too shallow a hump to
interpret the observed rebrightening bumps in our sample.

The third model invokes a two-component jet that has a fast, narrow jet being encompassed by a slow, wider jet, with the fast one powering the prompt emission and early afterglow onset and the slow one powering the rebrightening bump (e.g. Racusin et al. 2008). This is likely the best model to interpret the rebrightening bumps. Since the rebrightening is interpreted as deceleration of the second slow jet, the rising and decaying slopes and the $t_r/t_d$ ratio should be similar to those of the onset bumps.
This is indeed the case, as shown shown in Figure \ref{Onset_RB_Dis}.
On the other hand, the properties of the re-brightening humps are not correlated with the prompt gamma-ray properties. This suggests that the slow component responsible for the re-brightening bump is likely an independent component that is not related to the prompt gamma-ray emission.

One may estimate the initial Lorentz factor of the slow component if it is decelerated by the same medium as the fast component. The  emission in the re-brightening component should be in the spectral regime $\nu_m < \nu < \nu_c$. The luminosity of the radiation from external shocks at $t$ can be given by (e.g., Sari et al. 1998),
\begin{equation}
L\propto E_{\rm K}^{(p+3)/4}t^{(3-3p)/4},
\end{equation}
where $E_{\rm K}$ is the kinetic energy of the slow component. Assuming that the microphysical parameters in the two components are the same, one can estimate the ratio of $E_{K}$ for the two components
\begin{equation}
\frac{E^{\rm re}_{\rm K}}{E^{\rm on}_{\rm K}}=\left(\frac{L^{\rm re}}{L^{\rm on}}\right)^{4/(p+3)}\left(\frac{t^{\rm re}}{t^{\rm on}}\right)^{3(p-1)/(p+3)}.
\end{equation}
Therefore, the ratio of the initial Lorentz factors of the two jet components can be estimated as
\begin{equation}
\frac{\Gamma^{\rm re}_{0}}{\Gamma^{\rm on}_{0}}=\left(\frac{L^{\rm re}}{L^{\rm on}}\right)^{1/2(p+3)}\left(\frac{t^{\rm re}}{t^{\rm on}}\right)^{3(p-1)/8(p+3)}\left(\frac{t^{\rm on}_{\rm p}}{t^{\rm re}_{\rm p}}\right)^{3/8} ~.
\end{equation}
Taking typical values from our analysis results, i.e., $p=2.5$, $L^{\rm on}_{p}\sim 10^{47}$ erg s$^{-1}$, $L^{\rm re}_{p}\sim 5.8\times 10^{45}$ erg s$^{-1}$, $t^{\rm on}_{\rm p} \sim 200$s and $t^{\rm re}_{\rm p}\sim 5\times 10^4$s, one can get ${E^{\rm re}_{\rm K}}/{E^{\rm on}_{\rm K}}\sim 11.55$ and $\Gamma^{\rm re}_0\sim 34$ for $\Gamma^{\rm on}_{0}\sim 200$. Therefore, the typical kinetic energy of the slow component is about one order of magnitude larger than the fast component but with a much smaller  initial Lorentz factor.

\section{Conclusions}
We have presented a detailed analysis of the optical afterglow onset and late re-brightening bumps with a sample of 146 well-sampled optical lightcurves. We made empirical fits to the lightcurves to identify these bumps and studied the statistical properties of various parameters.  We summarize our results in the following.

\begin{itemize}
\item A smooth onset bump is observed in 38 GRBs. Among them 27 show a clear onset feature followed by a normal decay segment without detection of a late re-brightening bump, while 11 others are followed by a late re-brightening bump. Sixteen GRBs in our sample show a power-law decay segment followed by a late re-brightening hump. No X-ray onset bump was observed to coincide the optical onset bumps in our sample. However, an associated X-ray re-brightening bump was detected for half of the optical re-brightening bumps.

\item The distributions of $\alpha_1$ and $\alpha_2$ of both the onset and re-brighten bumps are consistent with each other. The $\alpha_1$ distribution is in the range of [0.3, 0.4] with a typical value of $\sim 1.5$, while $\alpha_2$ is narrowly distributed in the range [-1.8, -0.6]with a typical value about $-1.15$. The distribution of the ratio $R_{\rm td}$ is clustered around $ 0.4$. The $t^{\rm on}_{\rm p}$ distribution ranges from 30 to 3000 seconds, while $t^{\rm re}_{\rm p}$ is distributed in a much wider range from several hundreds of seconds to days.

\item Both the onset and re-brighten bumps share the same width-peak luminosity relation, i.e., $\log w=(0.50\pm 0.11)+(0.94\pm 0.03) \log t_{\rm p}$, indicating that a bump peaking earlier tends to be narrower. The peak luminosity is anti-correlated with the peak time for both the onset and re-brightening bumps, but the power-law indices of the correlations are different, i.e., $-1.81\pm 0.32$ for the onset bumps and $-0.83\pm 0.17$ for the re-brightening bumps. No correlation between the peak luminosities of the onset and re-brighten bumps is found.

\item It is found that $L^{\rm on}_{\rm R, p}$ is proportional to $E_{\gamma, \rm iso}$, i.e., $\log L^{\rm on}_{\rm R, p,48}=(3.77\pm 0.43)+(1.00\pm 0.14) \log E_{\gamma, \rm iso,50}$, and the isotropic energy release during in the onset bump is also correlated with $E_{\gamma, \rm iso}$. These results indicate that the afterglow onset bumps is likely due to the deceleration of the GRB fireballs.

\item We take the onset bumps as a probe for the properties of the GRB fireball and the circumburst medium density profile. We find that the electron spectral index $p$ is distributed in the range from 2 to 3.5 among bursts, with a typical value of 2.5. The medium density profile is characterized by $n\propto r^{-k}$, with the $k$ values narrowly distributed in the range [0.75, 1.25], with a typical value $k=1$. This profile is intermediate between a constant density ISM and a stellar wind. The physical origin of this profile is subject to further study. With this medium density profile, we re-derive the initial Lorentz factor of the fireball and confirm the $\Gamma_0-E_{\rm iso, \gamma}$ correlation discovered in our previous work (Liang et al. 2010).

\item The peak time and peak luminosity of the rebrightening bumps are not correlated with the properties of the prompt gamma-ray emission and those of the onset bumps. Their rising and decaying slopes are not consistent with the prediction of off-axis two-component jet models. Rather, it seems to be consistent with an on-axis two-component jet model, with the rebrightening bump signaling the slow component. Within this interpretation, the typical kinetic energy of the slow component is about one order of magnitude larger that the fast component that is responsible for the prompt emission and onset afterglow component, for typical values of the observed peak luminosities of the two components. Its initial Lorentz factor is typically only a few tens, being much smaller than that of the fast component.
\end{itemize}

Our results indicate that the optical afterglow is a critical probe for the GRB external shock. Inspecting the X-ray lightcurves in Figures 1-5, one can find that the early X-ray afterglow lightcurves are usually dominated by flares and the tail emission of prompt gamma-rays. The optical afterglow lightcurves are much less contaminated by these emission components. The smooth onset bumps are a clean probe of the key parameters of the fireball and circumburst medium density profile.

\acknowledgments We acknowledge the use of the public data from the Swift data archive. We appreciate helpful suggestions by the referee. This work is supported by the ``973" Program of China (2009CB824800), the National Natural Science Foundation of China (Grants No. 11025313, 11203008, 11078008, 11063001, 11163001,11033002), Special Foundation for Distinguished Expert Program of Guangxi, the Guangxi Natural Science Foundation (2013GXNSFFA019001; 2010GXNSFA013112, 2011GXNSFB018063 and 2010GXNSFC013011), the special funding for national outstanding young scientist (Contract No. 2011-135). BZ acknowledges support from NSF (AST-0908362).

\clearpage
\begin{deluxetable}{lcccccccccccccccccc}
\rotate
\tablewidth{650pt}
\tabletypesize{\tiny}
\tablecaption{Parameters of the Onset Bumps}
\tablenum{1}
\tablehead{
\colhead{GRB(Band)$^{\rm ref}$}
&\colhead{$F_{\rm m}$\tablenotemark{a}}
&\colhead{$\alpha_1$}
&\colhead{$\alpha_2$}
&\colhead{$t_{\rm p}$\tablenotemark{b}}
&\colhead{$w$\tablenotemark{b}}
&\colhead{$t_{\rm r}$\tablenotemark{b}}
&\colhead{$t_{\rm d}$\tablenotemark{b}}
&\colhead{$R_{\rm rd}$}
&\colhead{$R_{\rm rp}$}
&\colhead{$z$}
&\colhead{$E_{\rm \gamma,iso}$\tablenotemark{c}}
&\colhead{$E_{\rm R,iso}$\tablenotemark{d}}
&\colhead{$L_{\rm R,p}$\tablenotemark{e}}
&\colhead{$\Gamma_0$}
&\colhead{$R_{\rm dec}$\tablenotemark{f}}
}
\startdata
030418(V)$^{(1)}$&3.33$\pm$0.07&0.81$\pm$0.12&-0.55$\pm$0.04&1190$\pm$109&6015&926&5089&0.18&0.71&...&...&...&...&...&...\\
050502A(V)$^{(2)}$&55.22$\pm$3.56&0.87&-1.30$\pm$0.07&58$\pm$5&96&34&62&0.55&0.63&$3.793^{(36)}$&...&162.5&5139.5$\pm$331.4&...&...\\
050820A(R)$^{(3)}$&17.38$\pm$0.11&1.26$\pm$0.04&-1.07$\pm$0.02&477$\pm$6&896&258&637&0.41&0.53&$2.612^{(36)}$&15924$\pm$1244&301.6&671.7$\pm$4.3&226&1.99\\
060110(V)$^{(4)}$&125.78$\pm$1.81&1.05$\pm$0.04&-0.77$\pm$0.03&50&132&30&101&0.30&0.58&...&320$\pm$60&...&...&190&0.51\\
060418(H)$^{(5)}$&52.99$\pm$1.51&1.24$\pm$0.14&-1.27$\pm$0.02&170$\pm$5&252&85&167&0.51&0.50&$1.489^{(37)}$&4859$\pm$1056&200.3&606.4$\pm$17.3&252&1.27\\
060605(R)$^{(6)}$&8.65$\pm$0.23&0.78$\pm$0.13&-1.47$\pm$0.04&590$\pm$45&1216&336&880&0.38&0.75&$3.78^{(38)}$&283$\pm$45&362.3&1277.7$\pm$33.2&133&0.62\\
060607A(H)$^{(7)}$&16.65$\pm$0.27&2.39$\pm$0.12&-1.40$\pm$0.02&179$\pm$3&220&65&155&0.42&0.35&$3.082^{(39)}$&2342$\pm$149&202.0&926.9$\pm$15.1&310&1.34\\
060904B(R)$^{(8)}$&8.35$\pm$0.16&2.40$\pm$0.18&-1.76$\pm$0.03&400&616&195&421&0.46&0.45&$0.703^{(40)}$&77$\pm$10&8.6&19.4$\pm$0.4&108&0.88\\
060906(R)$^{(9)}$&0.37$\pm$0.06&0.55$\pm$0.68&-1.14$\pm$0.34&1149$\pm$355&2177&788&1389&0.57&0.76&$3.686^{(41)}$&1727$\pm$139&27.2&23.9$\pm$4.0&120&0.98\\
061007(R)$^{(6)}$&1895.00$\pm$34.75&2.00&-1.67&77$\pm$1&130&41&90&0.45&0.50&$1.261^{(42)}$&42104$\pm$4190&1277.2&14688.1$\pm$269.3&473&2.37\\
061121(V)$^{(10)}$&603.77$\pm$35.17&1.23$\pm$0.29&-1.07$\pm$0.23&208$\pm$25&394&114&280&0.41&0.54&$1.341^{(43)}$&28468$\pm$3272&830.2&4371.4$\pm$254.6&210&1.15\\
070318(V)$^{(11)}$&16.01$\pm$0.52&0.54$\pm$0.11&-1.16$\pm$0.16&507$\pm$46&947&347&600&0.58&0.77&$0.836^{(44)}$&135$\pm$33&28.7&47.0$\pm$1.5&84&0.53\\
070411(R)$^{(12)}$&2.75$\pm$0.05&0.61&-1.45$\pm$0.02&739$\pm$10&1196&473&723&0.65&0.74&$2.954^{(40)}$&1000$\pm$200&79.8&145.6$\pm$2.4&129&0.87\\
070419A(R)$^{(13)}$&0.45$\pm$0.02&0.93$\pm$0.12&-1.26$\pm$0.03&765$\pm$30&1212&437&775&0.56&0.59&$0.97^{(40)}$&19$\pm$2&1.8&1.9$\pm$0.1&62&0.43\\
070420(R)$^{(14)}$&14.63$\pm$1.17&1.29$\pm$0.48&-0.90$\pm$0.08&202$\pm$22&424&107&317&0.34&0.51&...&3100$\pm$500&...&...&157&1.45\\
071010A(R)$^{(15)}$&2.51$\pm$0.64&1.50&-1.14$\pm$0.04&586$\pm$66&996&289&707&0.41&0.48&$0.98^{(40)}$&13$\pm$2&8.3&10.0$\pm$2.5&72&0.45\\
071010B(R)$^{(16)}$&2.72$\pm$0.27&0.34$\pm$0.19&-0.60$\pm$0.06&287$\pm$145&1296&215&1081&0.20&0.91&$0.947^{(45)}$&174$\pm$90&6.4&6.4$\pm$0.6&...&...\\
071025(J)$^{(17)}$&5.75$\pm$0.17&1.17$\pm$0.15&-1.03$\pm$0.04&548$\pm$29&1090&309&781&0.40&0.55&$5.2^{(40)}$&1500$\pm$300&...&...&193&0.95\\
071031(R)$^{(18)}$&0.72$\pm$0.0007&0.63$\pm$0.002&-0.84$\pm$0.0007&1213$\pm$2&3312&853&2459&0.35&0.75&$2.692^{(40)}$&390$\pm$60&40.9&26.9$\pm$0.025&89&0.71\\
071112C(R)$^{(19)}$&2.58$\pm$0.14&1.64$\pm$1.18&-0.88$\pm$0.02&178$\pm$13&396&91&306&0.30&0.47&$0.823^{(46)}$&...&2.1&6.7$\pm$0.4&...&...\\
080310(R)$^{(20)}$&1.87$\pm$0.11&1.50&-1.16$\pm$0.26&184$\pm$12&286&85&200&0.43&0.45&$2.4266^{(40)}$&590$\pm$100&9.9&42.3$\pm$2.5&219&0.77\\
080319A(R)$^{(21)}$&0.18$\pm$0.02&1.80&-0.65$\pm$0.07&238$\pm$17&784&127&657&0.19&0.46&...&800$\pm$100&...&...&...&...\\
080330(R)$^{(22)}$&1.26$\pm$0.04&0.33&-1.12$\pm$0.25&578$\pm$25&1207&395&811&0.49&0.91&$1.51^{(40)}$&41$\pm$6&10.7&11.4$\pm$0.4&74&0.33\\
080603A(R)$^{(23)}$&0.57$\pm$0.03&1.82$\pm$0.21&-0.99$\pm$0.09&1044$\pm$167&6296&1130&5167&0.22&0.70&$1.67842^{(47)}$&...&...&...&95&1.08\\
080710(R)$^{(19)}$&2.59$\pm$0.04&1.60$\pm$0.07&-1.38$\pm$0.06&1934$\pm$46&4212&1174&3038&0.39&0.58&$0.845^{(40)}$&80$\pm$40&23.8&7.9$\pm$0.1&57&1.03\\
080810(R)$^{(24)}$&109.00$\pm$1.71&1.26$\pm$0.04&-1.21$\pm$0.003&117$\pm$2&308&78&230&0.34&0.65&$3.35^{(40)}$&3000$\pm$2000&836.7&4823.2$\pm$75.4&337&0.89\\
081008(R)$^{(26)}$&6.12$\pm$0.08&2.84$\pm$0.16&-0.96$\pm$0.004&163$\pm$2&290&59&230&0.26&0.33&$1.967^{(40)}$&...&16.9&127.0$\pm$1.6&250&1.12\\
081109A(H)$^{(27)}$&1.04$\pm$0.13&0.19$\pm$0.18&-0.94$\pm$0.03&559$\pm$127&1348&344&1005&0.34&0.98&$0.98^{(40)}$&530$\pm$80&...&...&68&0.68\\
081126(R)$^{(28)}$&12.30$\pm$0.04&1.14$\pm$0.02&-0.39$\pm$0.01&159$\pm$2&1450&127&1323&0.10&0.63&...&900$\pm$200&...&...&...&...\\
081203A(U)$^{(29)}$&146.60$\pm$0.30&2.58$\pm$0.02&-1.61$\pm$0.004&295$\pm$2&766&205&561&0.36&0.55&$2.1^{(40)}$&1700$\pm$400&904.0&3022.3$\pm$6.2&226&1.57\\
090102(R)$^{(30)}$&59.05$\pm$1.98&5.22&-1.57$\pm$0.04&50$\pm$1&45&11&34&0.31&0.20&$1.547^{(40)}$&1400$\pm$500&26.6&711.8$\pm$23.8&61&0.84\\
090313(R)$^{(31)}$&6.75$\pm$0.43&1.23$\pm$0.16&-1.25$\pm$0.09&1315$\pm$109&2126&690&1436&0.48&0.53&$3.375^{(40)}$&460$\pm$50&445.7&451.7$\pm$28.9&107&1.02\\
090510(R)$^{(32)}$&0.03$\pm$0.003&0.47$\pm$0.14&-0.98$\pm$0.12&1579$\pm$650&5046&853&4194&0.20&0.90&$0.903^{(49)}$&42$\pm$4&0.3&0.1$\pm$0.01&61&0.84\\
090812(R)$^{(33)}$&14.21$\pm$0.87&1.36$\pm$0.32&-1.37$\pm$0.29&71$\pm$8&104&35&69&0.51&0.49&$2.452^{(40)}$&4586$\pm$597&27.6&305.4$\pm$18.8&399&0.98\\
100901A(R)$^{(34)}$&1.45$\pm$0.04&1.87$\pm$0.31&-1.00$\pm$0.10&1260$\pm$76&2349&577&1772&0.33&0.43&$1.408^{(50)}$&245&22.2&14.2$\pm$0.4&87&1.21\\
100906A(V)$^{(34)}$&95.73$\pm$3.06&3.40$\pm$0.23&-1.07$\pm$0.02&101$\pm$4&245&55&189&0.29&0.43&$1.727^{(51)}$&...&124.2&936.2$\pm$30.0&...&...\\
110205A(R)$^{(35)}$&25.24$\pm$0.11&4.00&-1.47$\pm$0.00&948$\pm$3&971&246&725&0.34&0.24&...&...&...&...&...&...\\
110213A(V)$^{(35)}$&53.19$\pm$2.12&1.54$\pm$0.09&-0.91$\pm$0.08&293$\pm$12&641&153&488&0.31&0.48&$1.46^{(52)}$&...&...&...&165&0.97\\
\enddata
\tablenotetext{a}{In units of $10^{-12}$ erg cm$^{-2}$s$^{-1}$.}
\tablenotetext{b}{In units of seconds.}
\tablenotetext{c}{In units of $10^{50}$ erg.}
\tablenotetext{d}{In units of $10^{48}$ erg.}
\tablenotetext{e}{In units of $10^{45}$ erg s$^{-1}$.}
\tablenotetext{f}{In units of $10^{17}$ cm .}
\tablerefs{
(1)  Rykoff et al. 2004;
(2)  Yost   et al.  2006;
(3)  Cenko  et al. 2006;
(4)  Cenko  et al. 2008;
(5)  Molinari  et al. 2007;
(6)  Rykoff   et al. 2009 ;
(7)  Molinari   et al. 2007;
(8)  Klotz   et al. 2008;
(9)  Rana et al.  2009;
(10) Uehara  et al. 2011;
(11)  Chester et al. 2008 ;
(12)  Ferrero  et al. 2008;
(13)  Melandri   et al. 2009;
(14)  Klotz  et al 2008. ;
(15)  Covino   et al. 2008;
(16)  Huang   et al.2009;
(17)  Perley   et al. 2009 ;
(18)  Kr\"{u}hler    et al. 2009;
(19)  Huang     et al. 2009;
(20)  Littlejohns   et al.2012;
(21)  Malesani et al. 2008 ;
(22)  Guidorzi et al.2009;
(23)  Guidorzi et al.2011;
(24)  Page  et al.2009;
(25)  Rossi  et al.2010;
(26)  Yuan  et al.2010;
(27)   Jin et al.2009;
(28)   Klotz et al.2009;
(29)   Kuin et al.2009;
(30)   Gendre  et al.2010;
(31)   Melandri   et al.2010;
(32)   Pelassa   et al.2010;
(33)   Wren  et al. 2009(GCN 9778);
(34)   Gorbovskoy et al.2011;
(35)   Cucchiara et al.2011;
(36)   Liang \& Zhang. 2006;
(37)   Prochaska et al.2007;
(38)   Ferreroet al. 2009;
(39)  Fynbo et al. 2009;
 (40)  Robertson \&  Ellis 2012;
 (41)  Zafar  et al. 2011;
 (42)  Schady   et al. 2008;
 (43)  Golenetskii et al. 2006;
 (44)  Krimm et al. 2009;
 (45)  Golenetskii  et al. 2007;
 (46)  Kann  et al. 2010;
 (47)  Guidorzi  et al. 2011;
 (48)  Zafar  et al. 2011;
 (49)  He, Hao-Ning et al.  2011;
 (50)  Sakamoto  et al. 2010;
 (51)  Gorbovskoy et al. 2012;
 (52)  Cucchiara et al. 2011.
}
\end{deluxetable}

\clearpage
\begin{deluxetable}{lccccccccccccccccc}
\tablewidth{600pt} \tabletypesize{\tiny}
\rotate
\tablecaption{Properties of the optical rebrightening and prompt gamma-rays in our sample}
\tablenum{2}
\tablehead{
\colhead{GRB(Band)}
&\colhead{$F_{\rm m}$\tablenotemark{a}}
&\colhead{$\alpha_1$}
&\colhead{$\alpha_2$}
&\colhead{$t_{\rm p}$\tablenotemark{b}}
&\colhead{$w$\tablenotemark{b}}
&\colhead{$t_{\rm r}$\tablenotemark{b}}
&\colhead{$t_{\rm d}$\tablenotemark{b}}
&\colhead{$R_{\rm rd}$}
&\colhead{$R_{\rm rp}$}
&\colhead{$z$}
&\colhead{$E_{\rm \gamma,iso}$\tablenotemark{c}}
&\colhead{$E_{\rm R,iso}$\tablenotemark{d}}
&\colhead{$L_{\rm R,p}$\tablenotemark{e}}
}
\startdata
021211(R)$^{(1)}$&0.28$\pm$0.03&1.80&-0.40$\pm$0.06&0.50&3.88&0.31&3.56&0.09&0.50&$1.01^{(28)}$&111$\pm$10&1.4&1.21$\pm$0.14\\
021004(R)$^{(2)}$&0.09$\pm$0.01&2.50&-1.27$\pm$0.03&170.31$\pm$5.31&228.51&62.01&166.49&0.37&0.34&$2.335^{(29)}$&500$\pm$115&373.5&1.80$\pm$0.10\\
030329(R)$^{(3)}$&1.66$\pm$0.20&2.04$\pm$0.54&-1.35$\pm$0.03&198.55$\pm$1.82&265.42&80.05&185.37&0.43&0.39&$0.17^{(30)}$&155$\pm$14&43.1&0.12$\pm$0.01\\
050502A(V)$^{(4)}$&0.34$\pm$0.14&0.80&-1.40$\pm$0.33&2.76$\pm$0.72&4.40&1.65&2.74&0.60&0.65&$3.793^{(31)}$&...&46.3&31.90$\pm$13.03\\
050721(R)$^{(5)}$&0.04$\pm$0.001&0.50&-1.06&98.86$\pm$7.85&214.89&67.97&146.92&0.46&0.81&...&460$\pm$90&0&...\\
050820A(R)$^{(6)}$&0.39$\pm$0.02&0.86$\pm$0.27&-1.05$\pm$0.02&19.39$\pm$1.01&39.87&12.09&27.78&0.44&0.65&$2.612^{(32)}$&15924$\pm$1244&287.2&15.00$\pm$0.85\\
060526(R)$^{(7)}$&0.06$\pm$0.002&1.60&-2.28&112.46$\pm$1.89&104.72&45.73&58.99&0.78&0.42&$3.21^{(33)}$&606$\pm$303&183.7&2.64$\pm$0.08\\
060729(U)$^{(8)}$&4.76$\pm$0.15&1.20$\pm$0.28&-1.37$\pm$0.06&30.04$\pm$2.67&67.93&18.61&49.32&0.38&0.65&$0.54^{(29)}$&65$\pm$5&167.8&5.00$\pm$0.16\\
060904B(R)$^{(9)}$&1.66$\pm$0.03&8.00&-0.98$\pm$0.02&1.77$\pm$0.02&3.46&0.60&2.86&0.21&0.27&$0.703^{(30)}$&77$\pm$10&12.3&3.85$\pm$0.08\\
060906(R)$^{(10)}$&0.29$\pm$0.02&2.91$\pm$0.91&-1.20&9.72$\pm$0.54&13.32&3.28&10.04&0.33&0.31&$3.686^{(29)}$&1727$\pm$139&151.3&18.39$\pm$1.46\\
060927(V)$^{(11)}$&$4.20$&1.40&-1.20&0.48&0.79&0.24&0.55&0.44&0.49&$5.6^{(29)}$&5815$\pm$862&183.8&1095.82\\
070318(V)$^{(12)}$&0.26$\pm$0.66&0.95$\pm$5.14&-1.60&52.82$\pm$104.81&73.03&29.10&43.93&0.66&0.59&$0.836^{(31)}$&135$\pm$33&38.9&0.77$\pm$1.94\\
071003(R)$^{(13)}$&0.99$\pm$0.03&1.00&-1.10&20.00&38.31&11.75&26.56&0.44&0.60&$1.605^{(32)}$&1800$\pm$600&306.1&21.21$\pm$0.55\\
080310(R)$^{(14)}$&2.08$\pm$0.12&0.51&-1.07&1.72$\pm$0.06&3.48&1.21&2.28&0.53&0.79&$2.4266^{(35)}$&590$\pm$100&119.0&47.09$\pm$2.67\\
080319C(V)$^{(15)}$&2.93$\pm$0.23&3.40$\pm$0.61&-0.94$\pm$0.05&0.28$\pm$0.01&0.56&0.11&0.45&0.25&0.34&$1.949^{(29)}$&5206$\pm$1041&16.5&62.05$\pm$4.89\\
080330(R)$^{(16)}$&0.68$\pm$0.15&1.20&-1.14$\pm$0.15&1.77$\pm$0.10&3.15&0.96&2.19&0.44&0.54&$1.51^{(35)}$&41$\pm$6&15.8&6.12$\pm$1.34\\
080413A(R)$^{(17)}$&23.18$\pm$1.80&1.00&-1.21$\pm$0.07&0.15&0.26&0.09&0.17&0.49&0.59&$2.433^{(35)}$&1855$\pm$397&105.6&719.01$\pm$55.90\\
080710(R)$^{(18)}$&0.79$\pm$0.05&0.32$\pm$0.05&-1.58$\pm$0.08&8.63$\pm$0.19&13.96&5.90&8.06&0.73&0.90&$0.845^{(35)}$&80$\pm$40&26.4&2.42$\pm$0.14\\
080913(J)$^{(19)}$&0.004$\pm$0.001&2.30&-0.95&78.93$\pm$33.97&173.07&38.53&134.54&0.29&0.43&$6.7^{(35)}$&711$\pm$89&51.3&1.22$\pm$0.48\\
081029(R)$^{(21)}$&0.77$\pm$0.19&1.77&-1.94$\pm$0.11&14.35$\pm$0.89&14.57&5.71&8.86&0.64&0.40&$3.85^{(33)}$&...&462.7&108.26$\pm$26.65\\
090102(R)$^{(22)}$&0.09$\pm$0.01&0.80&-0.96$\pm$0.02&5.73$\pm$0.40&13.19&3.73&9.46&0.39&0.67&$1.547^{(35)}$&1400$\pm$500&8.2&1.08$\pm$0.12\\
090902B(R)$^{(23)}$&0.02$\pm$0.002&1.20&-0.81&137.28$\pm$22.35&360.45&82.29&278.15&0.30&0.56&$2.452^{(34)}$&4586$\pm$597&77.5&0.31$\pm$0.05\\
100219A(R)$^{(24)}$&0.03$\pm$0.003&0.91&-2.24&20.47$\pm$1.08&22.71&10.93&11.79&0.93&0.59&$4.6667^{(35)}$&...&34.6&1.17$\pm$0.10\\
100901A(R)$^{(25)}$&1.63$\pm$0.02&2.50$\pm$0.13&-1.50&24.16$\pm$0.60&31.58&9.57&22.01&0.43&0.37&$1.408^{(36)}$&245&371.2&8.26$\pm$0.12\\
101024A(R)$^{(26)}$&0.16$\pm$0.07&0.30&-1.09$\pm$0.11&2.71$\pm$1.31&5.78&1.84&3.94&0.47&0.92&...&...&...&...\\
110213A(V)$^{(27)}$&12.06$\pm$1.16&1.12$\pm$0.19&-1.72$\pm$0.55&6.14$\pm$0.72&7.72&3.12&4.60&0.68&0.53&$1.46^{(37)}$&...&685.7&65.34$\pm$6.28
\enddata
\tablenotetext{a}{In units of $10^{-12}$ erg cm$^{-2}$s$^{-1}$.}
\tablenotetext{b}{In units of kilo seconds.}
\tablenotetext{c}{In units of $10^{50}$ erg.}
\tablenotetext{d}{In units of $10^{48}$ erg.}
\tablenotetext{e}{In units of $10^{45}$ erg s $^{-1}$.}
\tablerefs{
(1)  Li et al.2003;
(2)  Lazzati  et al. 2003;
(3)  Resmi et al. 2005;
(4)  Yost   et al. 2006;
(5)  Antonelli  et al. 2006;
(6)  Cenko  et al. 2006;
(7)  Th$\ddot{o}$ne  et al.2010;
(8)  Grupeet et al. 2007 ;
(9)  Klotz  et al.2008;
(10) Rana et al. 2009 ;
(11) Ruiz-Velasco et al. 2007;
(12) Chester et al. 2008 ;
(13) Perley   et al. 2008;
(14)  Littlejohns   et al. 2012;
(15)  Li, W.   et al.(2008);
(16)  Guidorzi et al. 2009  ;
(17)  Yuan  et al. 2008 ;
(18)  Kr\"{u}hler et al. 2009 ;
(19)  Greiner  et al. 2009 ;
(20)  Rossi et al. 2010 ;
(21)  Nardini et al. 2011 ;
(22)  Gendre  et al. 2010 ;
(23)  Cenko et al. 2011;
(24)  Mao et al. 2011;
(25)  Gorbovskoy  et al. 2011;
(26)  Laas-Bourez  et al.(2010);
(27)  Cucchiara et al. 2011;
(28)  Liang \& Zhang 2006
(29)  Zafar et al. 2011
(30) Robertson \& Ellis 2012
(31) Krimm  et al. 2009
(32) Kr\"{u}hler et al. 2009
(33) Nardini, M.et al. 2011
(34) Pandey et al. 2010
(35) Mao et al. 2012
(36) Sakamoto et al. 2010
(37) Cucchiara  et al. 2011
}
\end{deluxetable}

\clearpage
\begin{figure*}
\includegraphics[angle=0,scale=0.35]{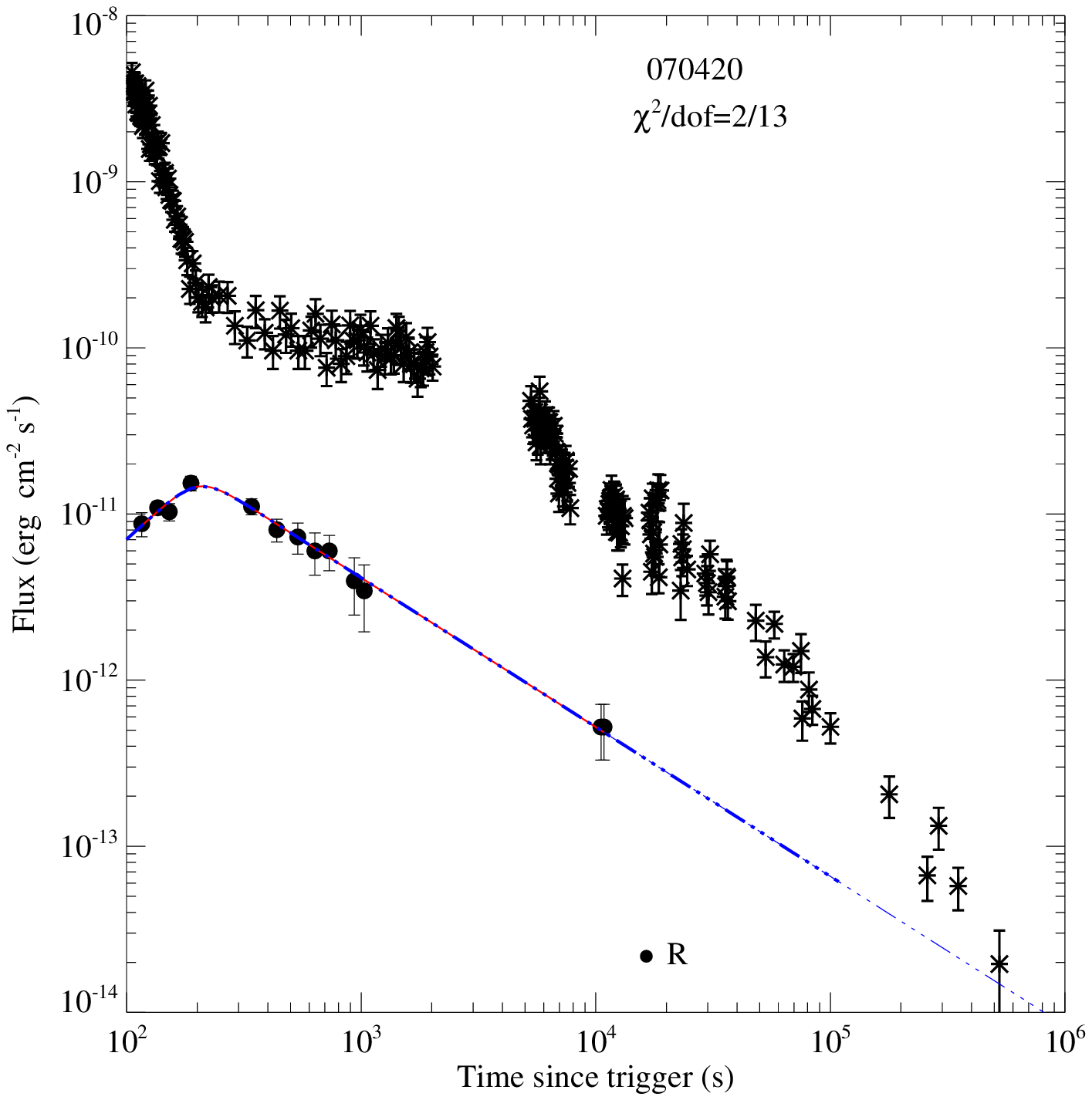}
\includegraphics[angle=0,scale=0.35]{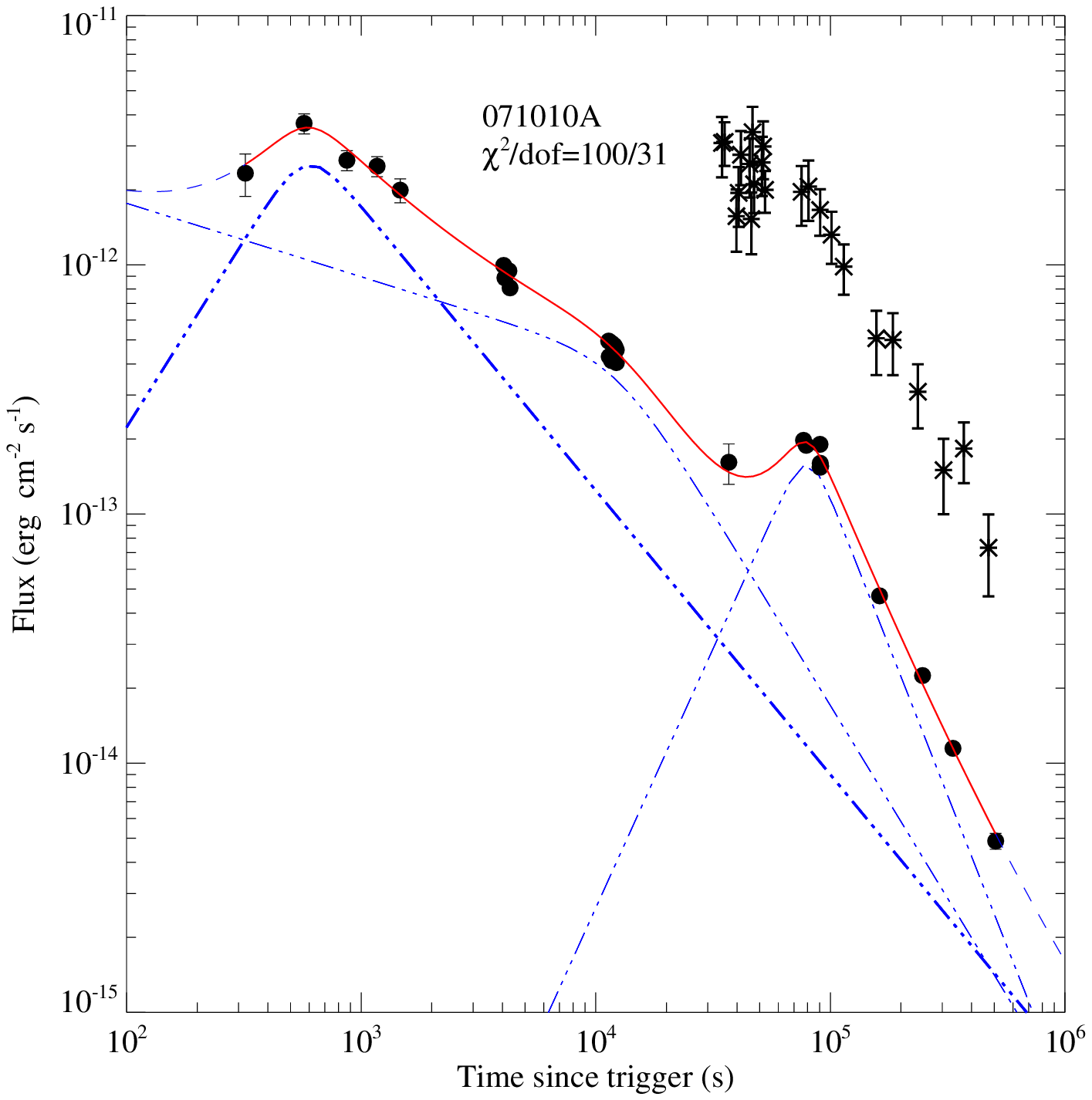}
\includegraphics[angle=0,scale=0.35]{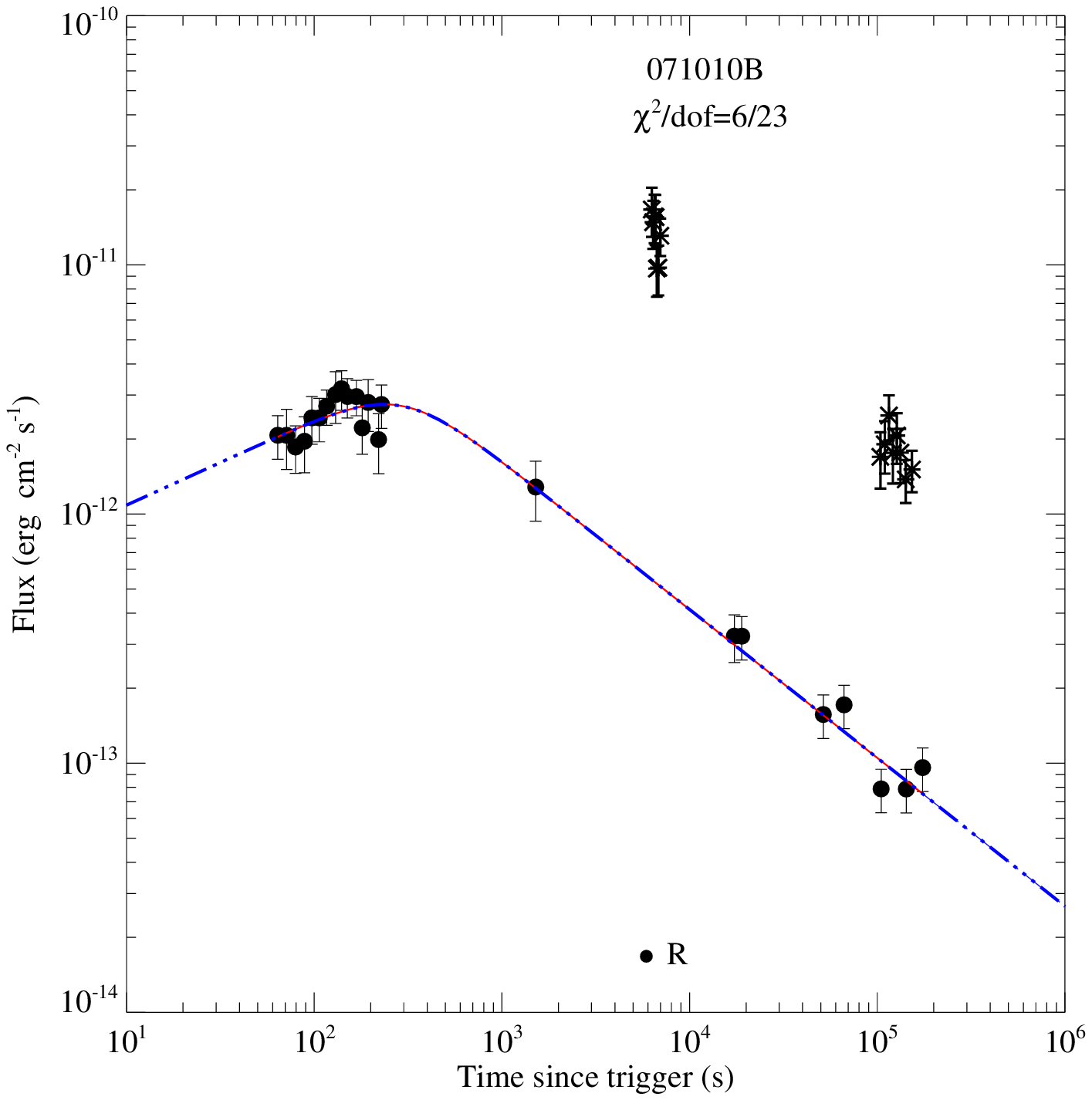}
\includegraphics[angle=0,scale=0.35]{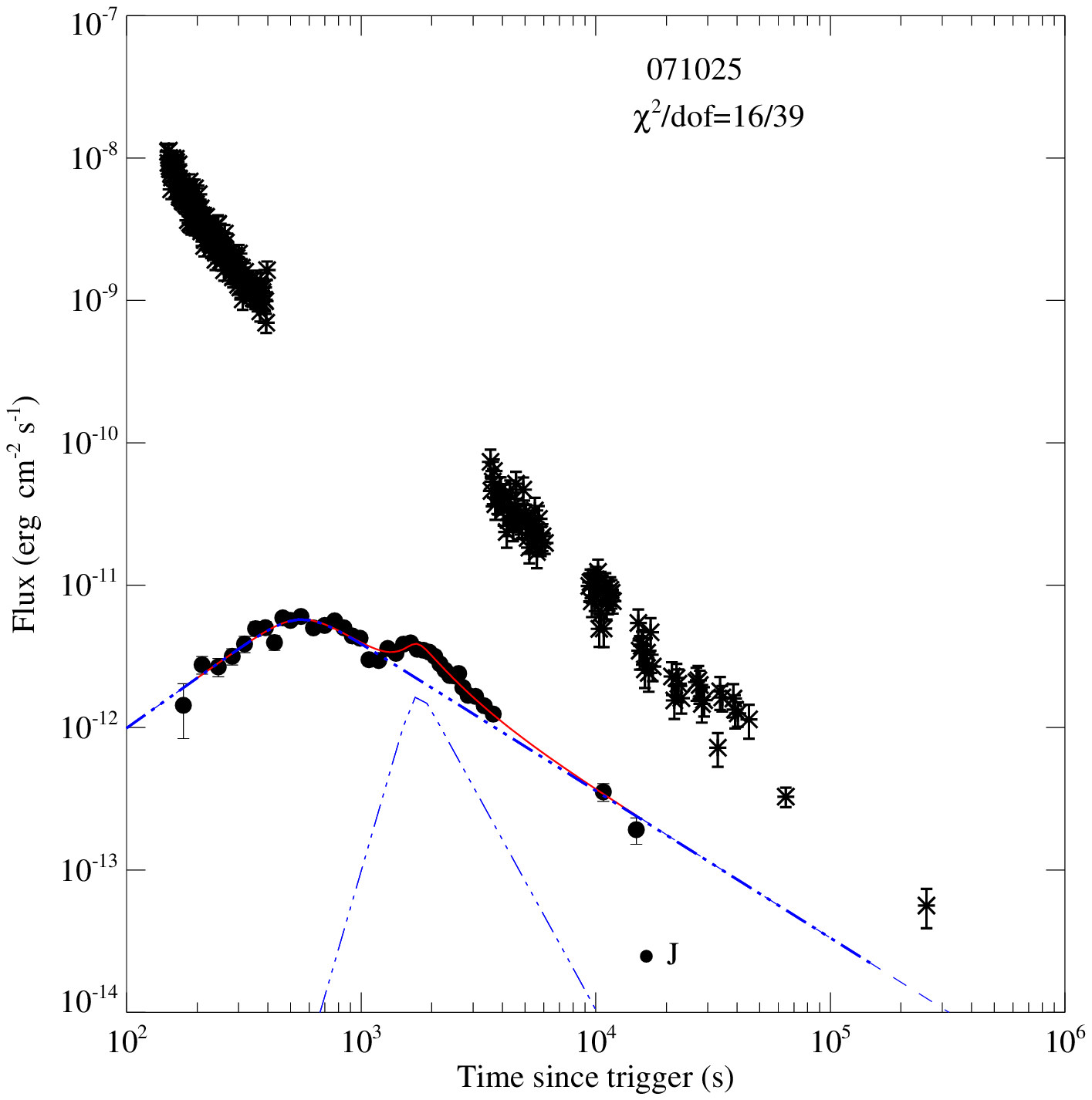}
\includegraphics[angle=0,scale=0.35]{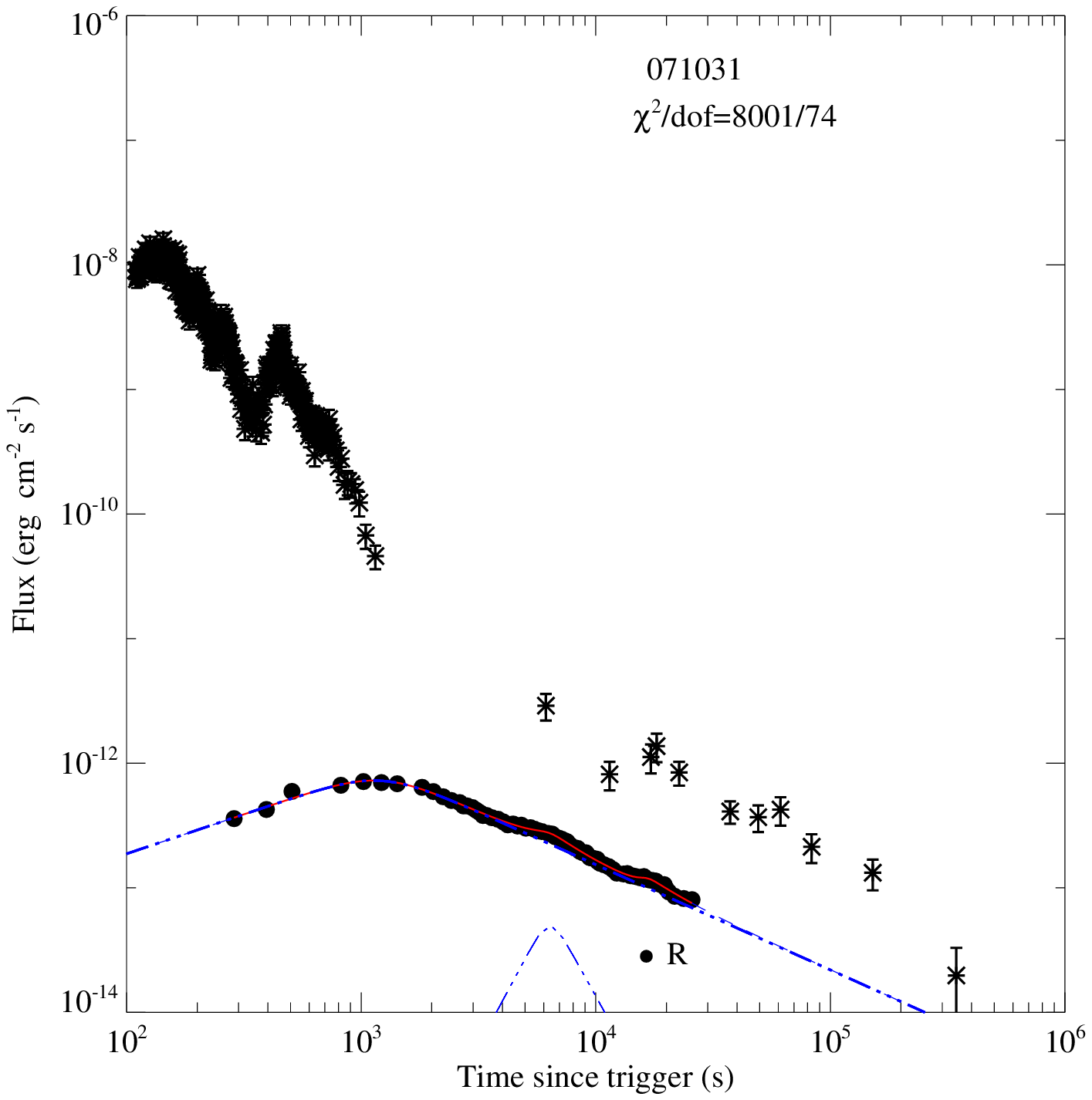}
\includegraphics[angle=0,scale=0.35]{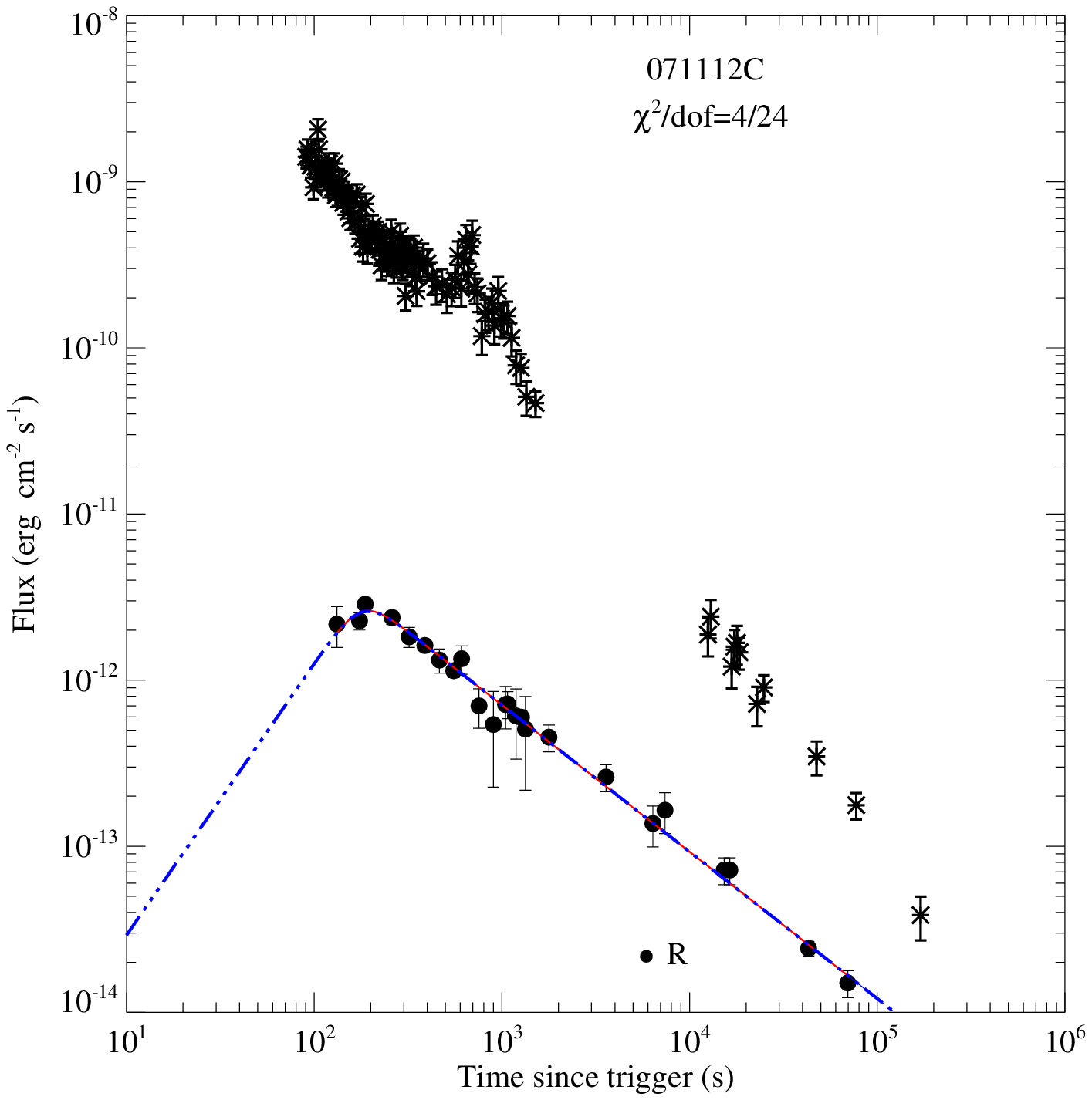}
\includegraphics[angle=0,scale=0.35]{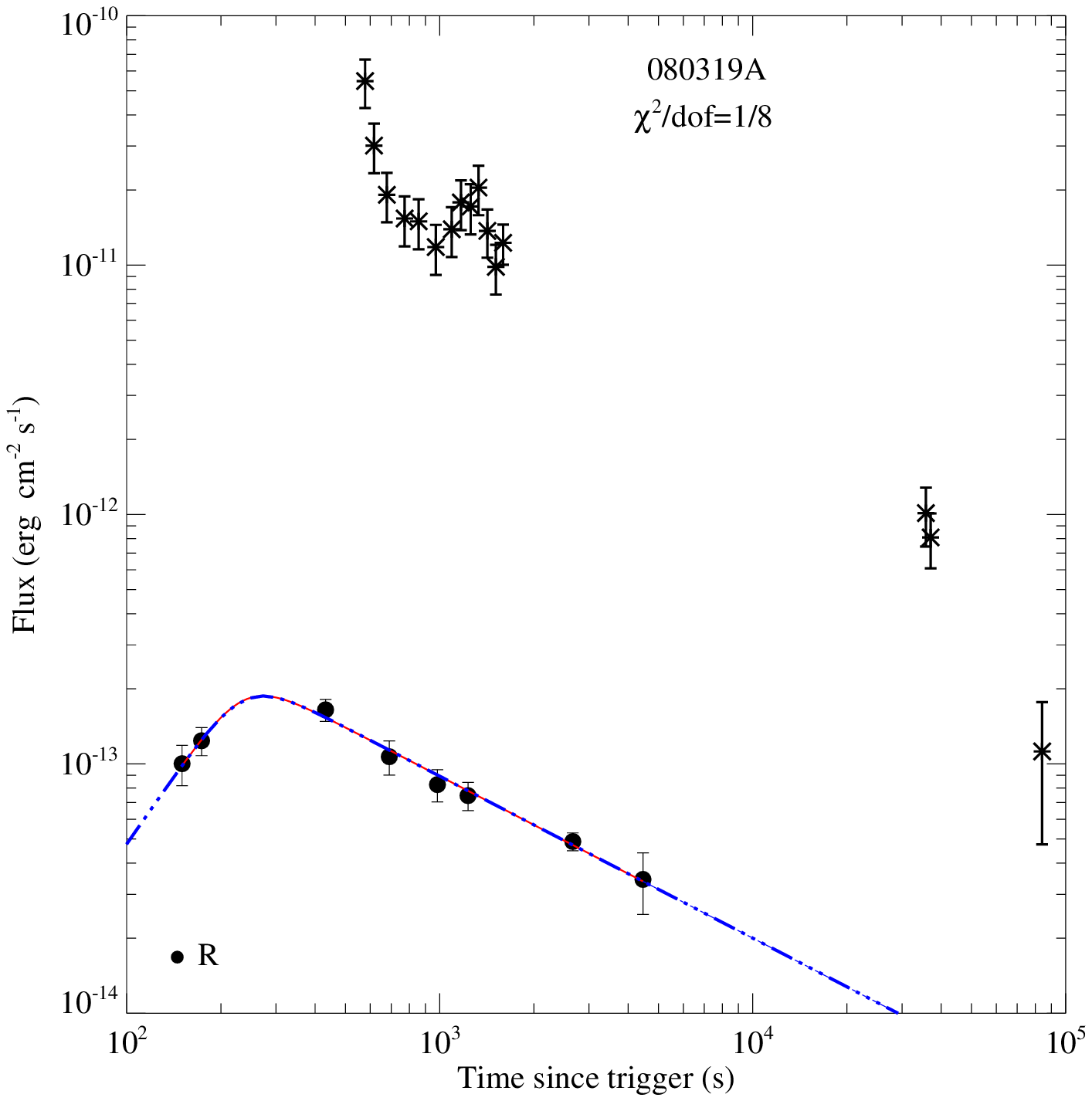}
\includegraphics[angle=0,scale=0.35]{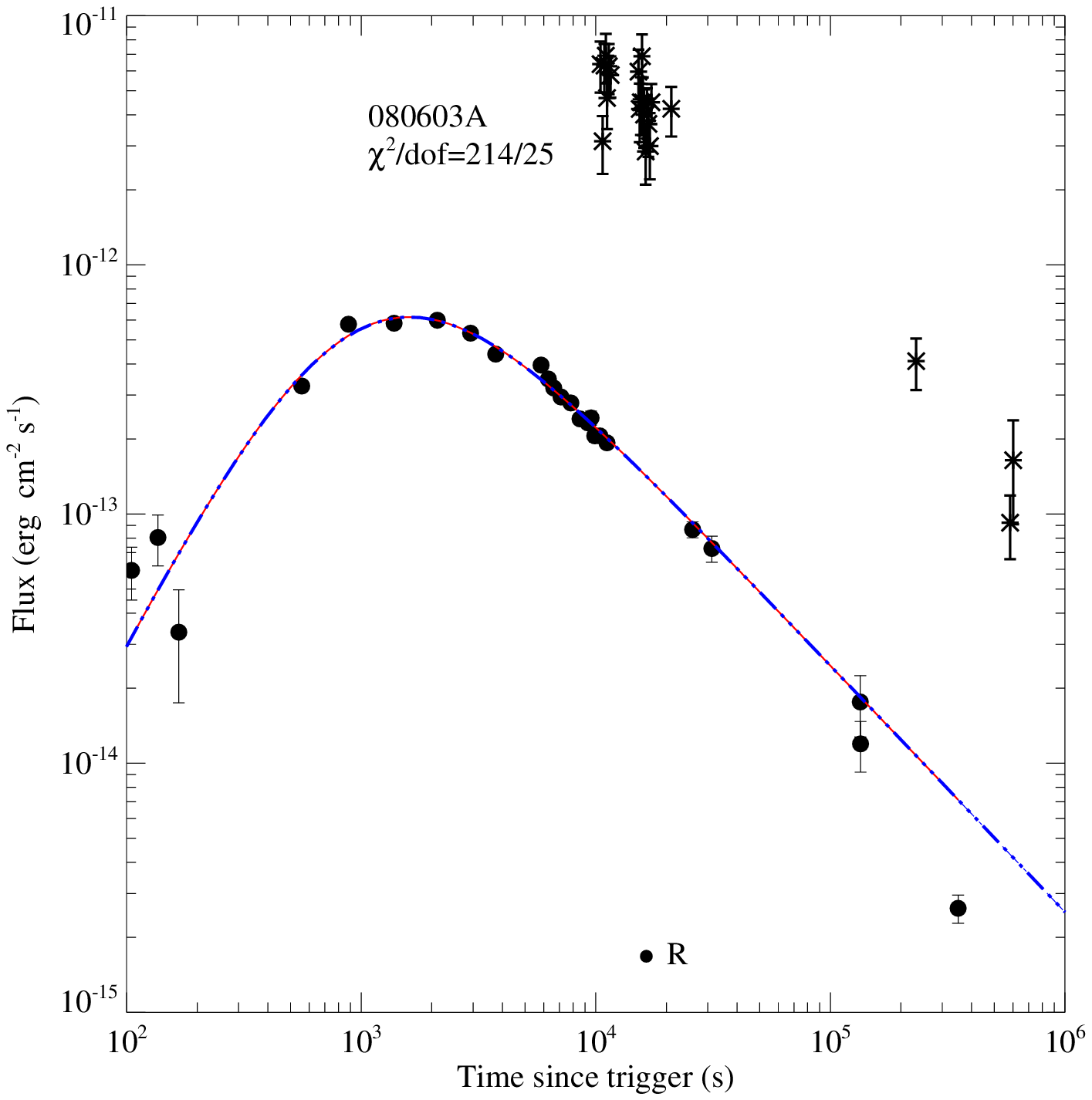}
\includegraphics[angle=0,scale=0.35]{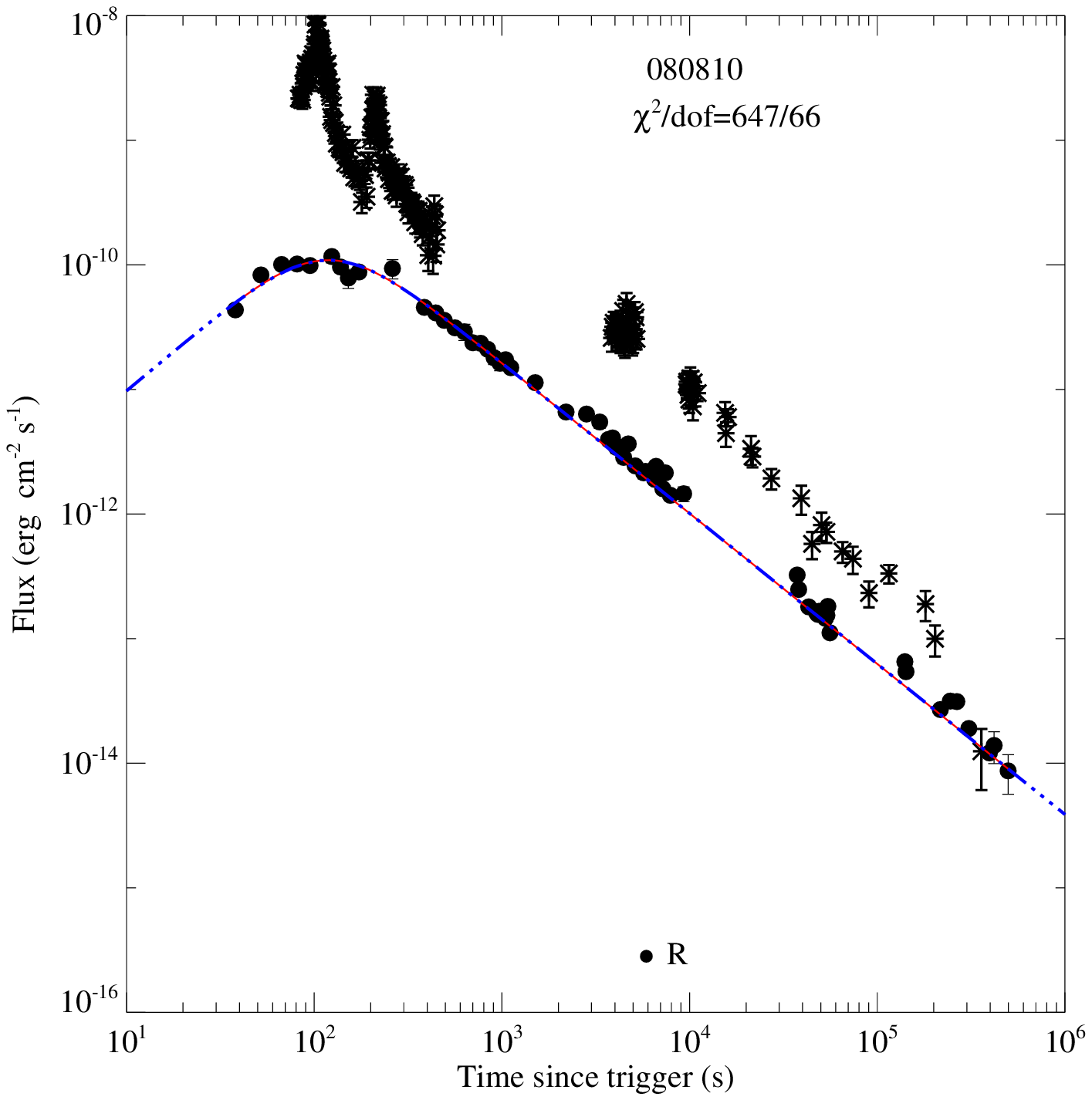}
\includegraphics[angle=0,scale=0.35]{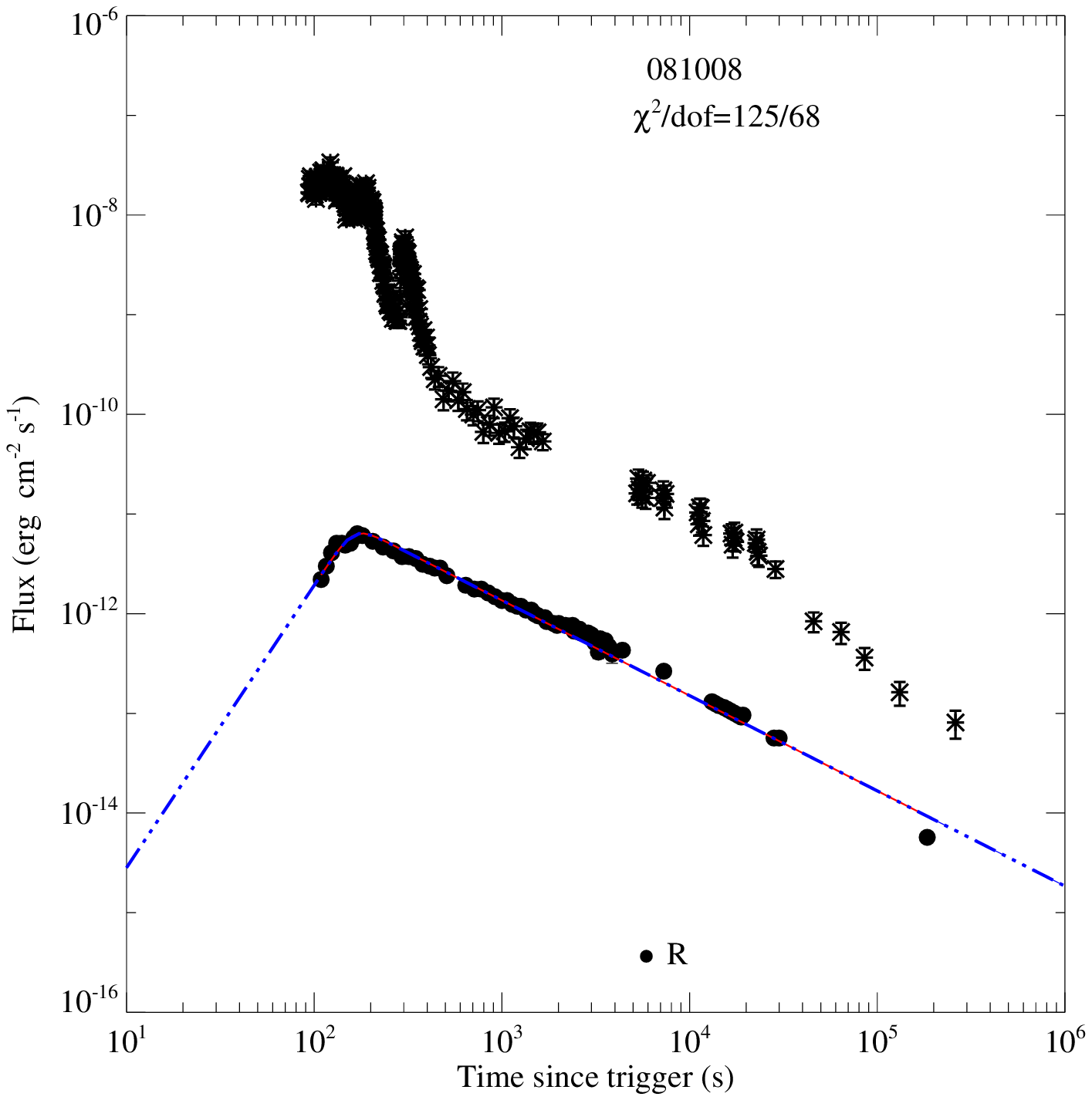}
\includegraphics[angle=0,scale=0.35]{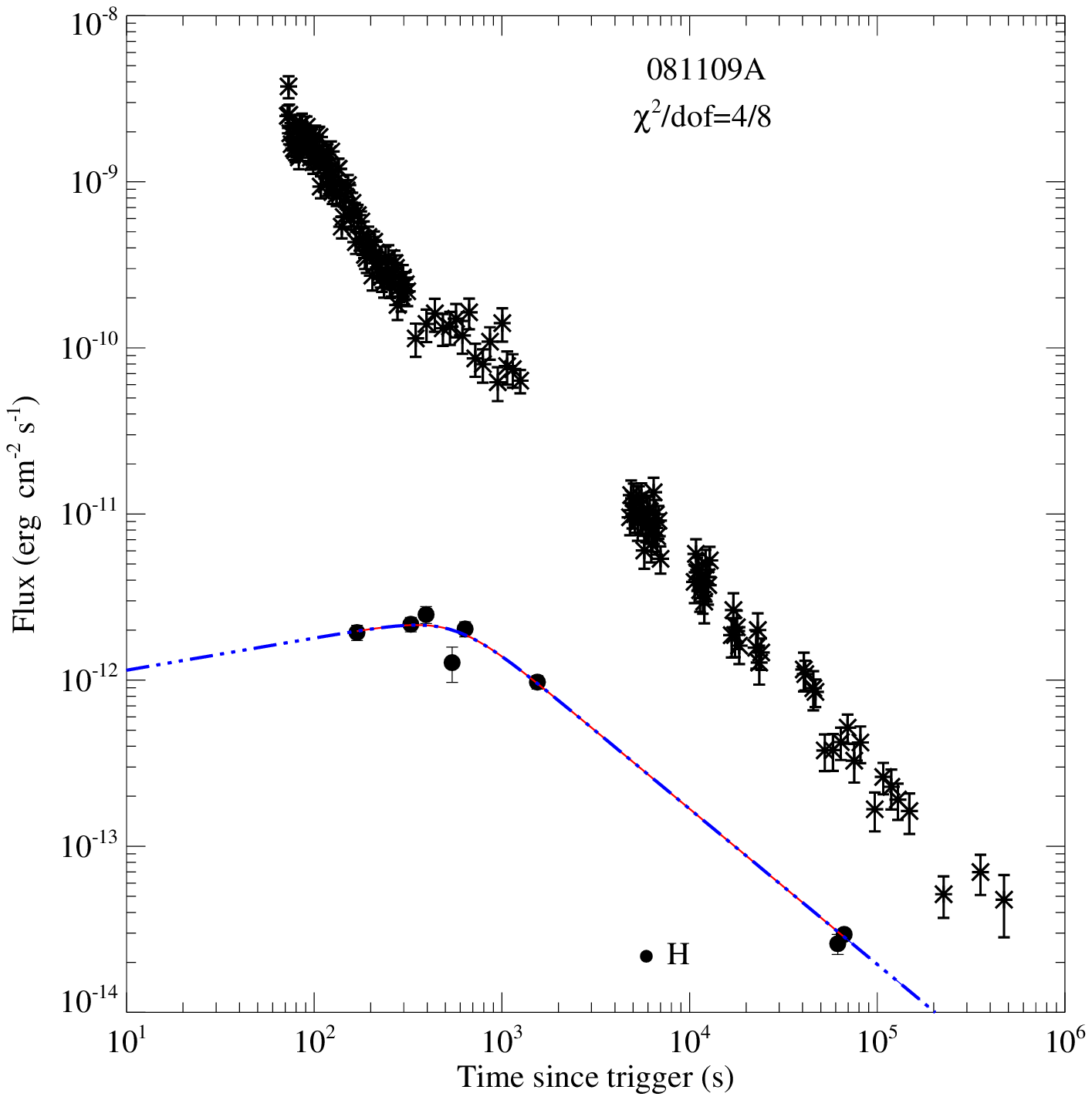}\hfill
\includegraphics[angle=0,scale=0.35]{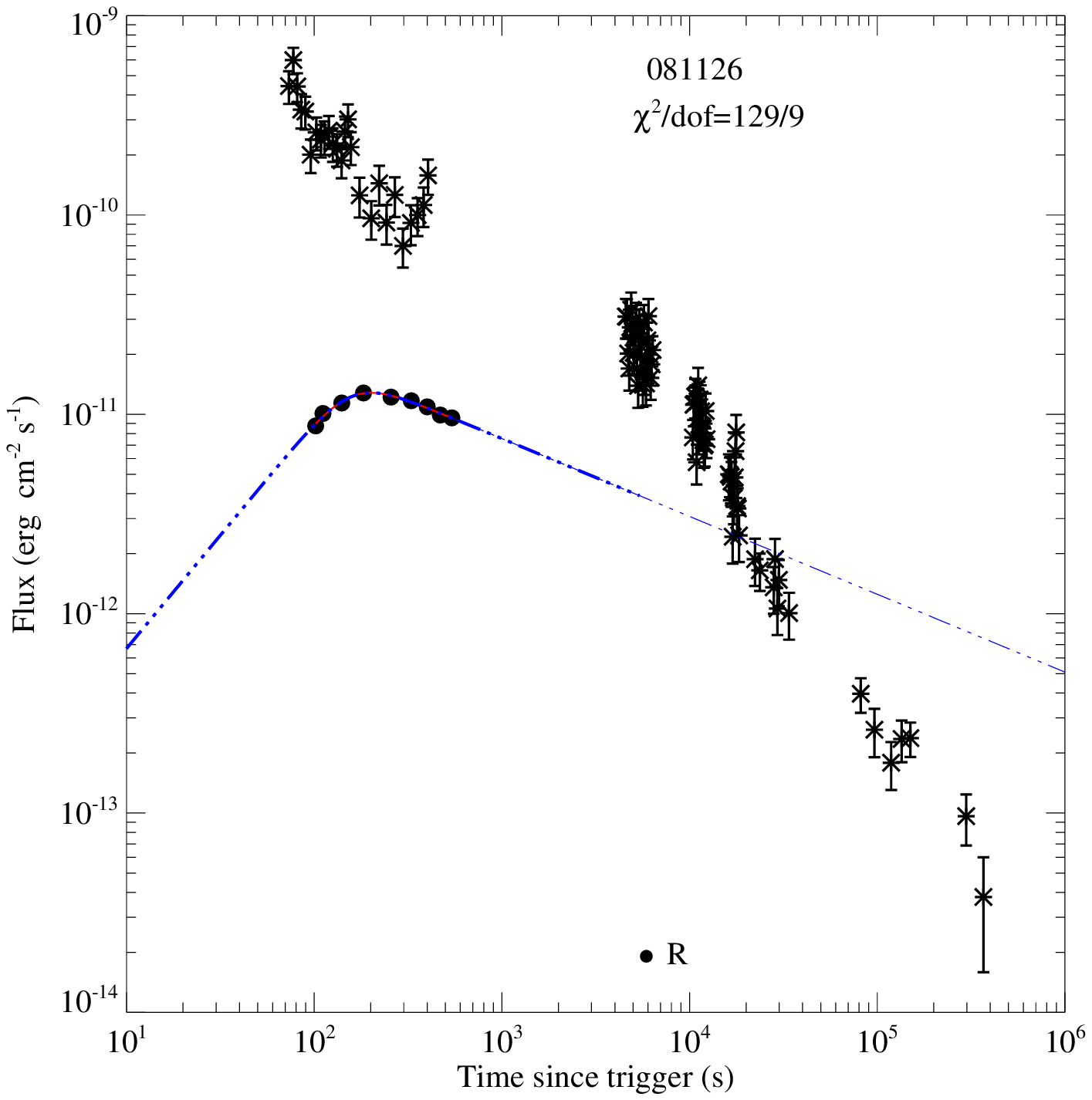}
\center{Fig. 1--- Continued}
\end{figure*}
\begin{figure*}
\includegraphics[angle=0,scale=0.35]{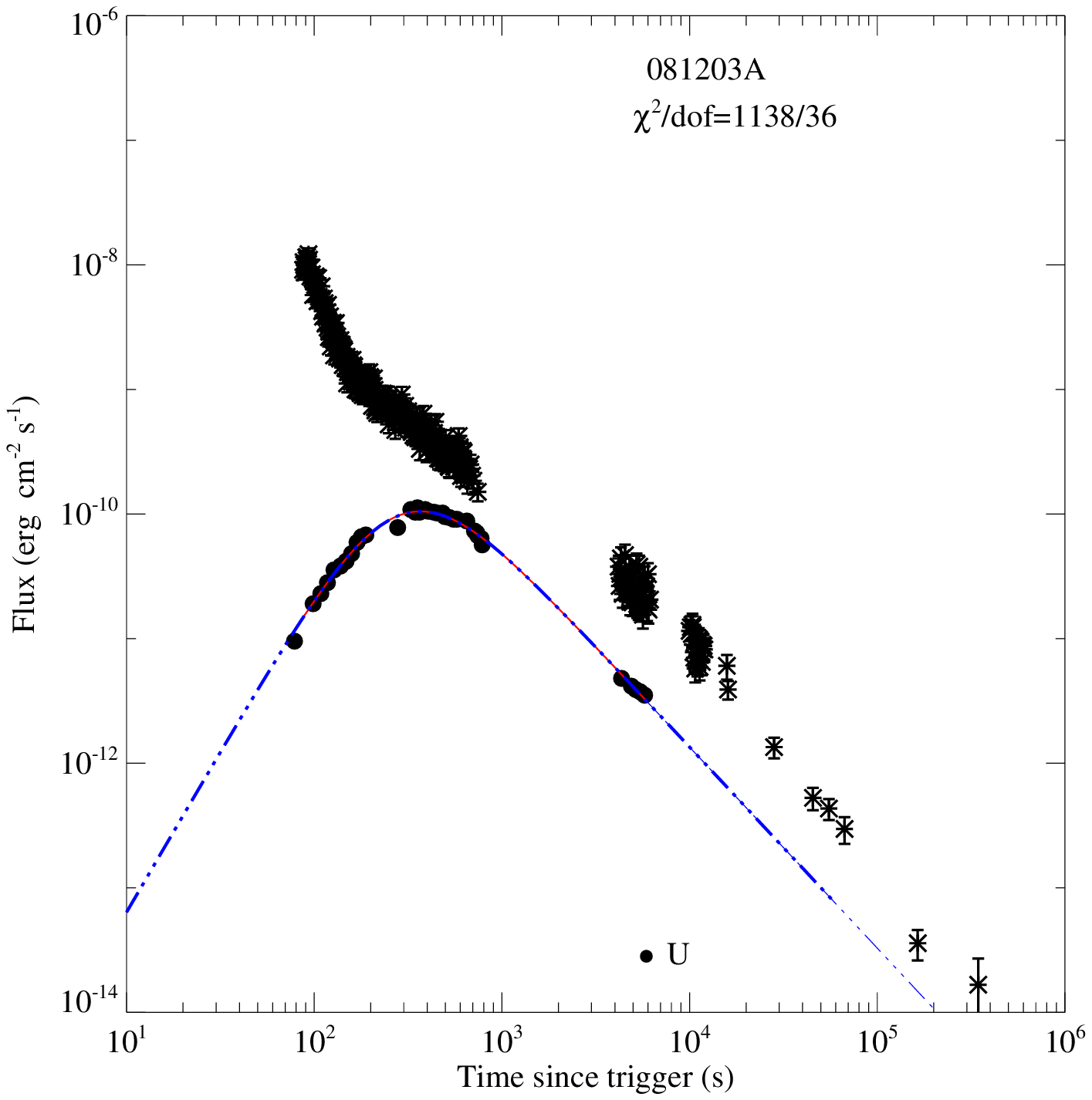}
\includegraphics[angle=0,scale=0.35]{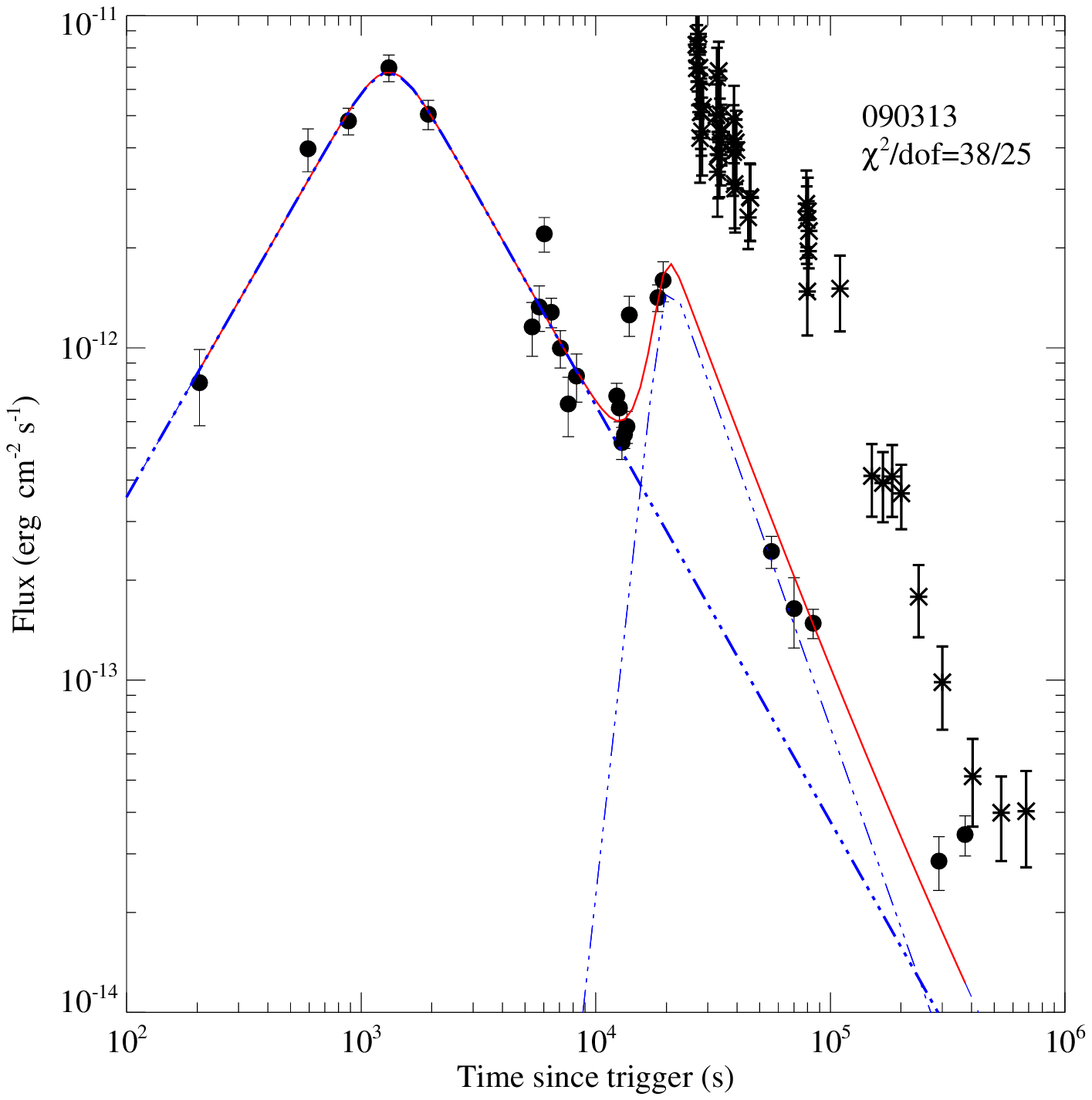}
\includegraphics[angle=0,scale=0.35]{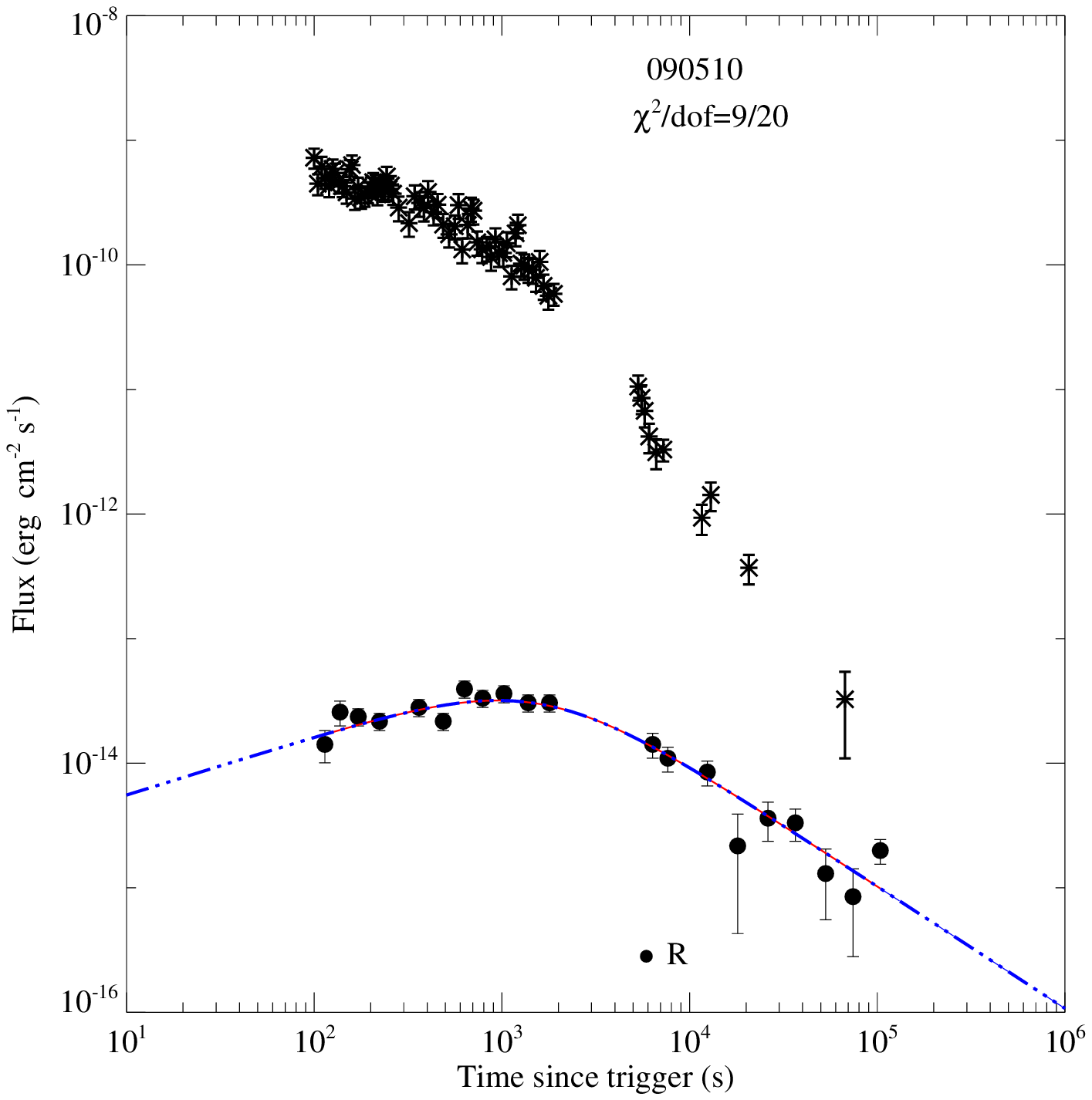}
\includegraphics[angle=0,scale=0.35]{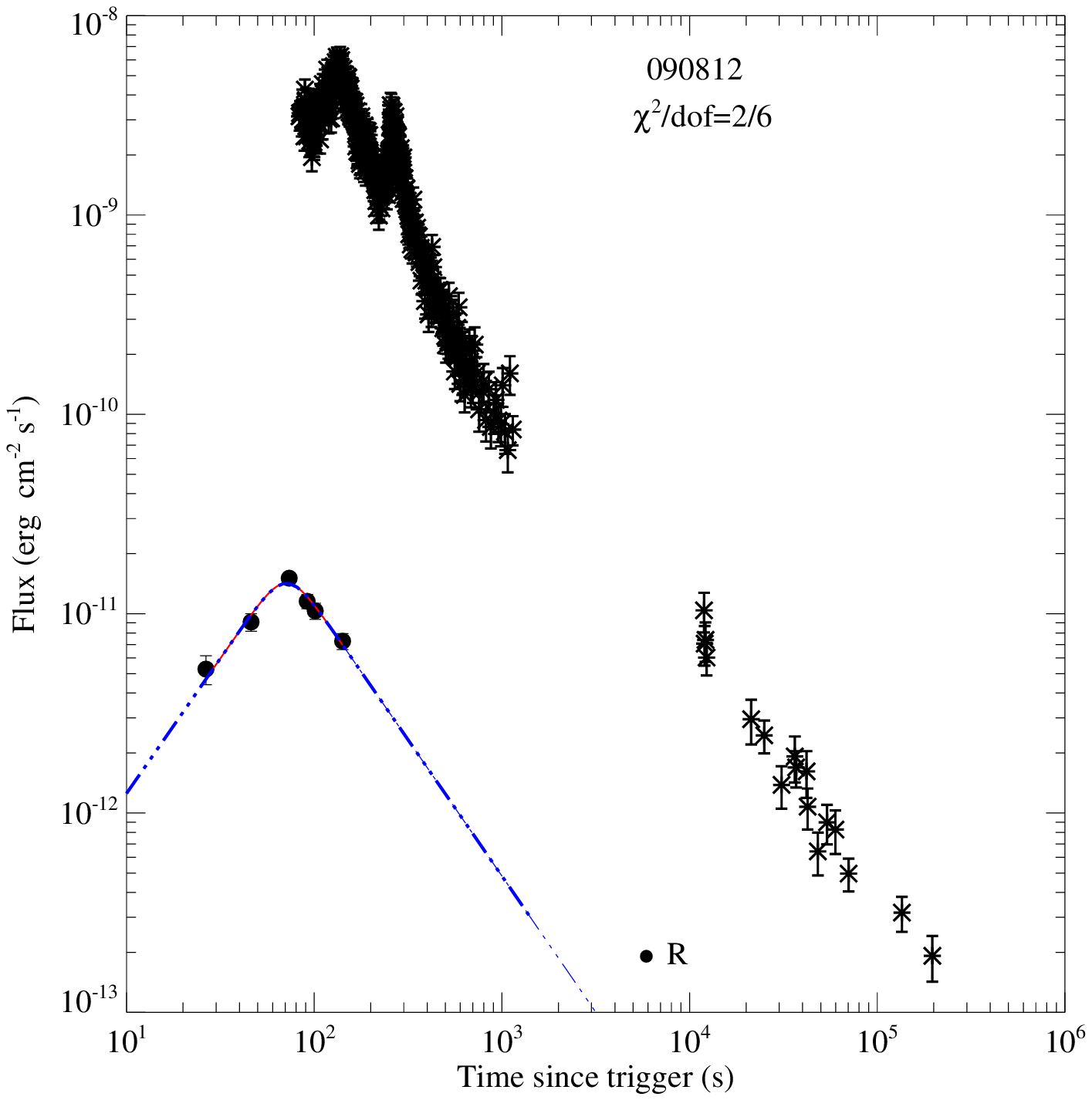}
\includegraphics[angle=0,scale=0.35]{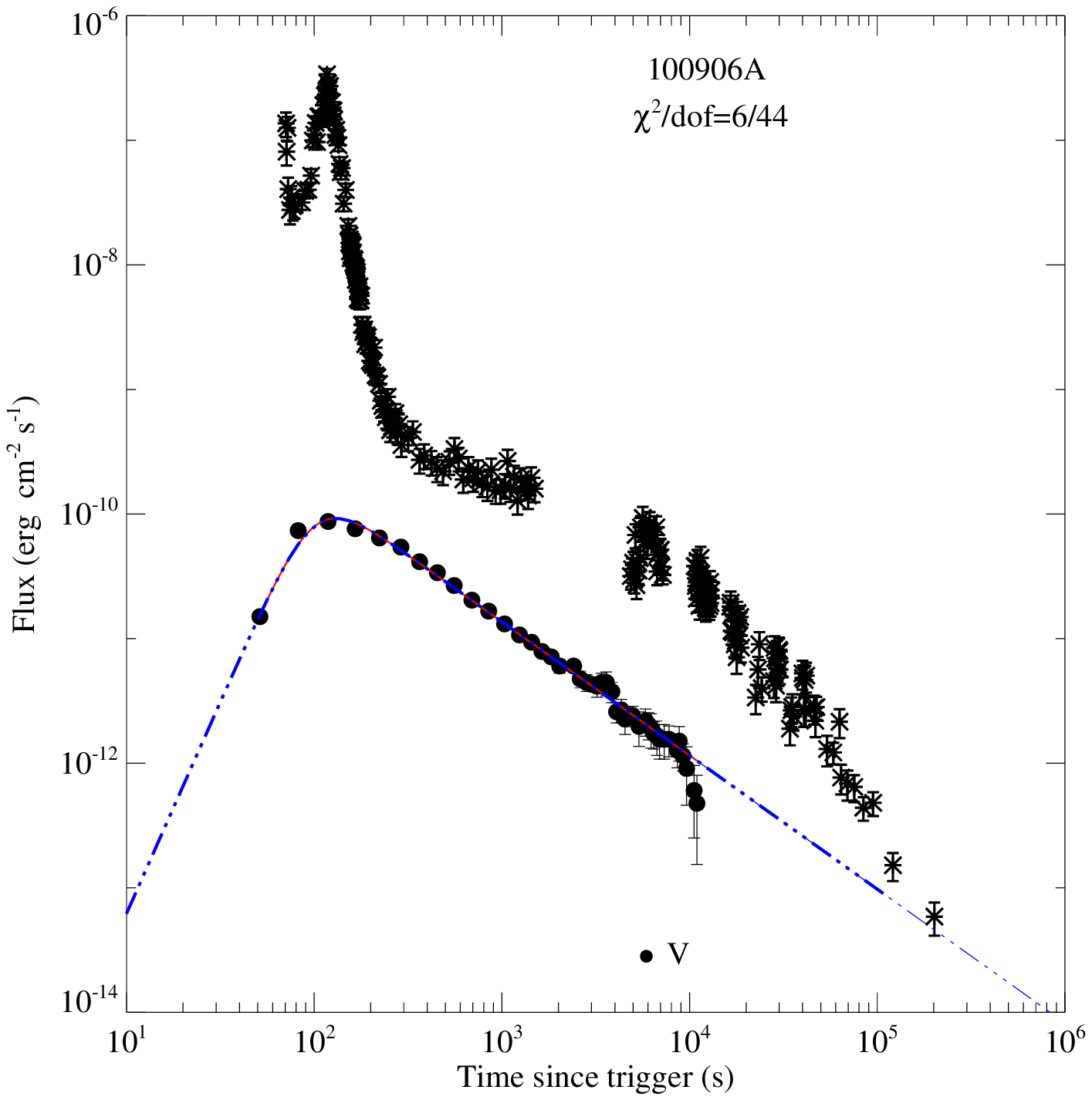}\hfill
\includegraphics[angle=0,scale=0.35]{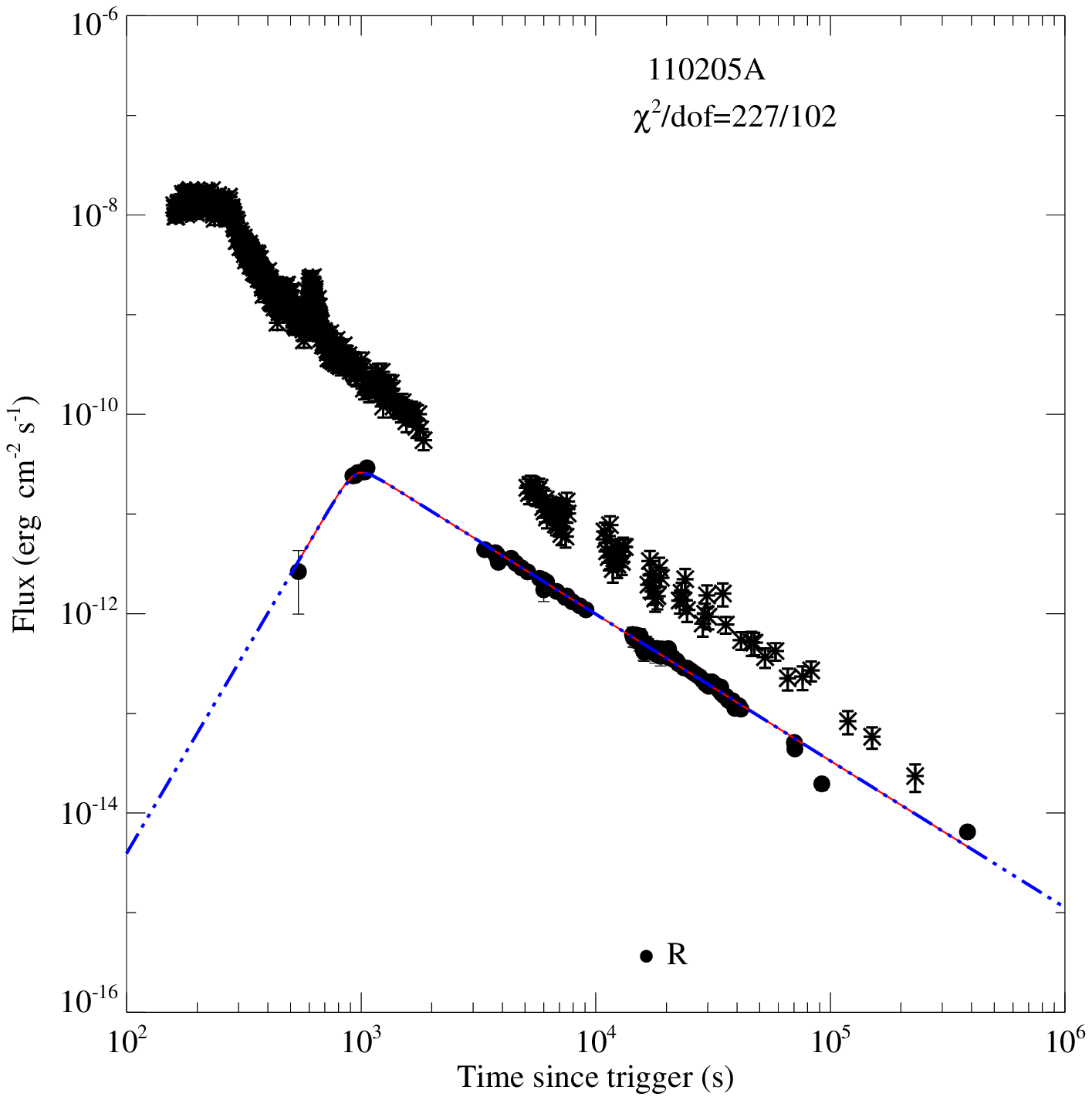}
\center{Fig. 1--- Continued}
\end{figure*}

\begin{figure*}
\includegraphics[angle=0,scale=0.35]{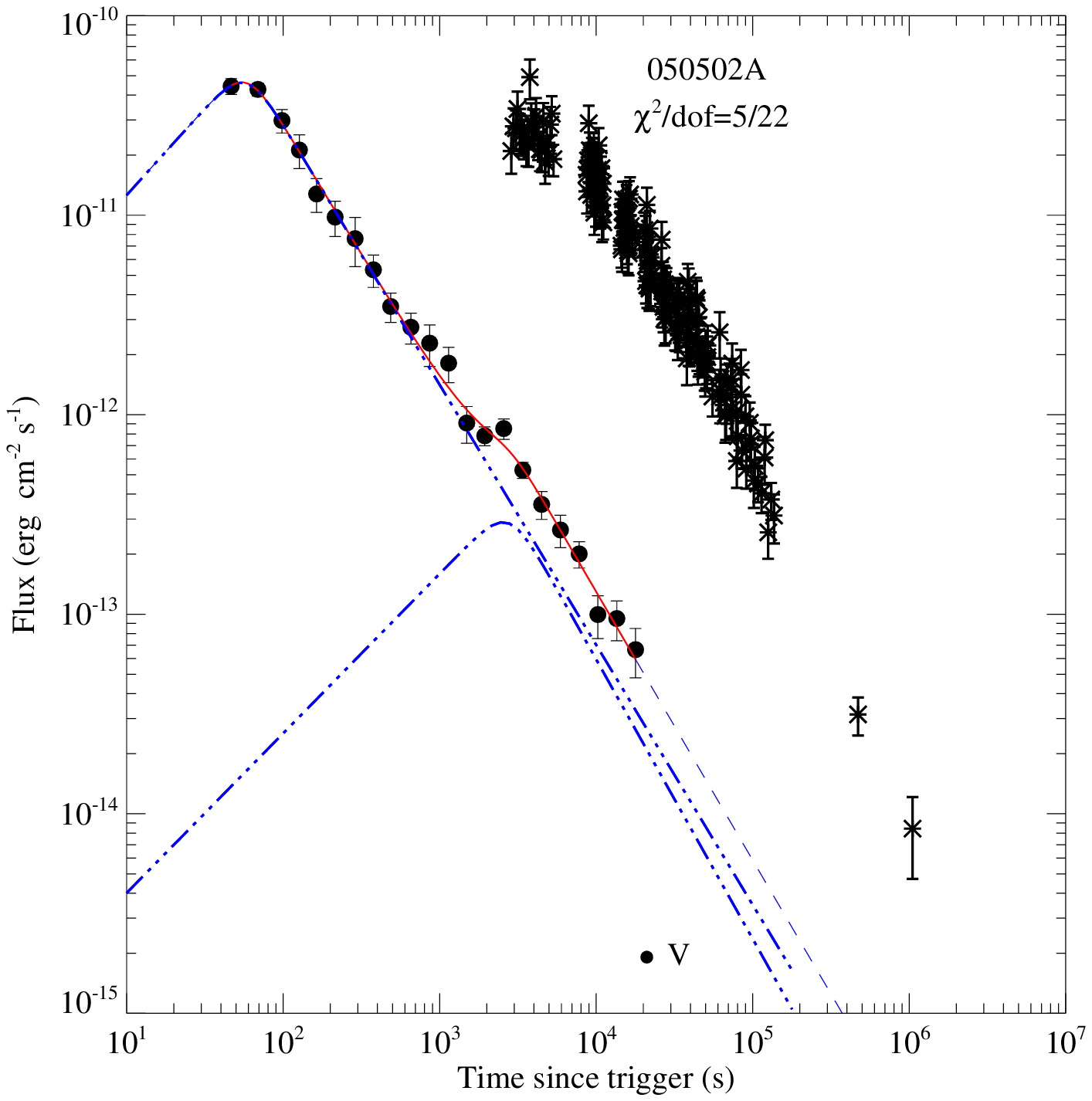}
\includegraphics[angle=0,scale=0.35]{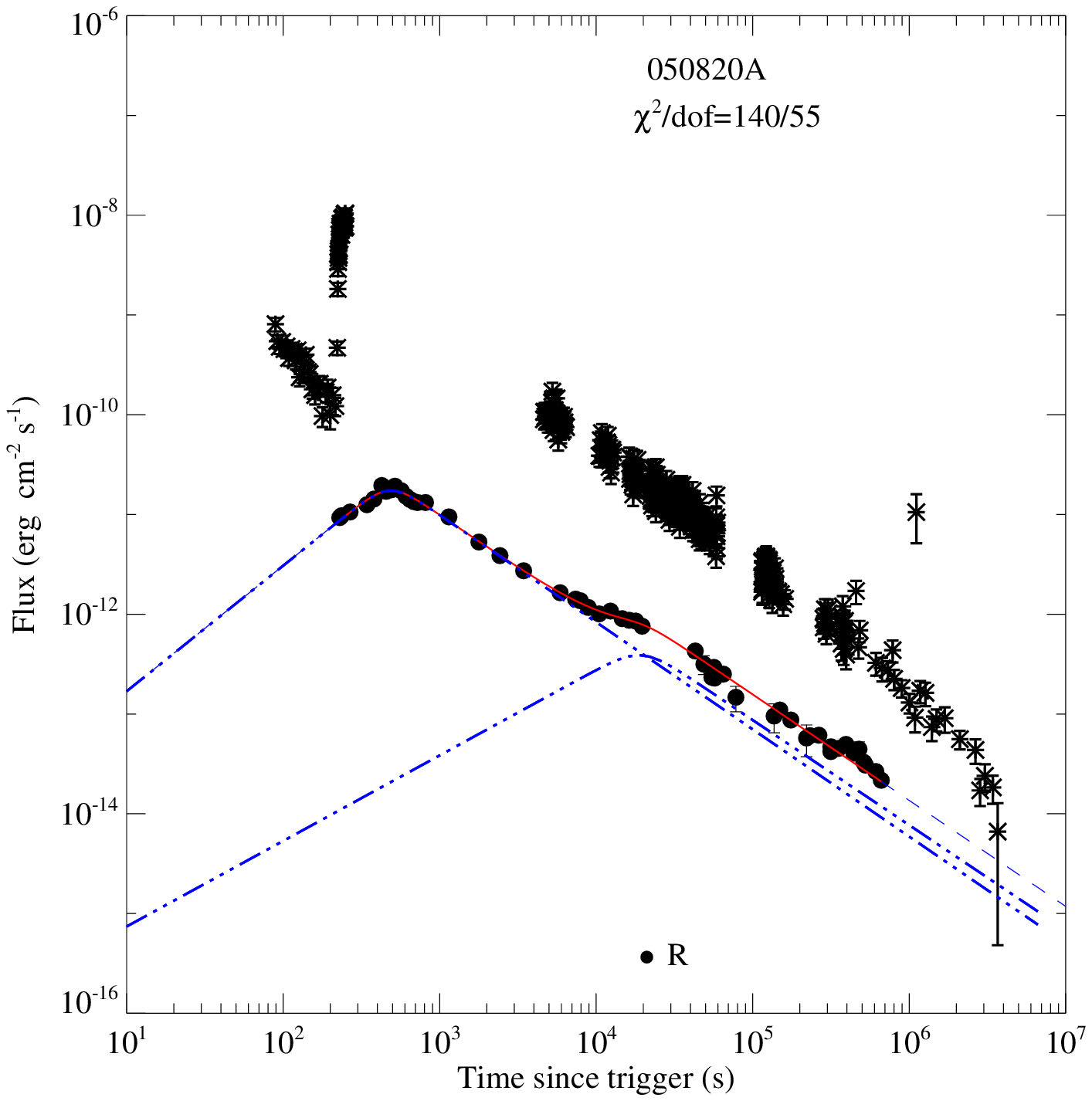}
\includegraphics[angle=0,scale=0.35]{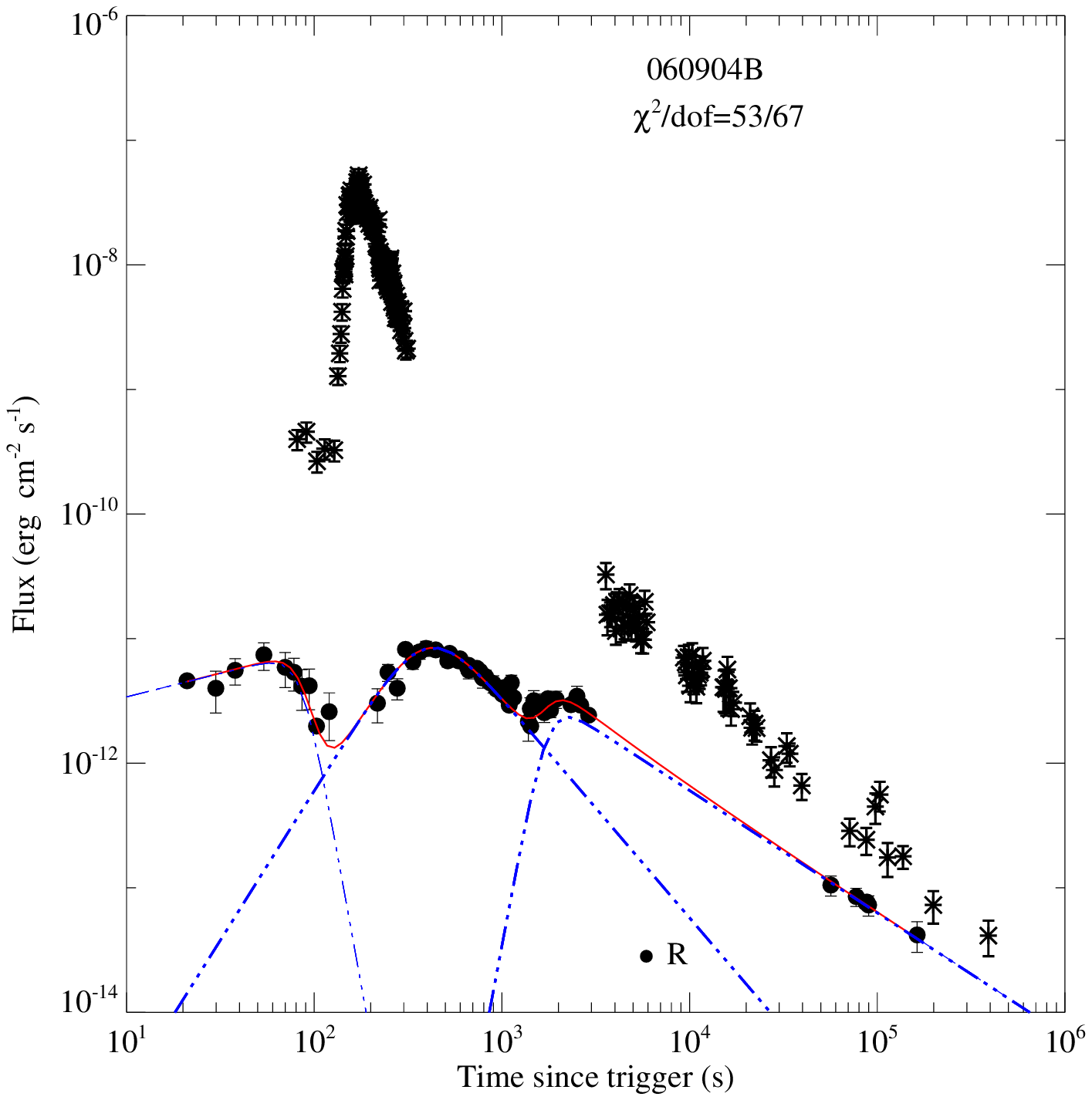}
\includegraphics[angle=0,scale=0.35]{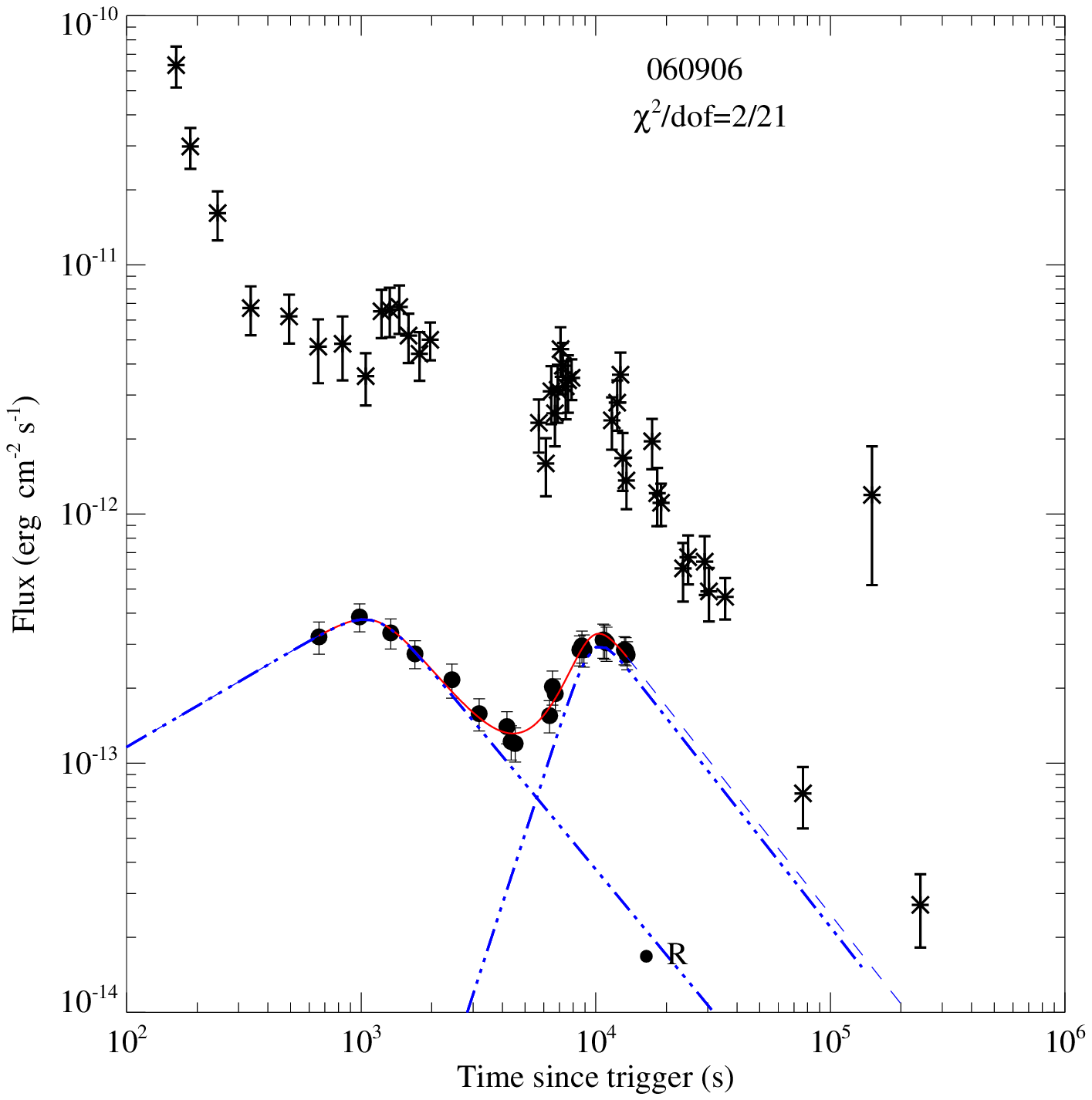}
\includegraphics[angle=0,scale=0.35]{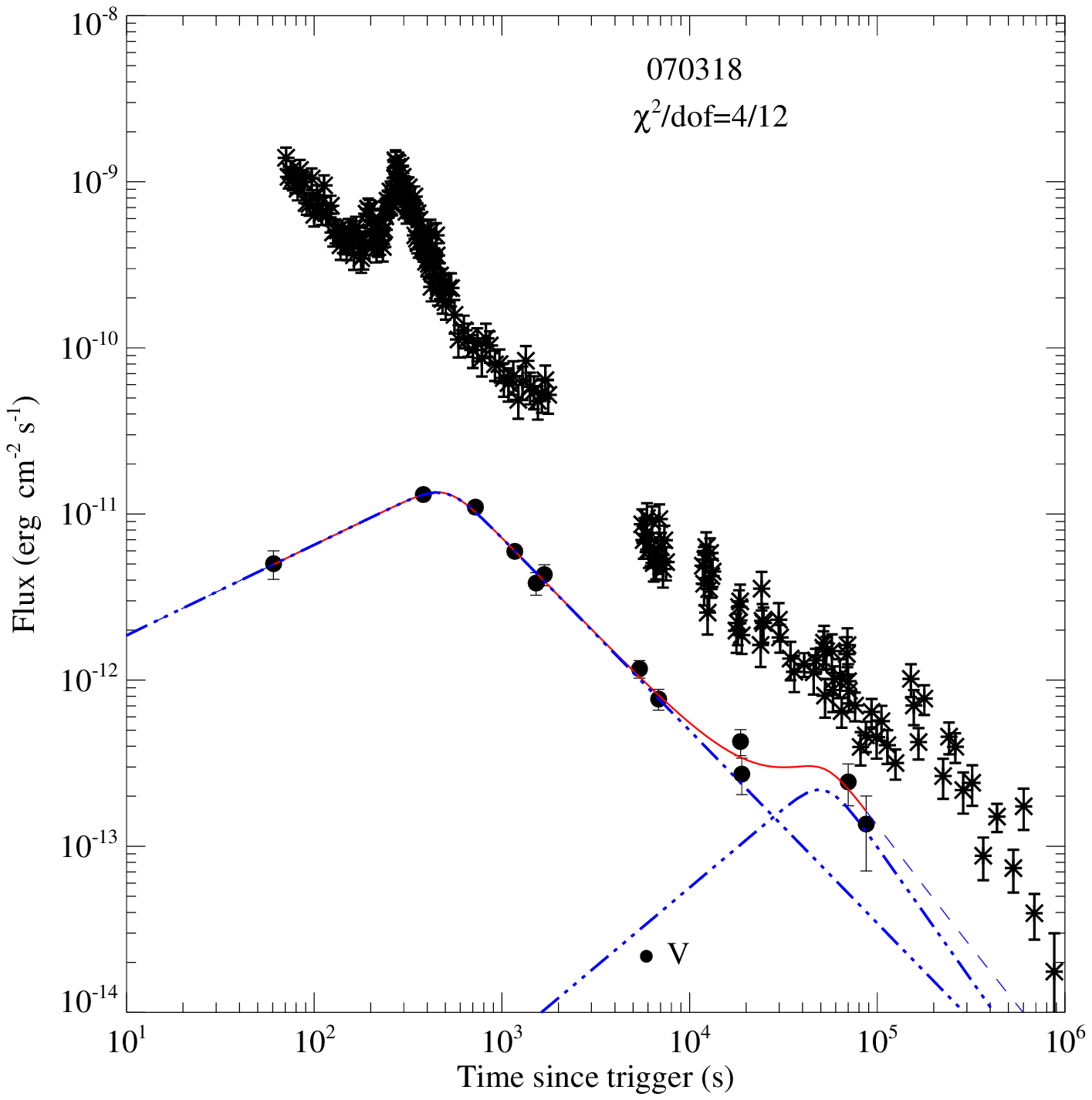}
\includegraphics[angle=0,scale=0.35]{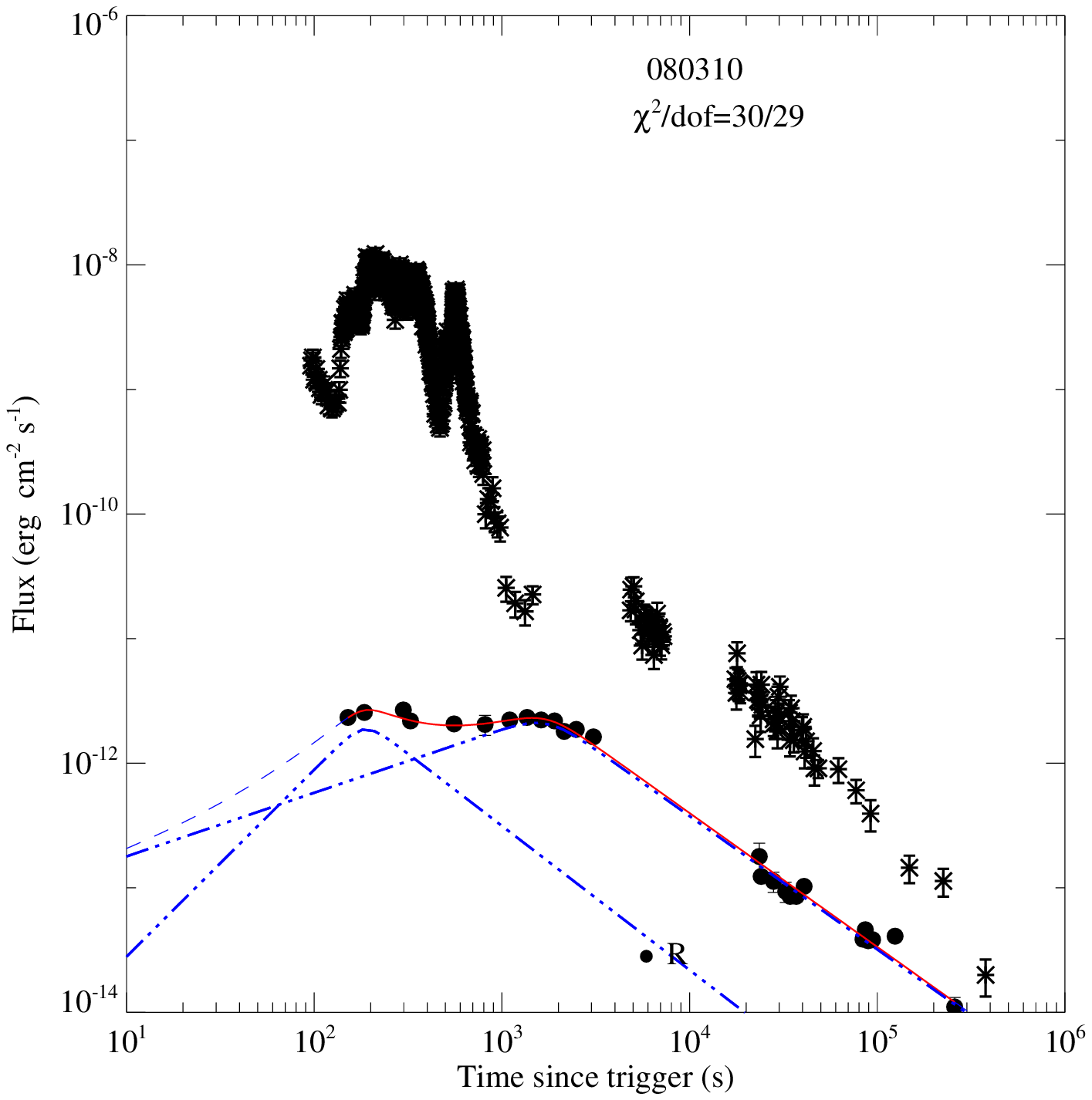}
\includegraphics[angle=0,scale=0.35]{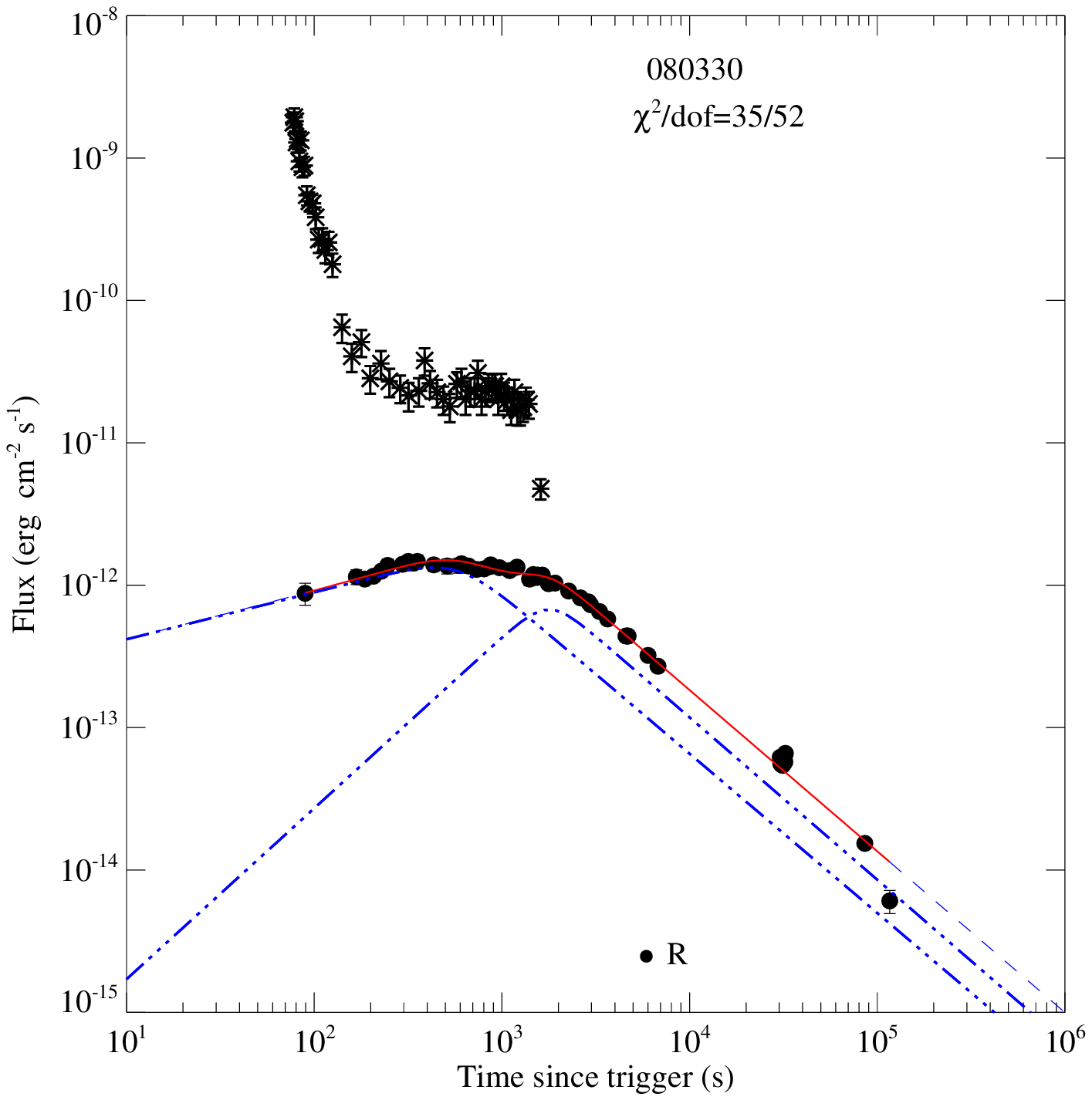}
\includegraphics[angle=0,scale=0.35]{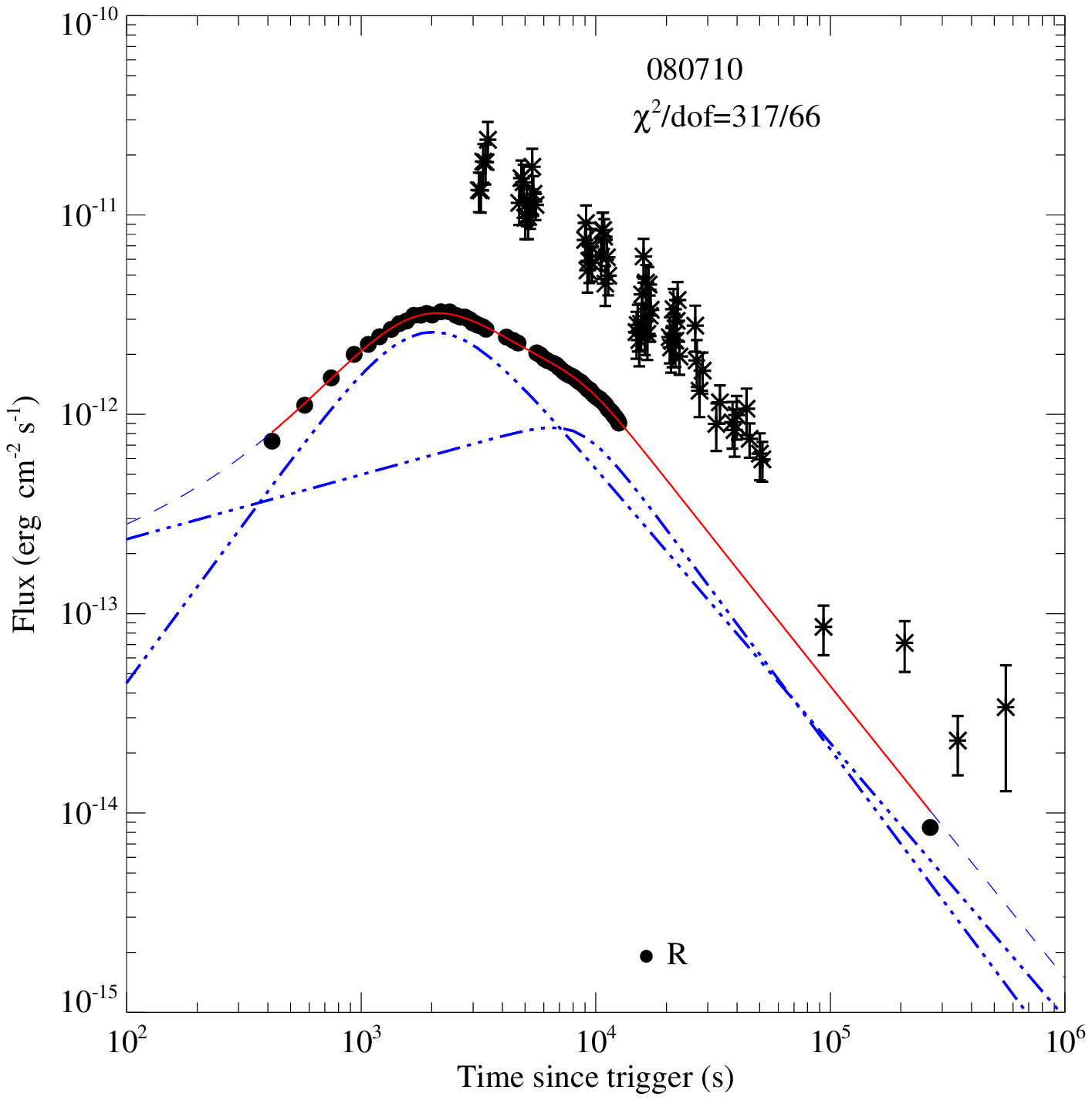}
\includegraphics[angle=0,scale=0.35]{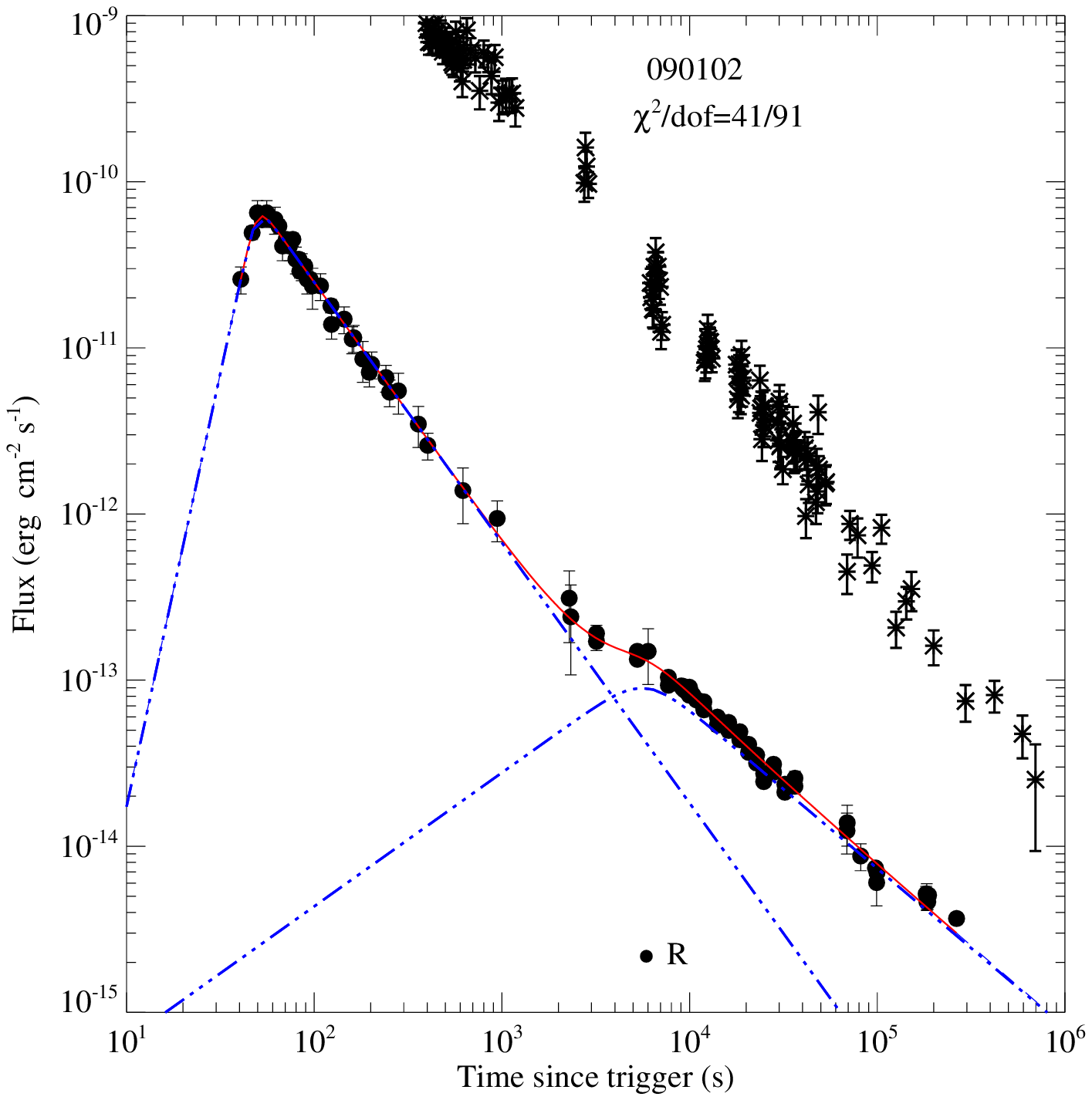}
\includegraphics[angle=0,scale=0.35]{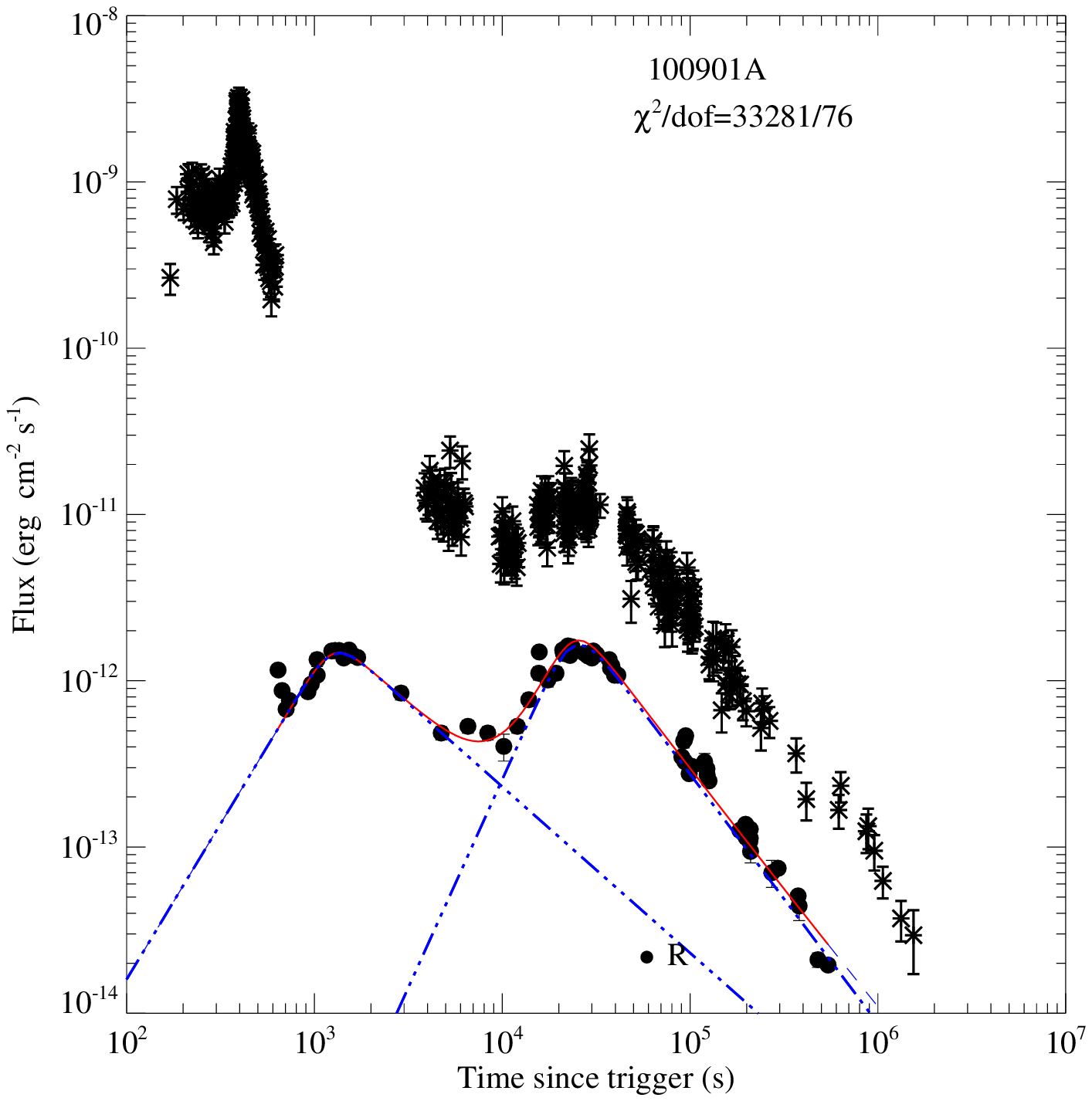}
\includegraphics[angle=0,scale=0.35]{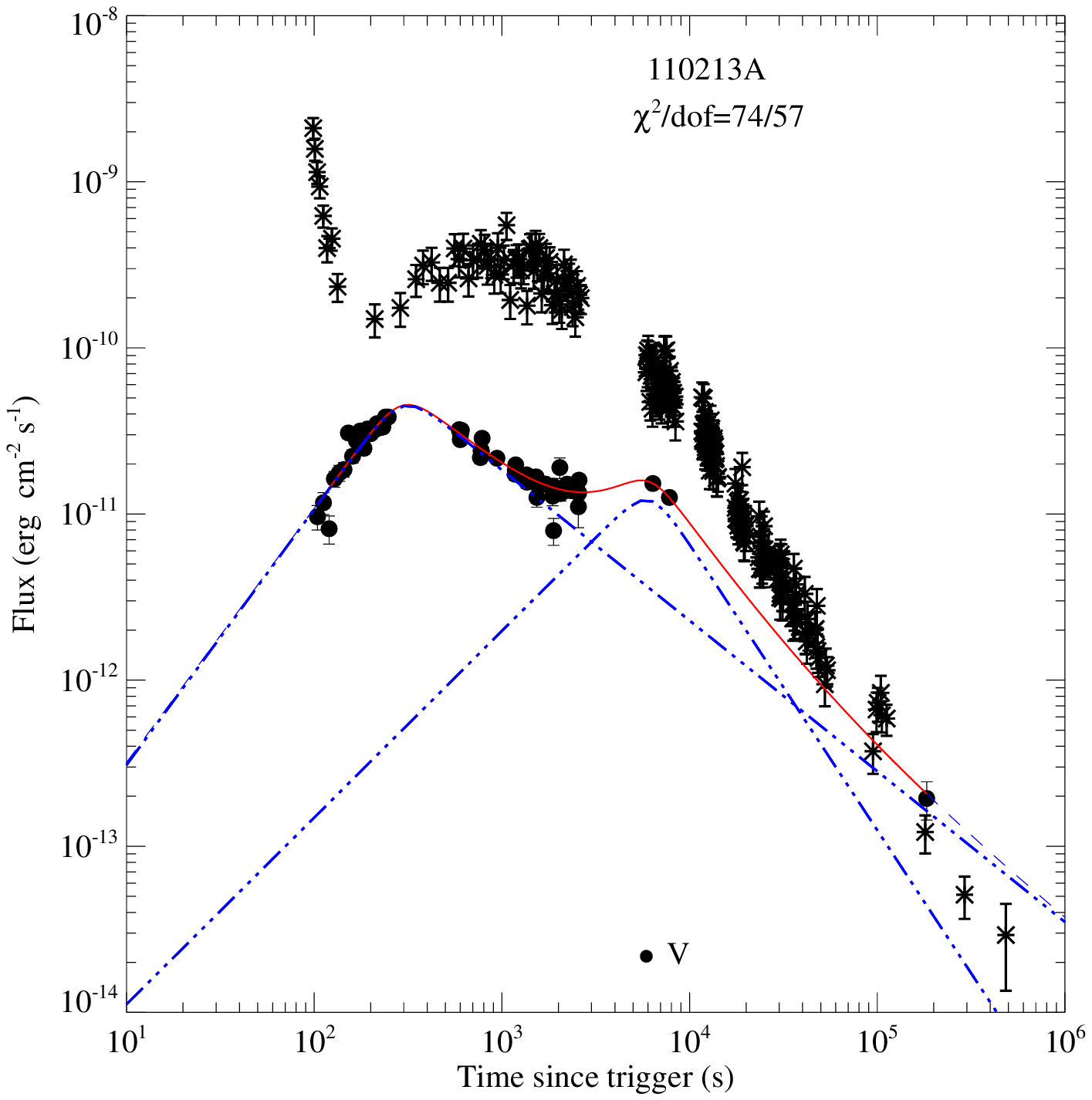}\hfill
\caption{Optical lightcurves with detections of both an initial afterglow onset bump and a late re-brightening hump(s). The symbols and line styles are the same as Figure \ref{Onset}.} \label{Onset_RB}
\end{figure*}

\begin{figure*}
\includegraphics[angle=0,scale=0.35]{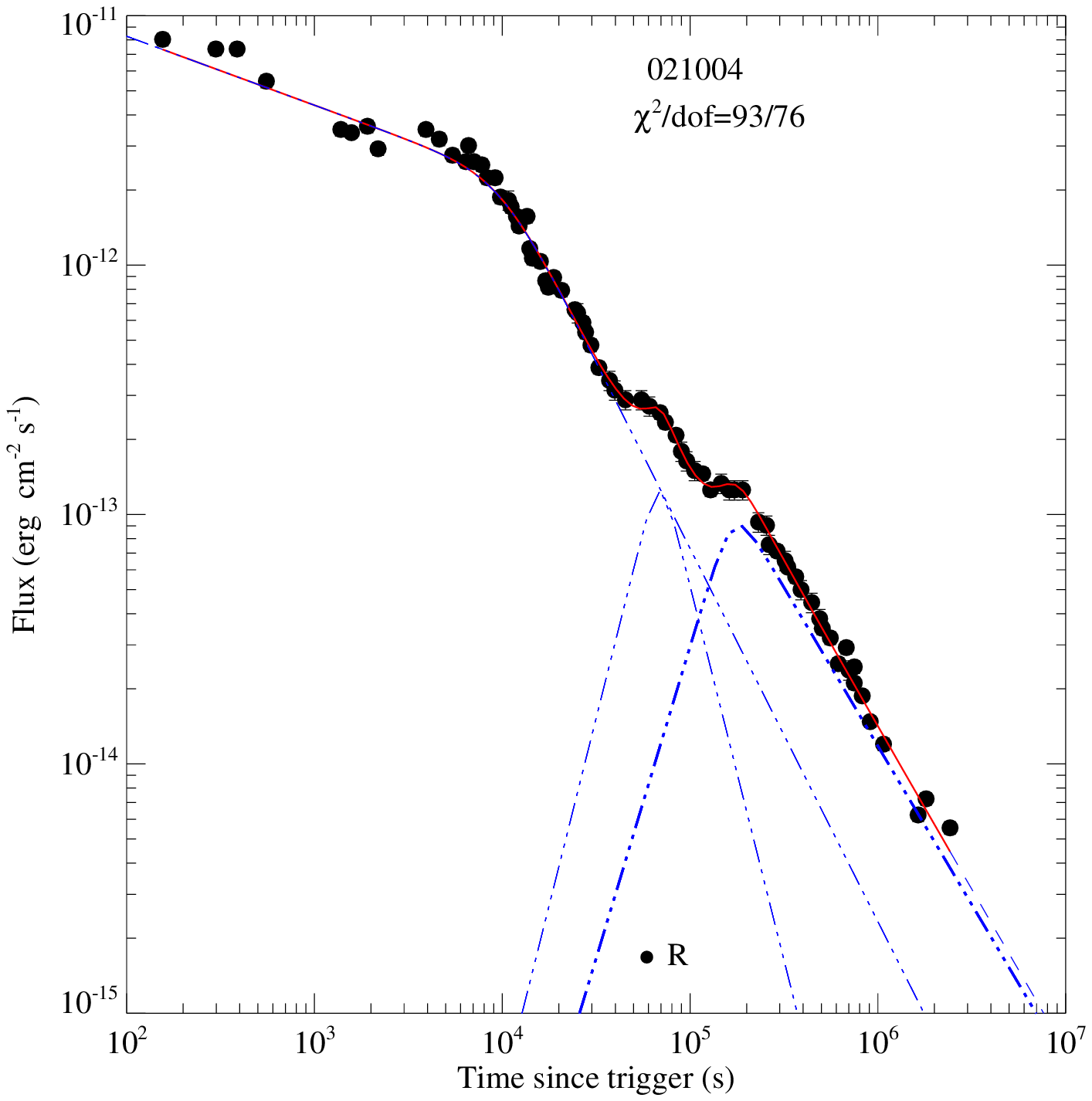}
\includegraphics[angle=0,scale=0.35]{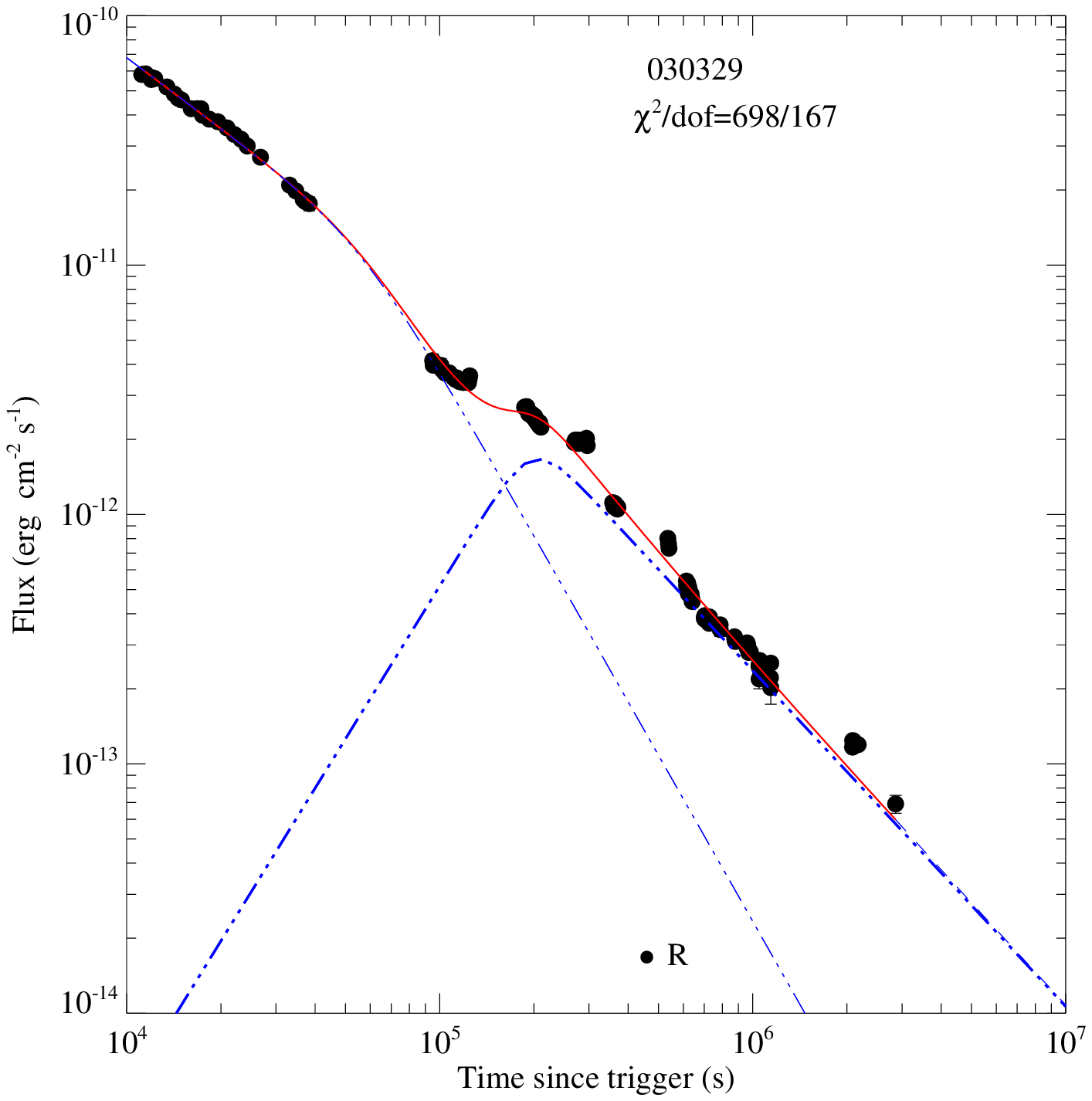}
\includegraphics[angle=0,scale=0.35]{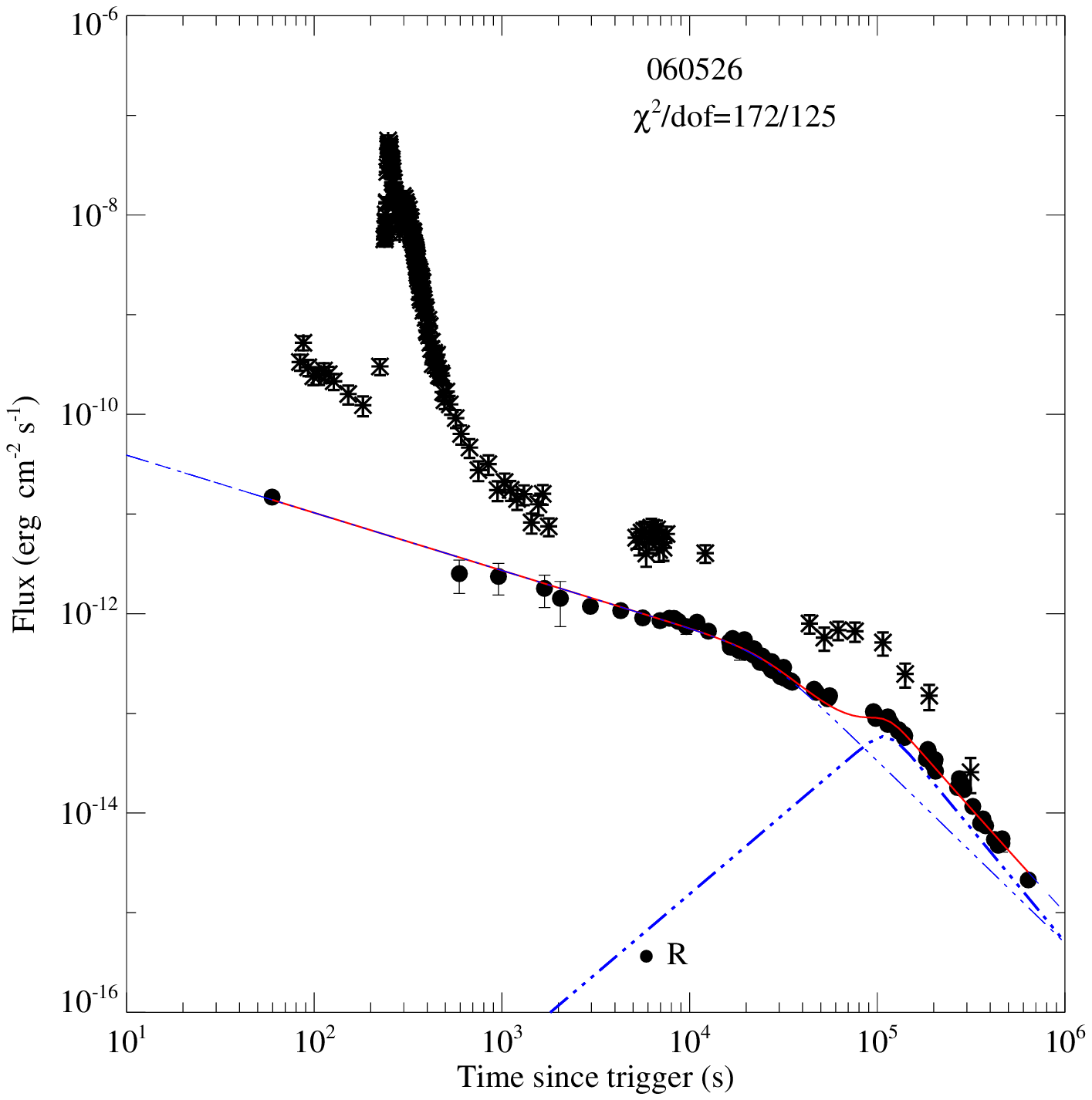}
\includegraphics[angle=0,scale=0.35]{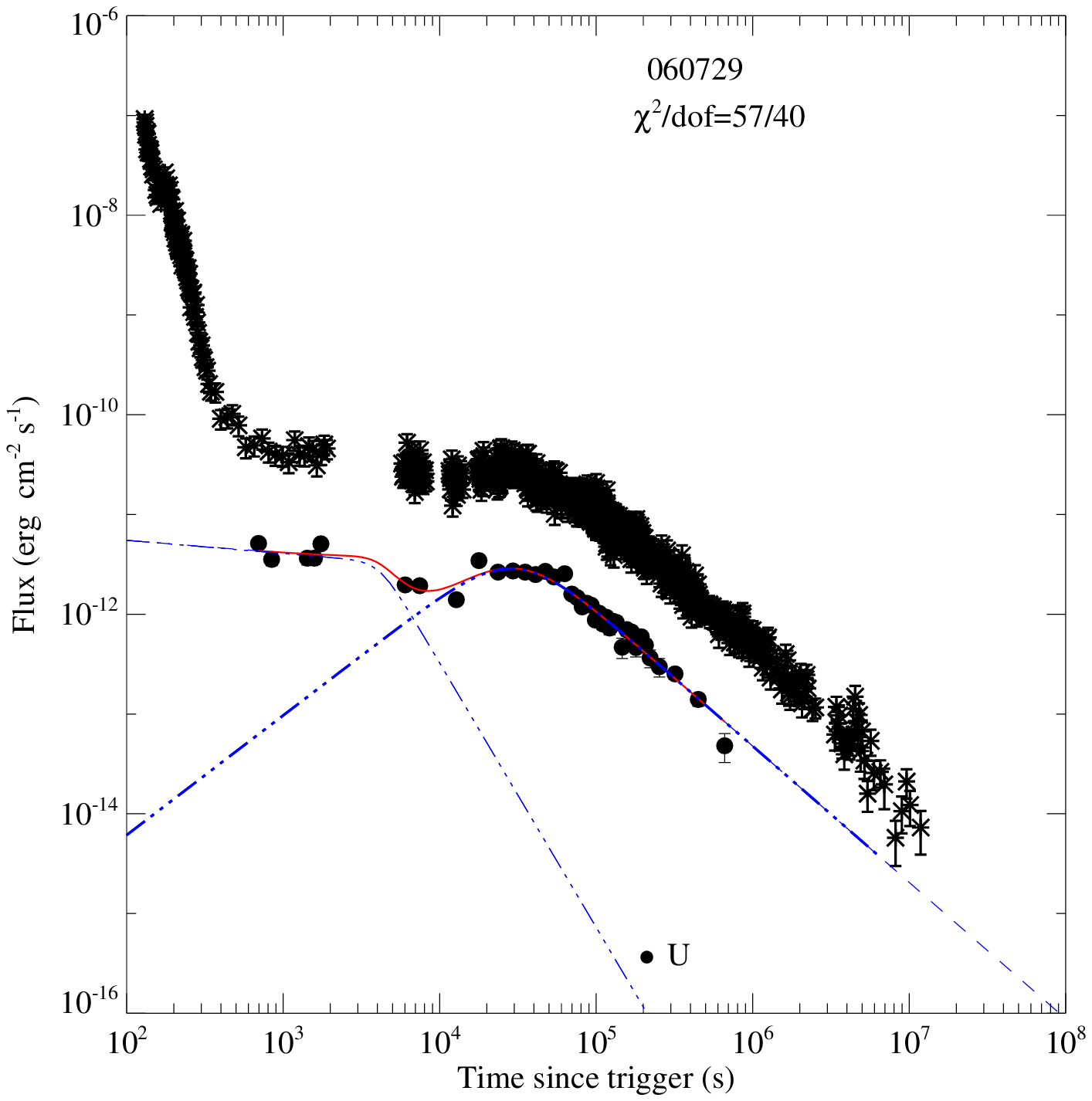}
\includegraphics[angle=0,scale=0.35]{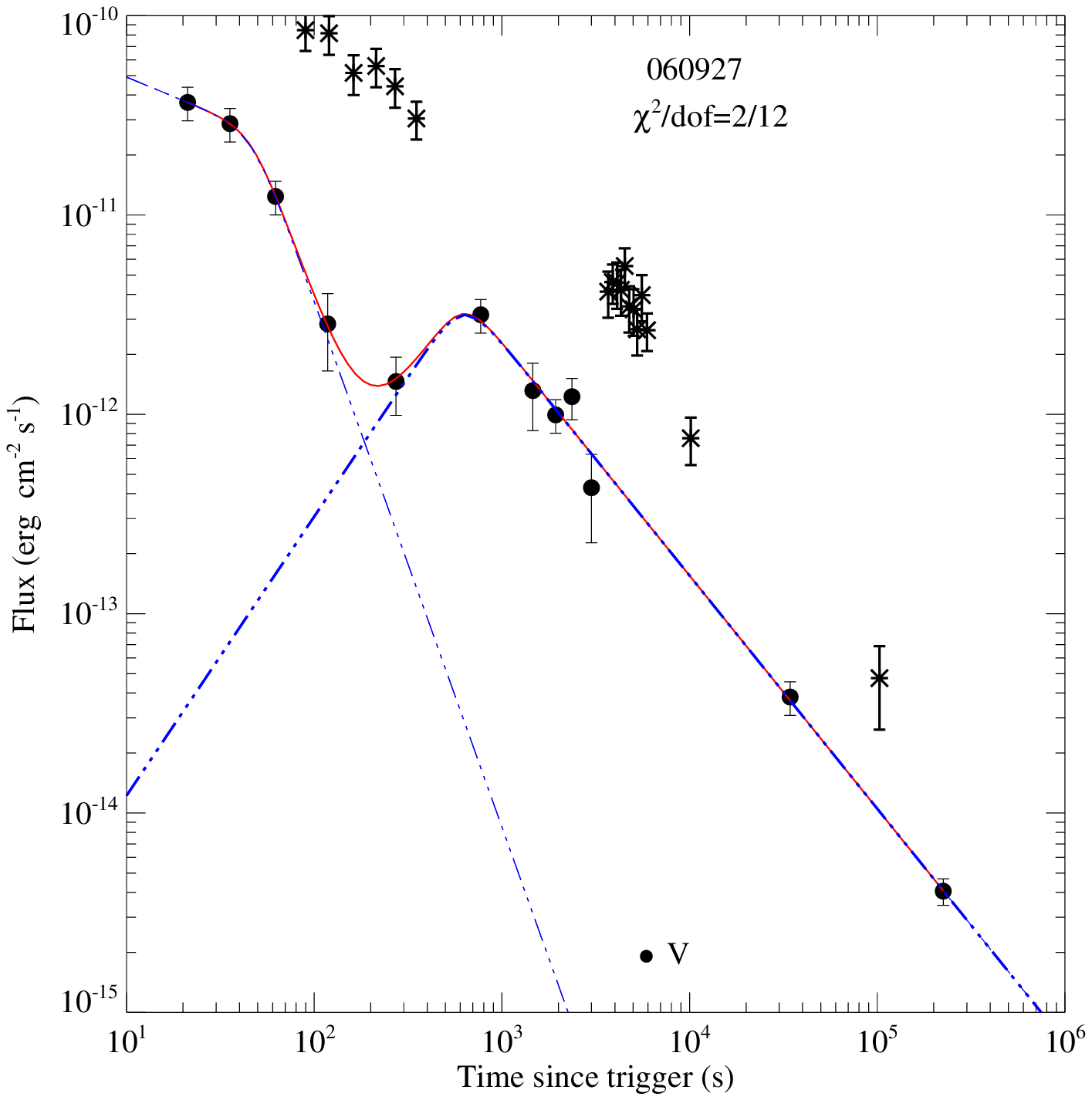}
\includegraphics[angle=0,scale=0.35]{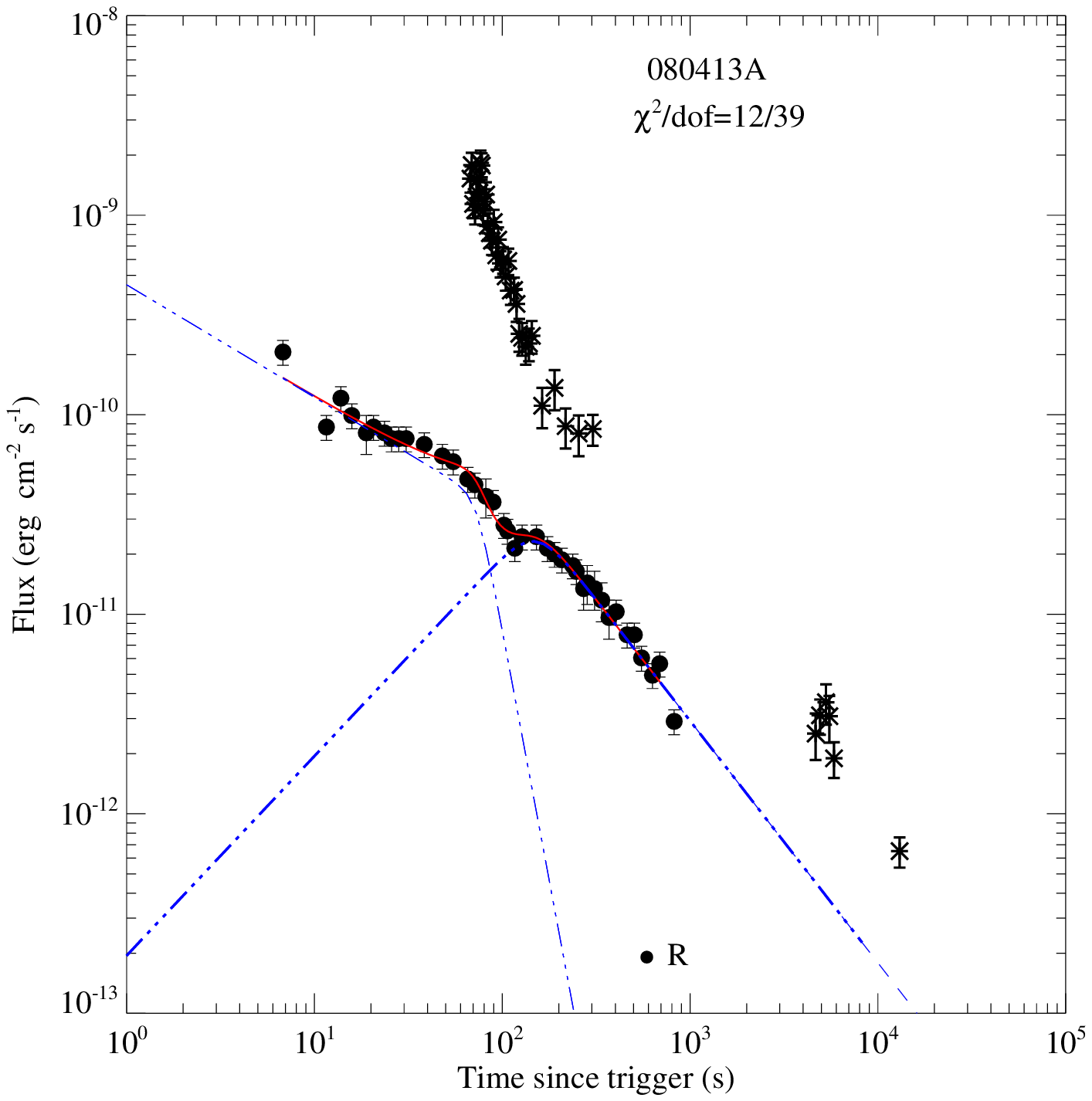}
\includegraphics[angle=0,scale=0.35]{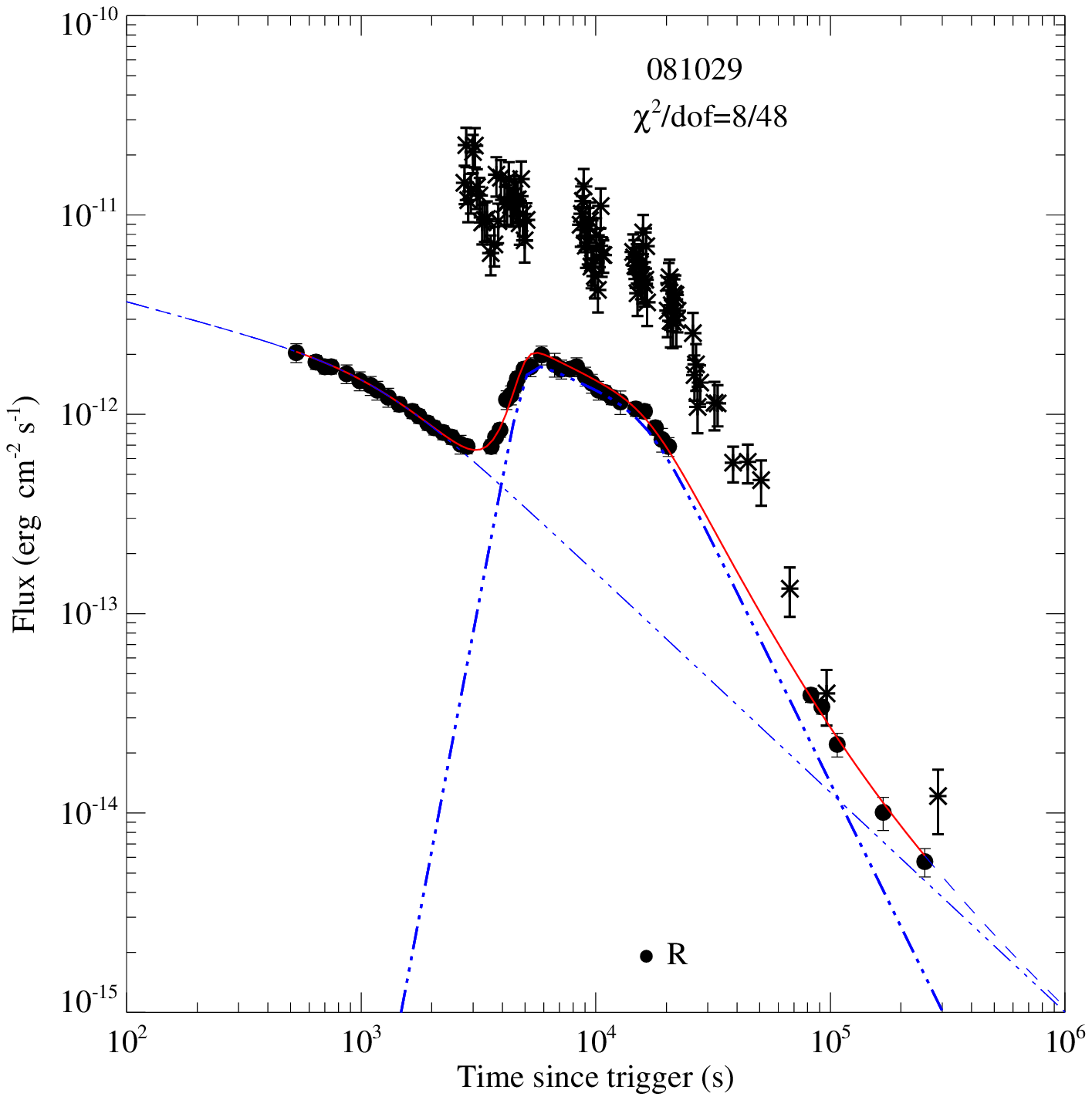}\hfill
\includegraphics[angle=0,scale=0.35]{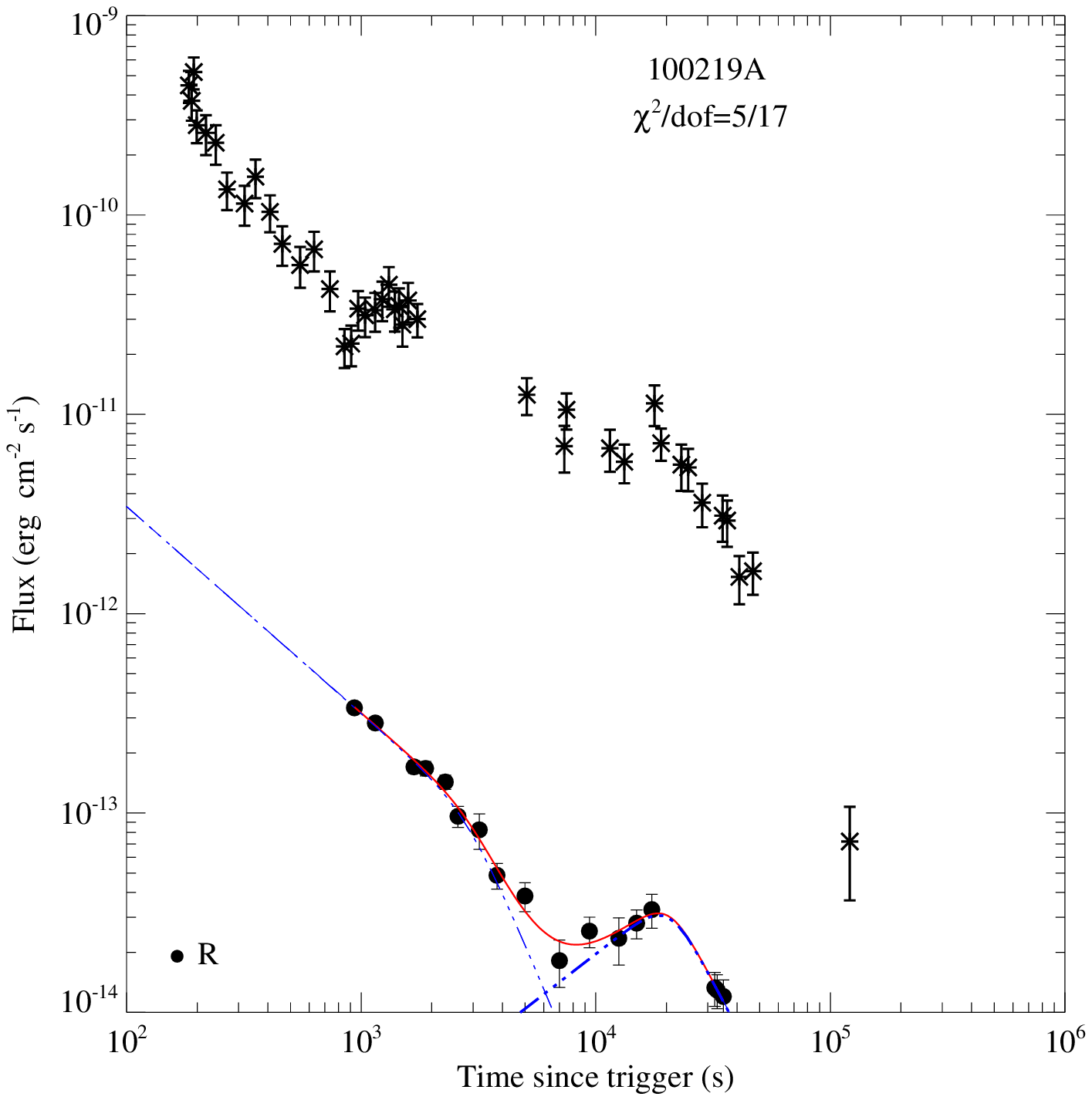}
\caption{Optical lightcurves with detections of both an initial shallow decay segment and a late re-brightening hump(s). The symbols and line styles are the same as Figure \ref{Onset}.\label{Shallow_RB}}
\end{figure*}

\begin{figure*}
\includegraphics[angle=0,scale=0.35]{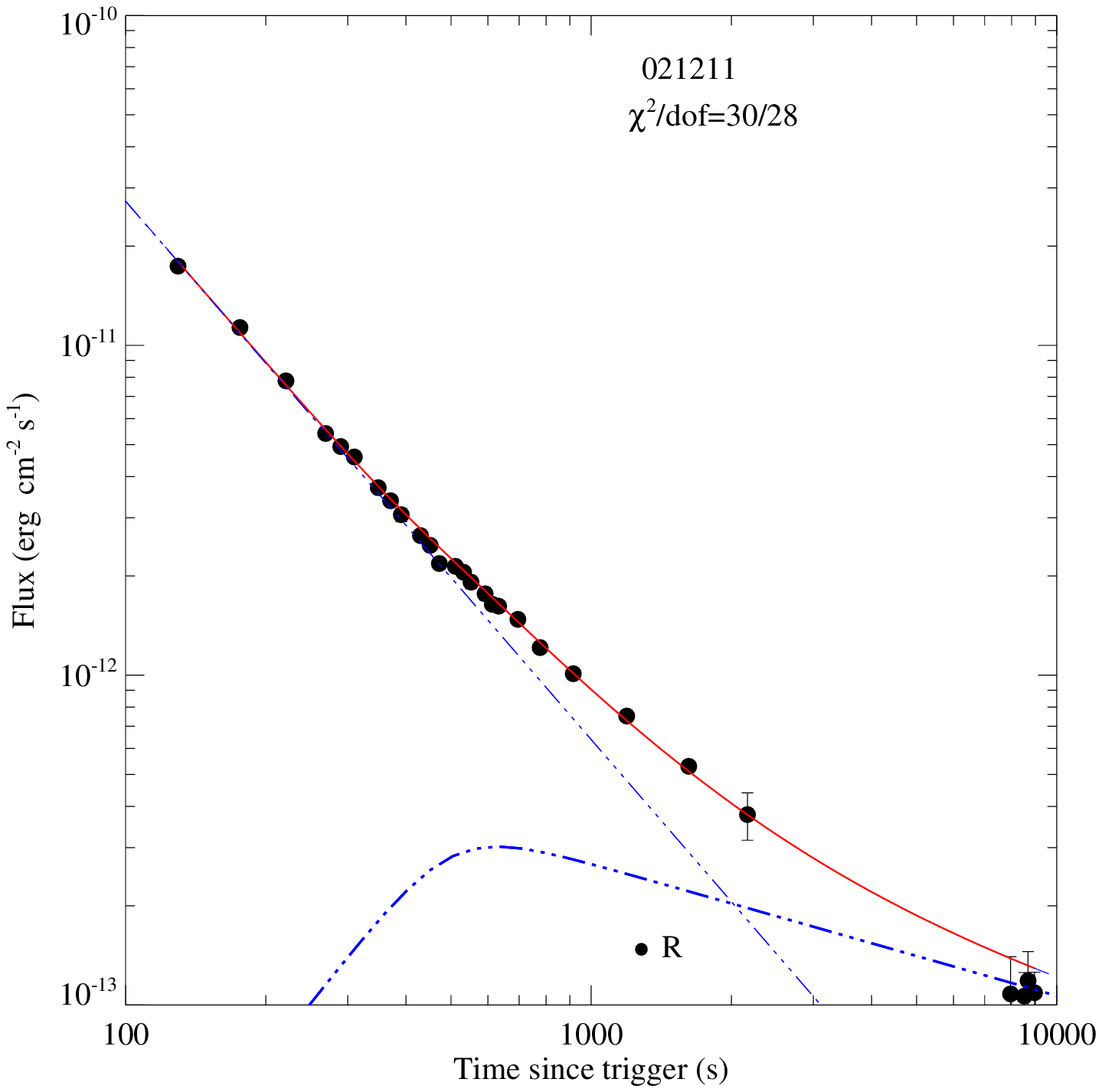}
\includegraphics[angle=0,scale=0.35]{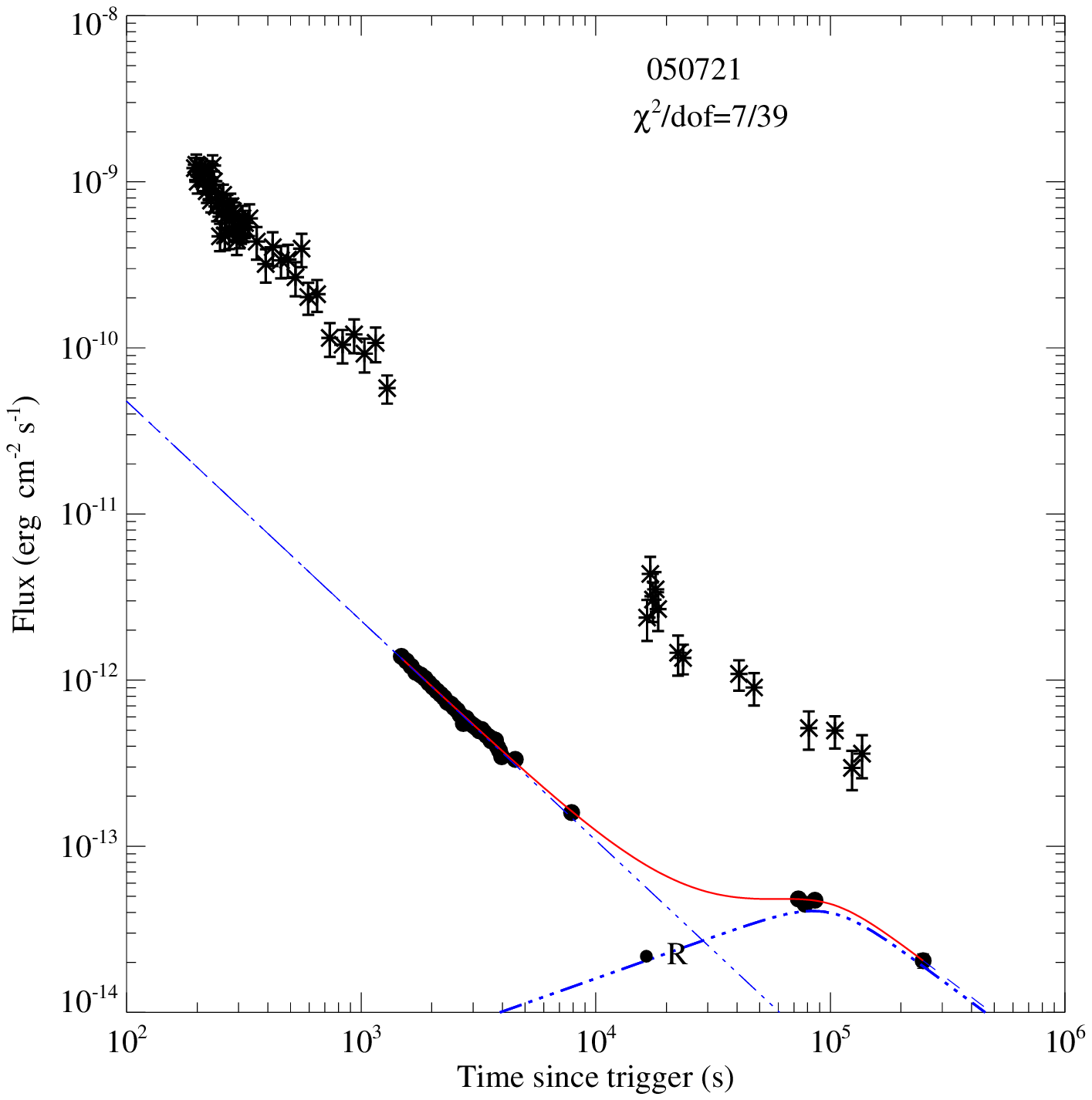}
\includegraphics[angle=0,scale=0.35]{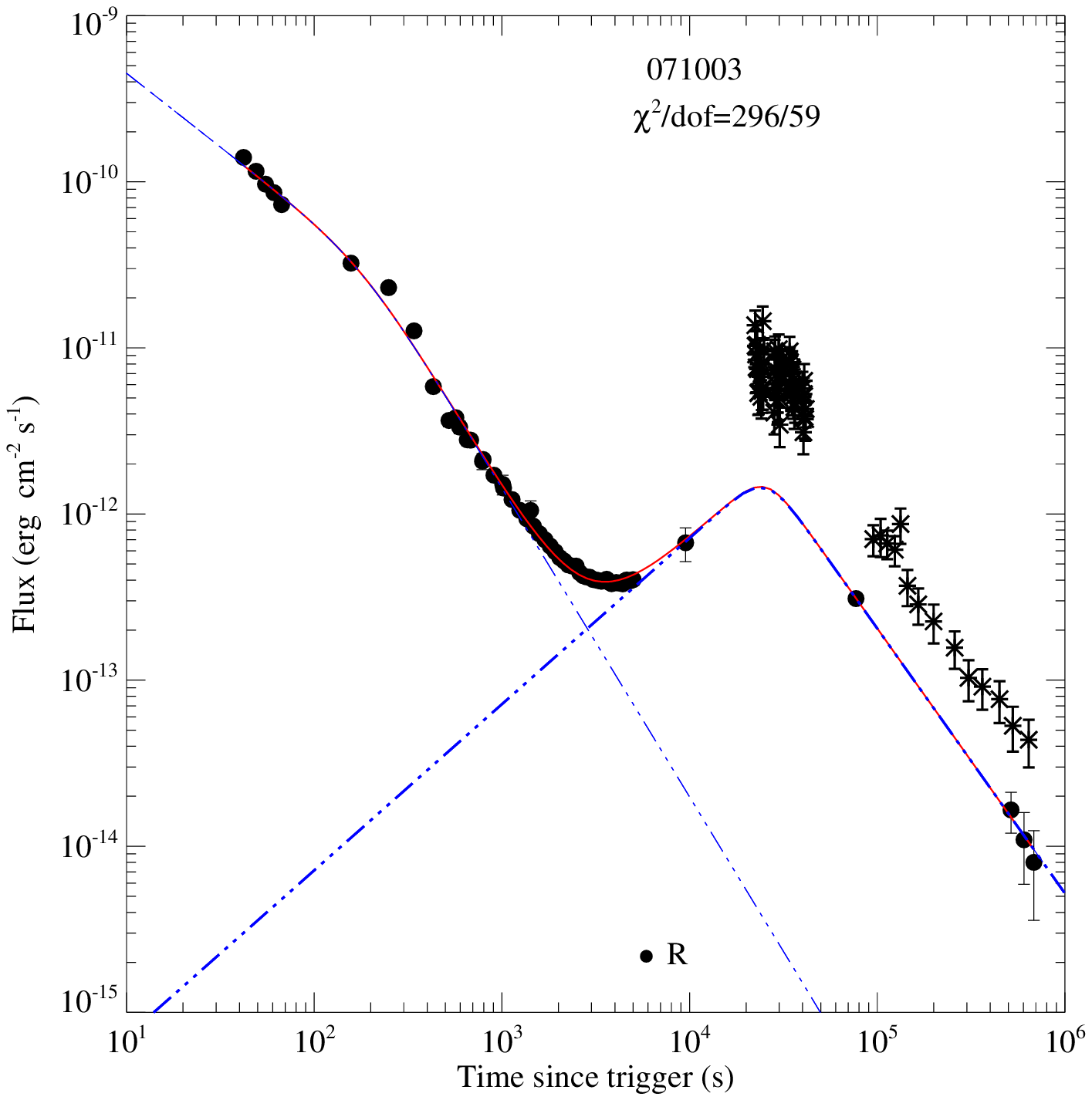}
\includegraphics[angle=0,scale=0.35]{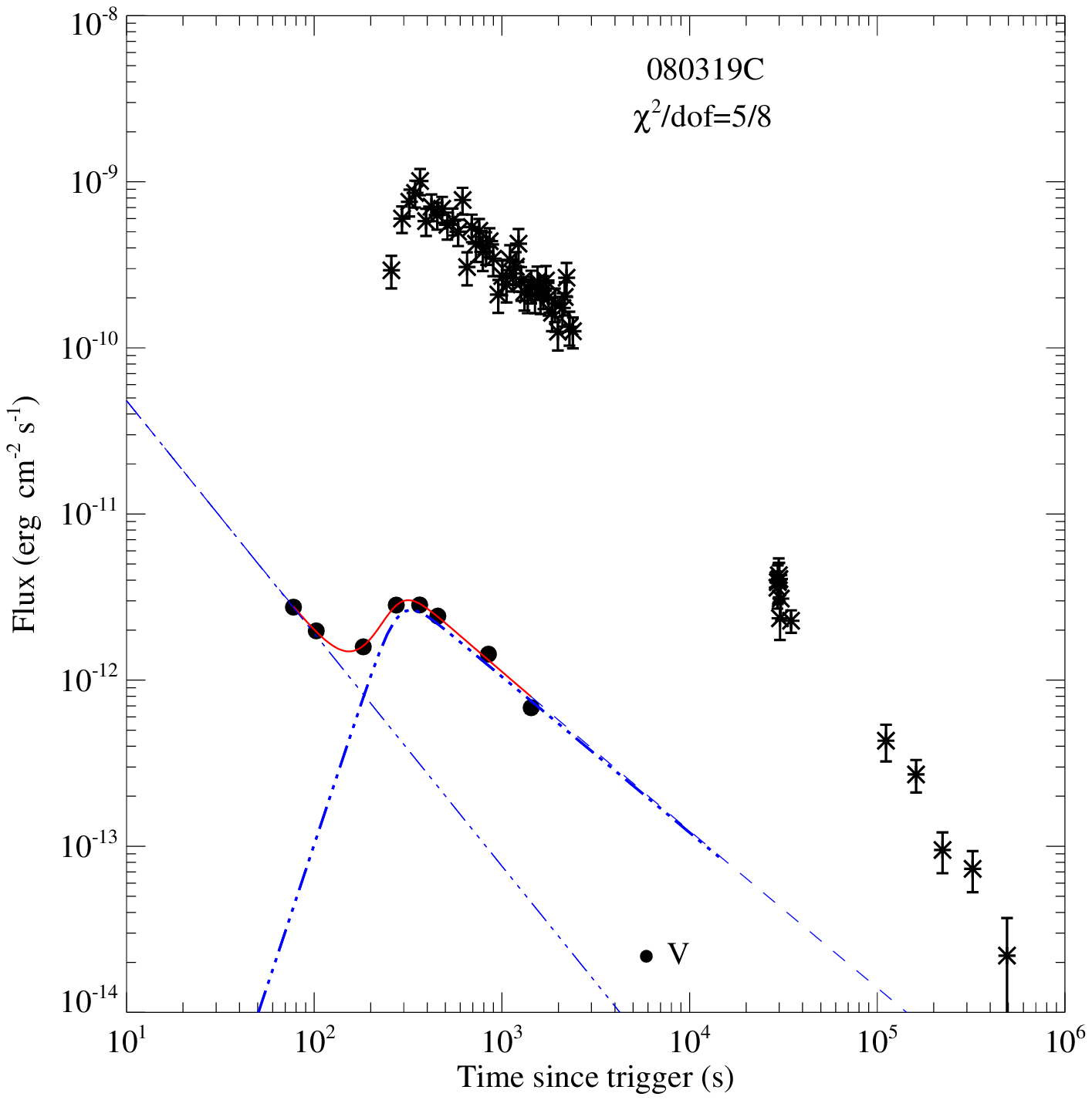}
\includegraphics[angle=0,scale=0.35]{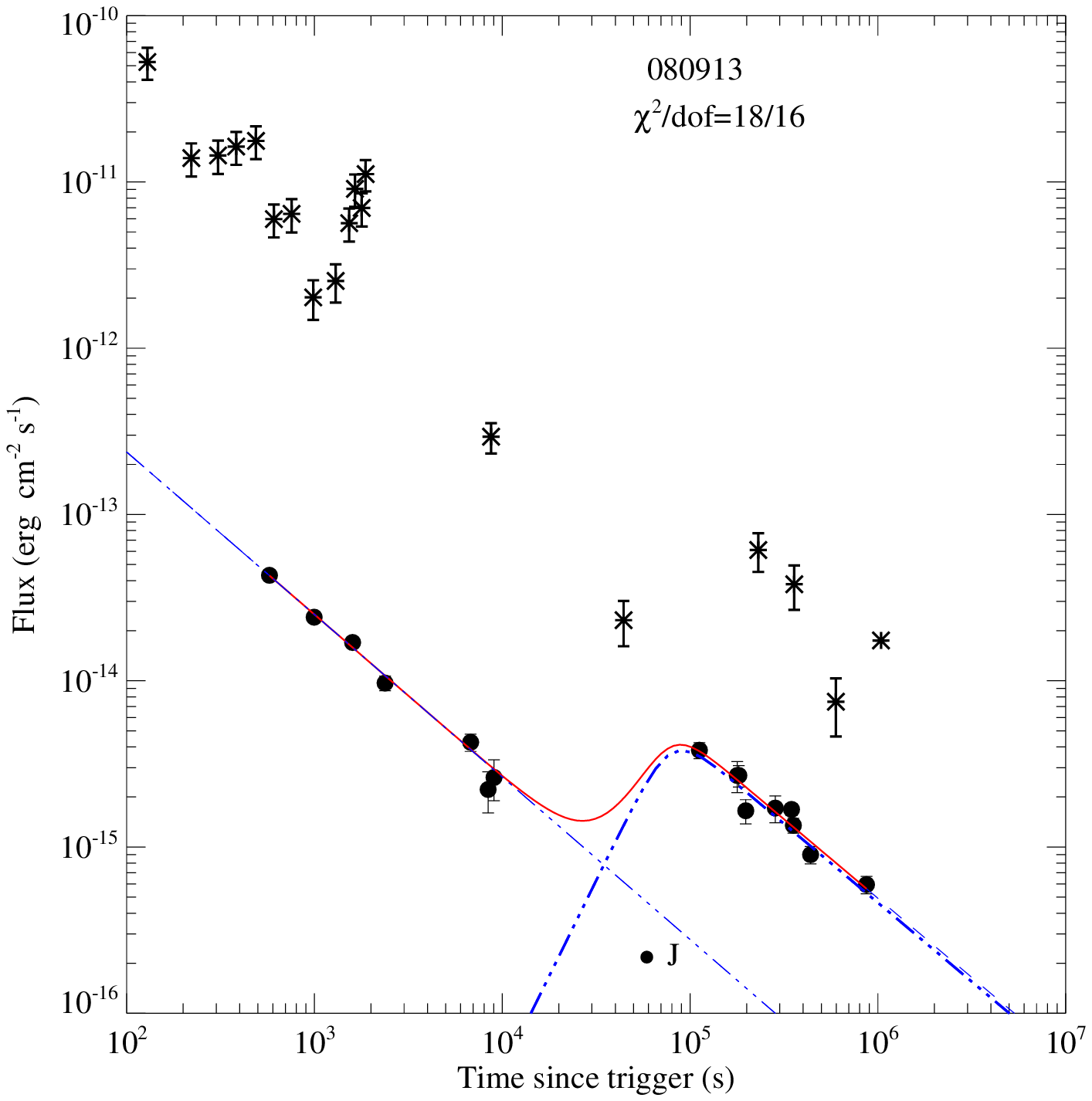}
\includegraphics[angle=0,scale=0.35]{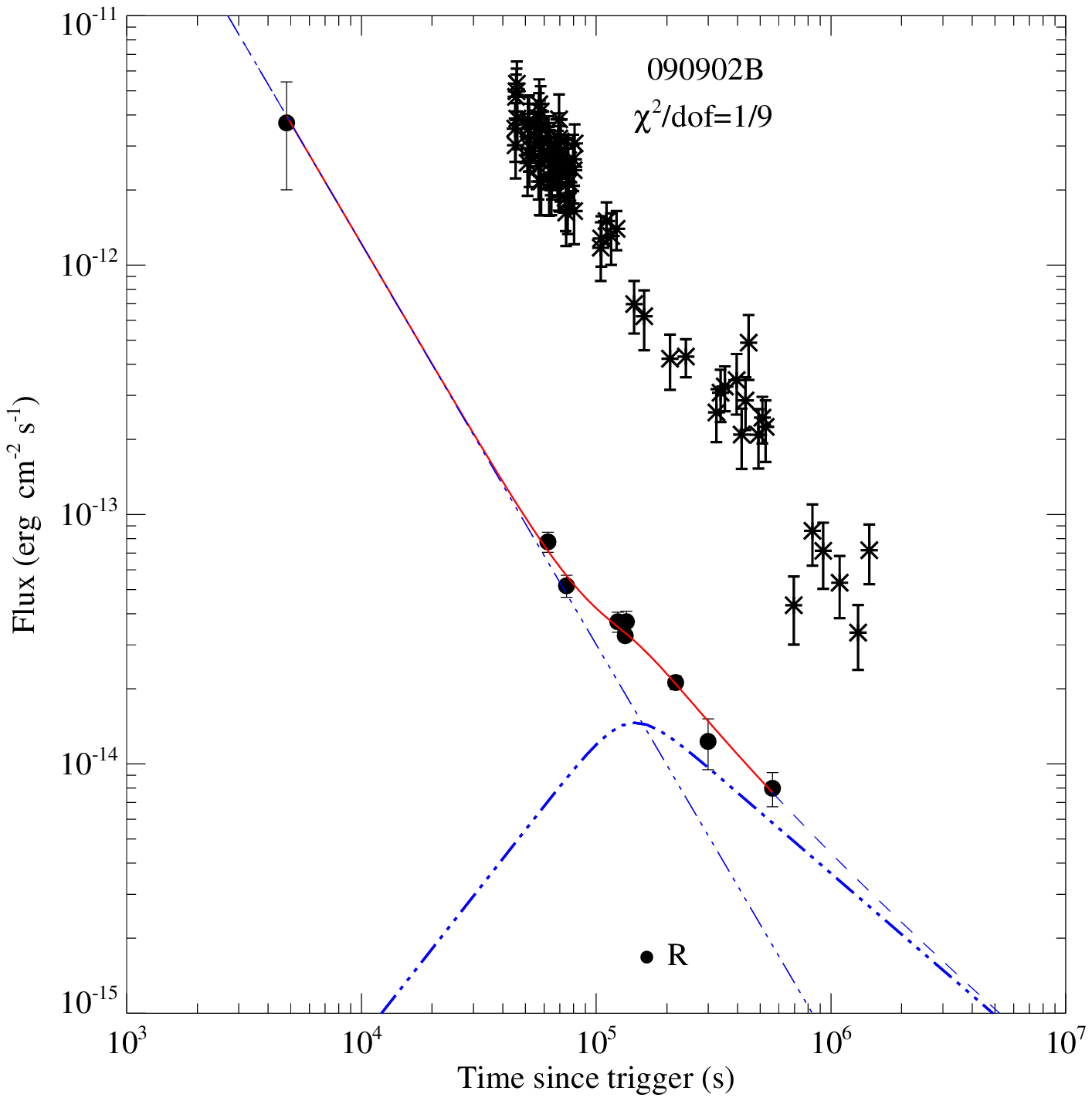}\hfill
\includegraphics[angle=0,scale=0.35]{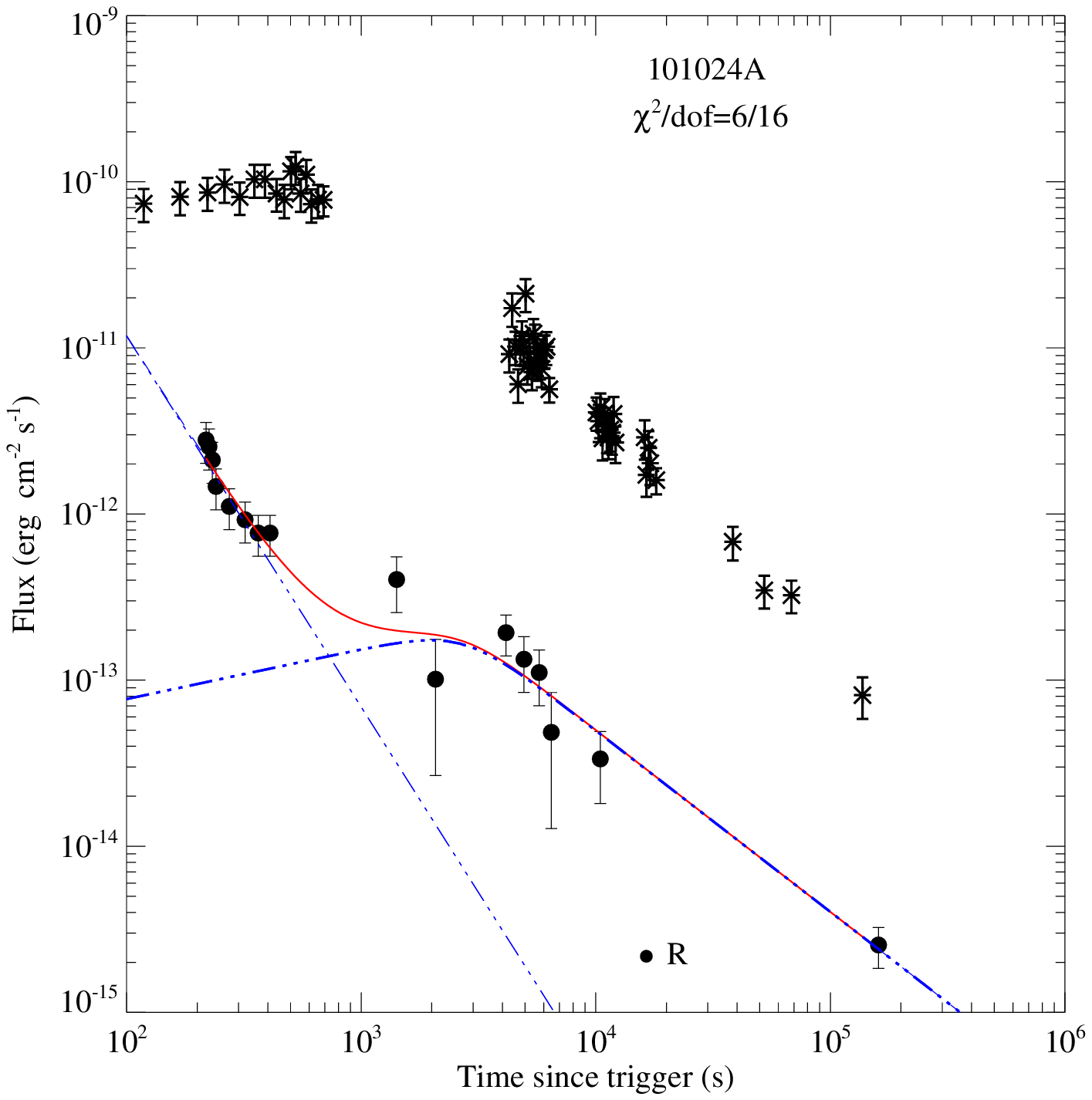}
\caption{Optical lightcurves with detections of both an initial normal decay segment and a late re-brightening hump(s). The symbols and line styles are the same as Figure \ref{Onset}\label{Normal_RB}.}
\end{figure*}

\begin{figure*}
\includegraphics[angle=0,scale=0.35]{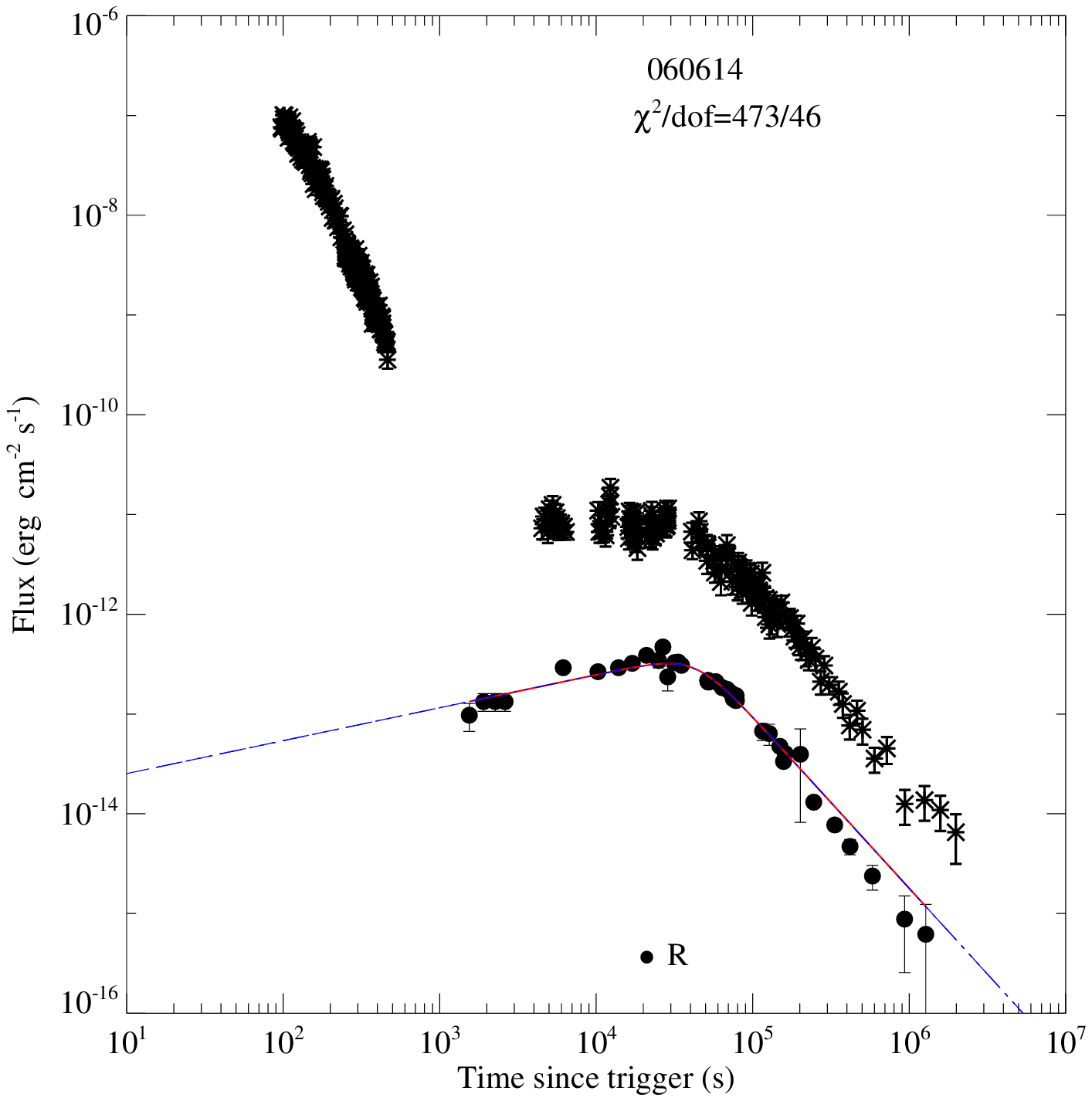}
\includegraphics[angle=0,scale=0.35]{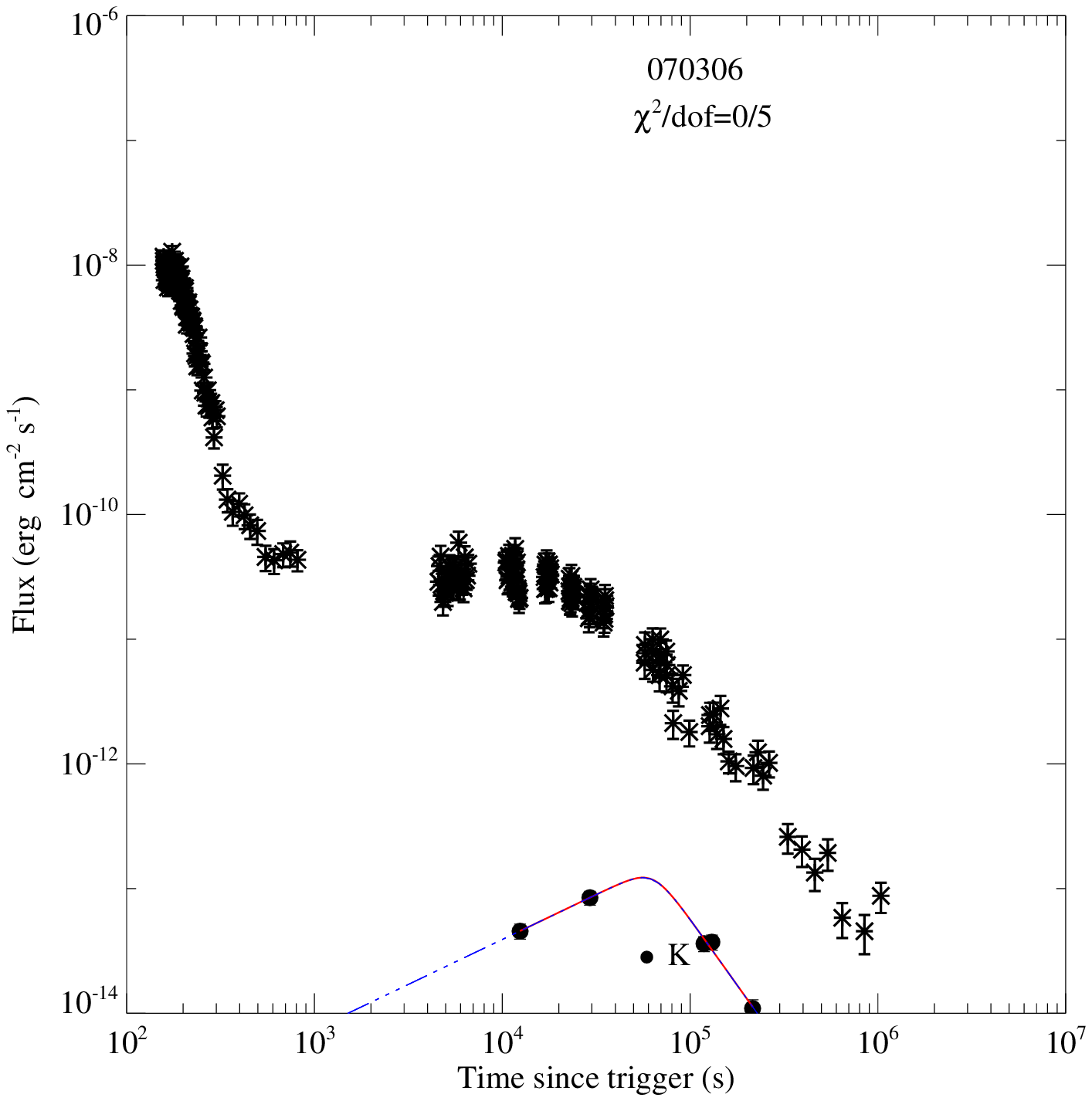}
\includegraphics[angle=0,scale=0.35]{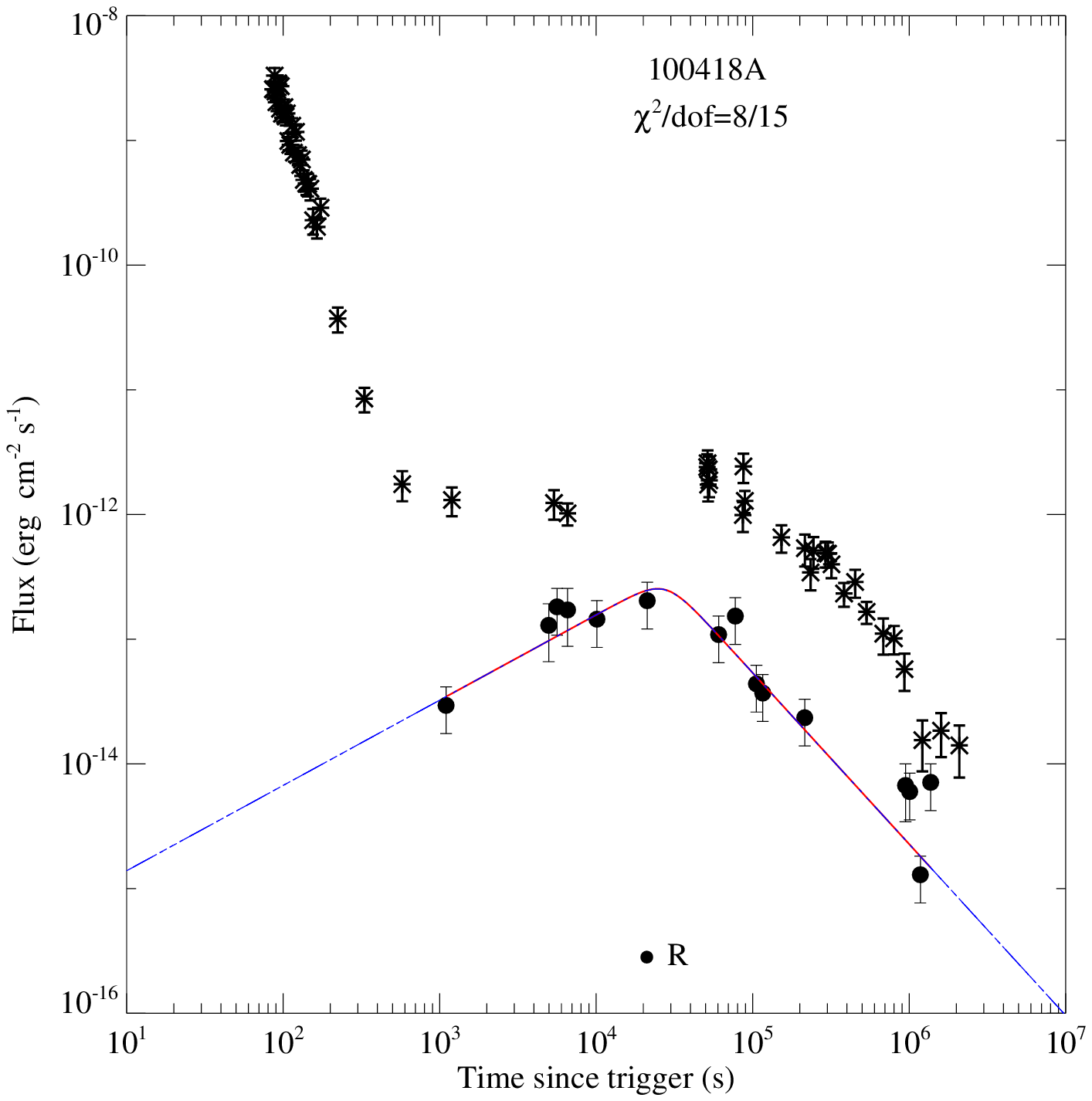}
\caption{Optical lightcurves with detection of a late bump peaking at $>10^4$ seconds post the GRB trigger. The symbols and line styles are the same as Figure \ref{Onset}\label{Late_bump}.}
\end{figure*}

\clearpage

\begin{figure}
\includegraphics[angle=0,scale=0.5]{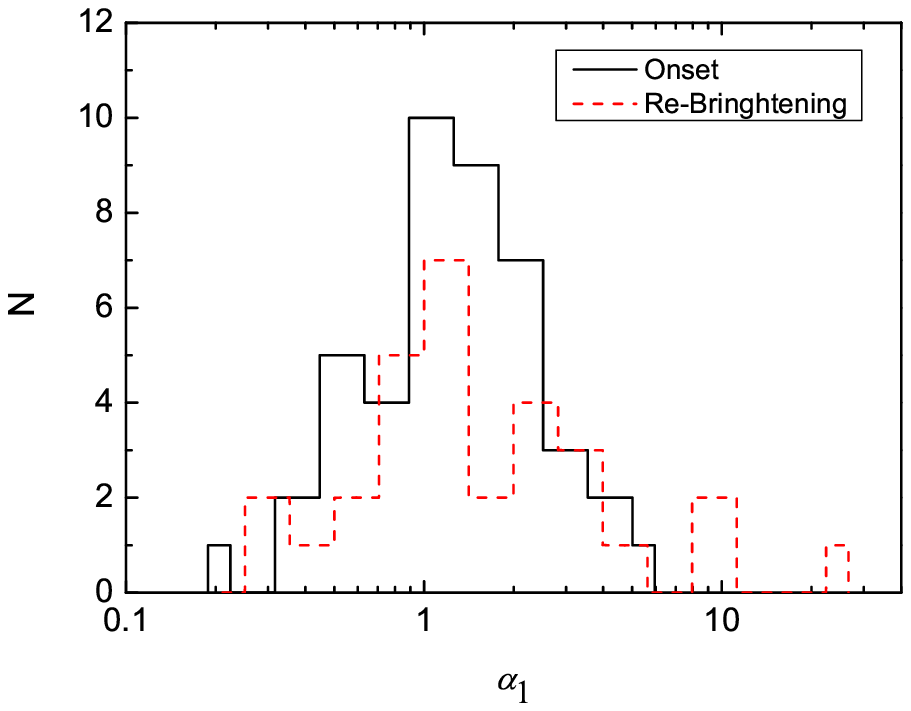}
\includegraphics[angle=0,scale=0.5]{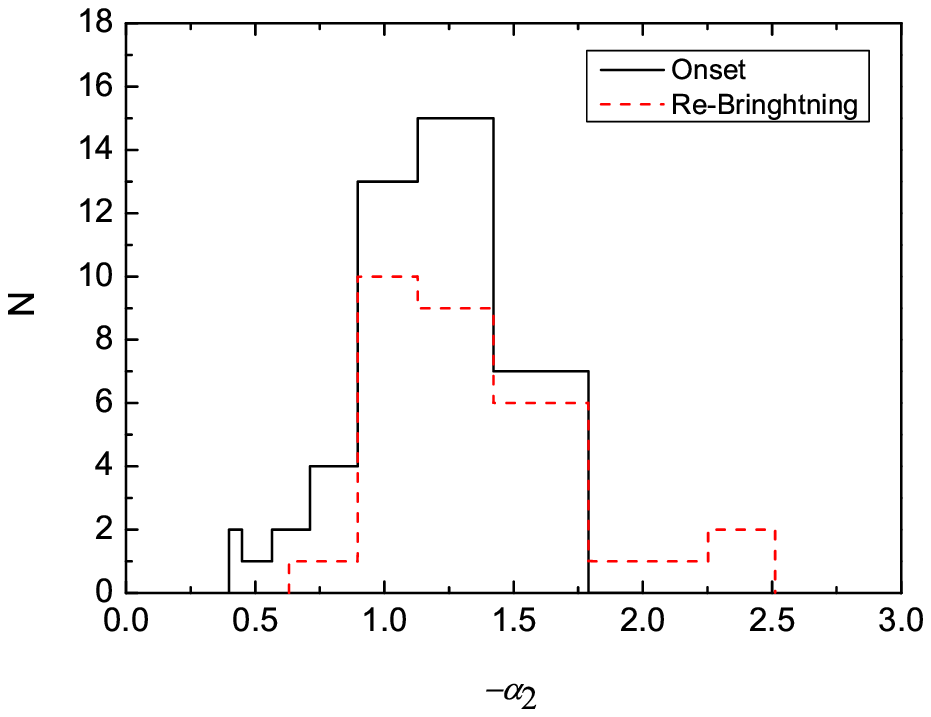}
\includegraphics[angle=0,scale=0.5]{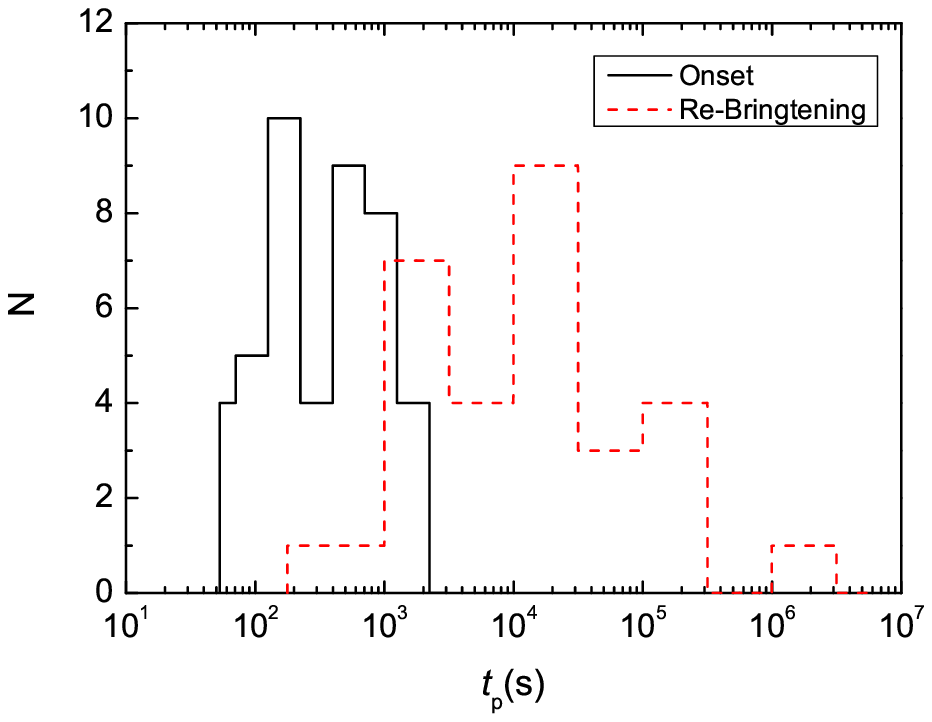}
\includegraphics[angle=0,scale=0.5]{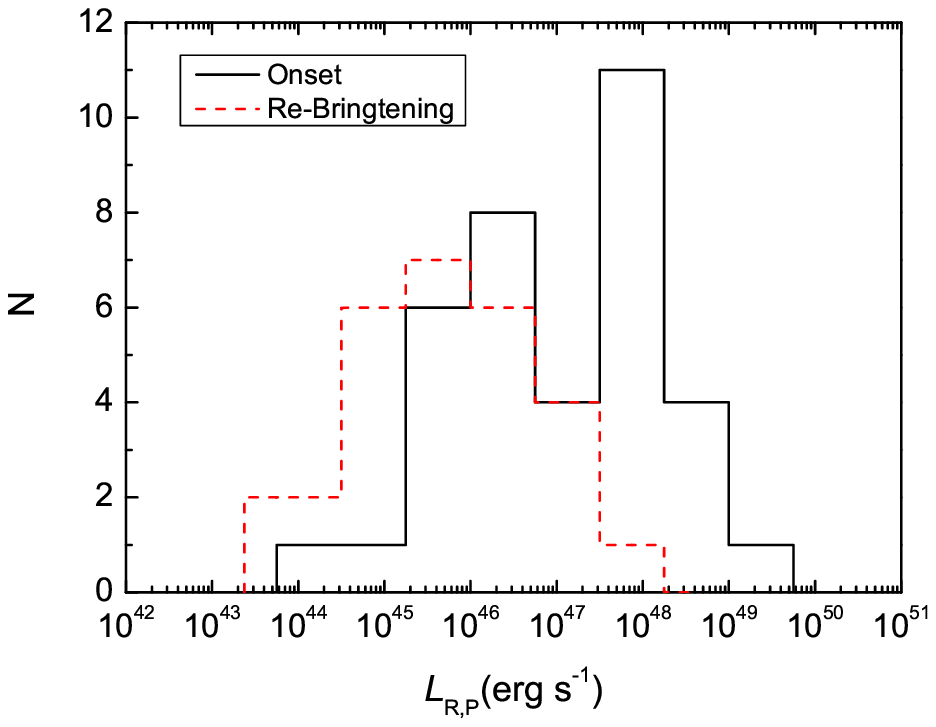}
\includegraphics[angle=0,scale=0.5]{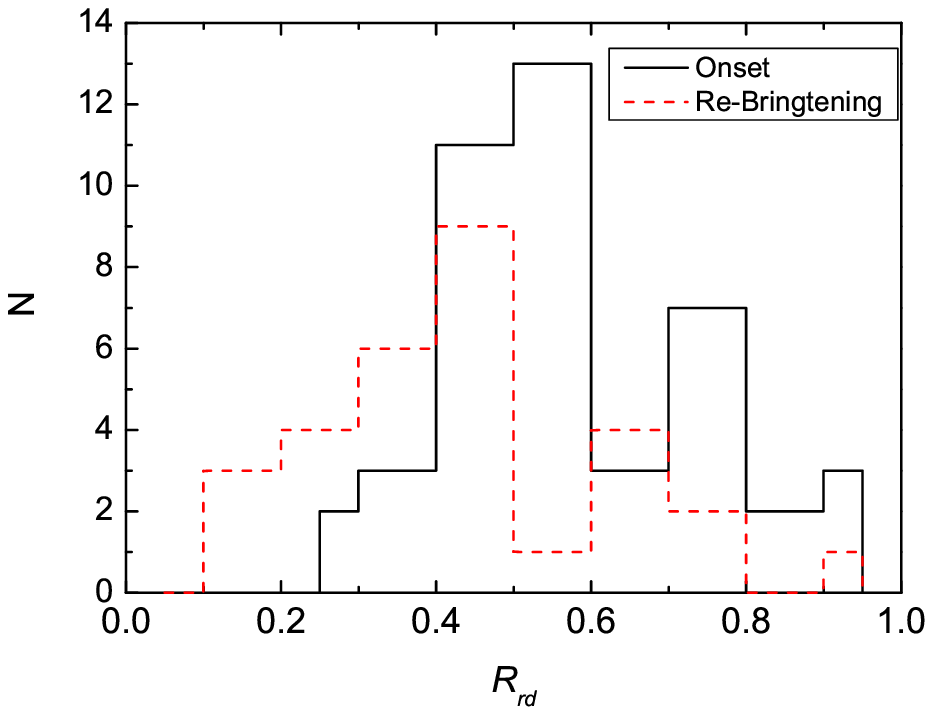}
\caption{Distributions of the characteristics of both the onset and re-brightening bumps.} \label{Onset_RB_Dis}
\end{figure}

\clearpage

\begin{figure}
\includegraphics[angle=0,scale=0.5]{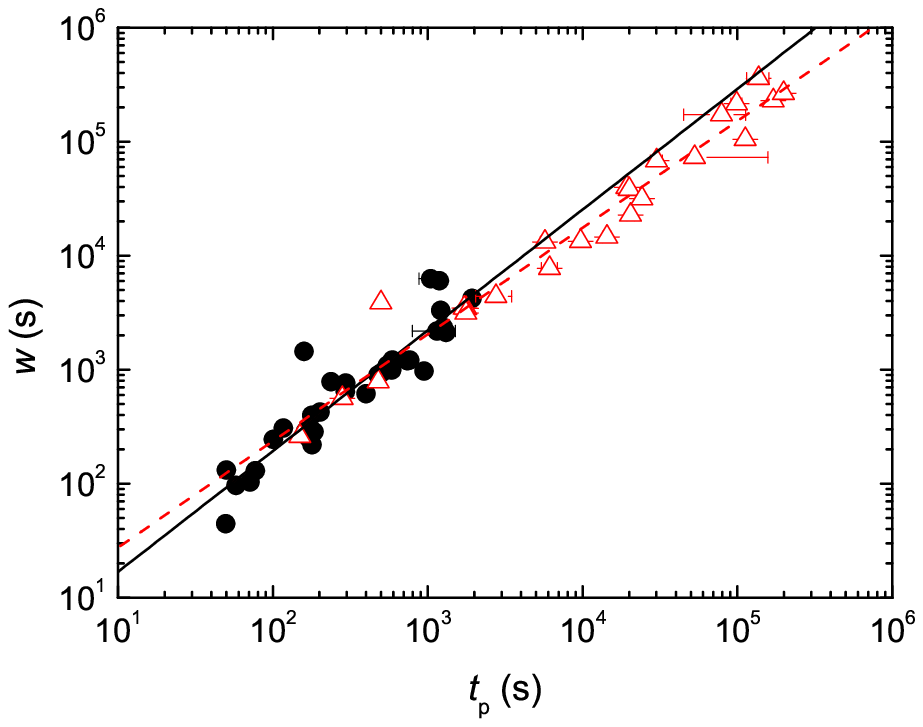}
\includegraphics[angle=0,scale=0.5]{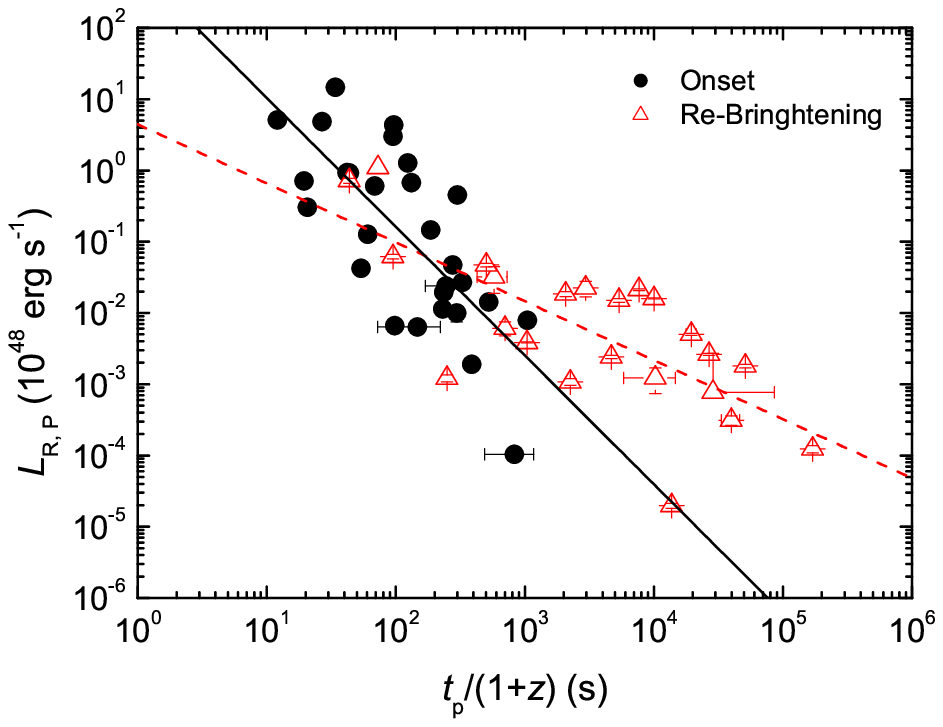}
\includegraphics[angle=0,scale=0.5]{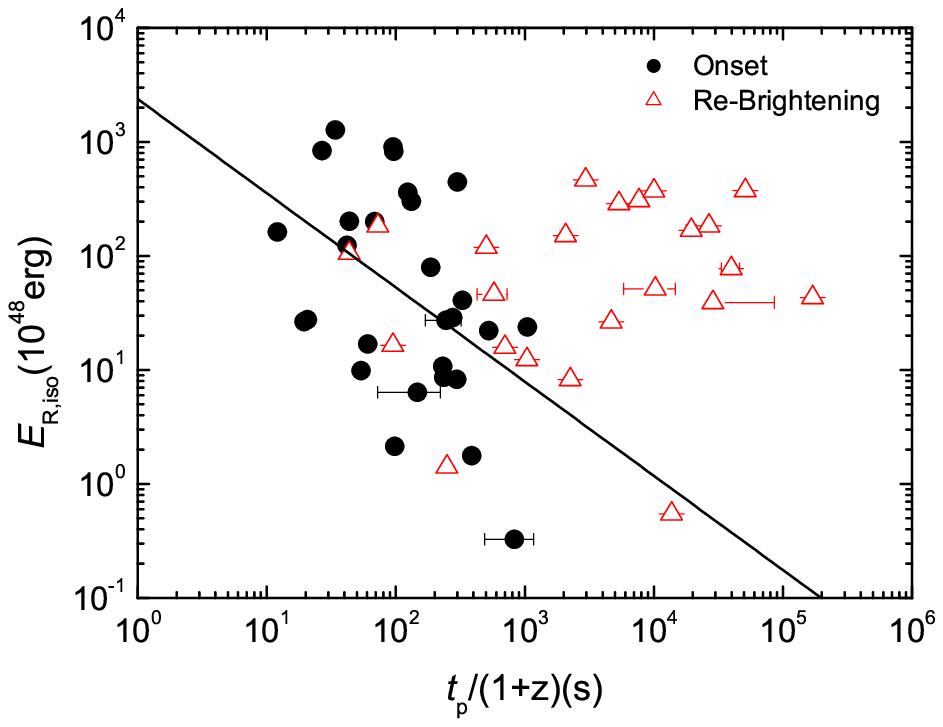}
\caption{Width, isotropic peak luminosity and energy release in the R band as a function of the peak time for the afterglow onset (black solid dots) and re-brightening (open triangles) humps for the GRBs in our sample. Lines are the best fits.}
\label{Onset_RB_Corr}
\end{figure}

\clearpage

\begin{figure}
\includegraphics[angle=0,scale=0.5]{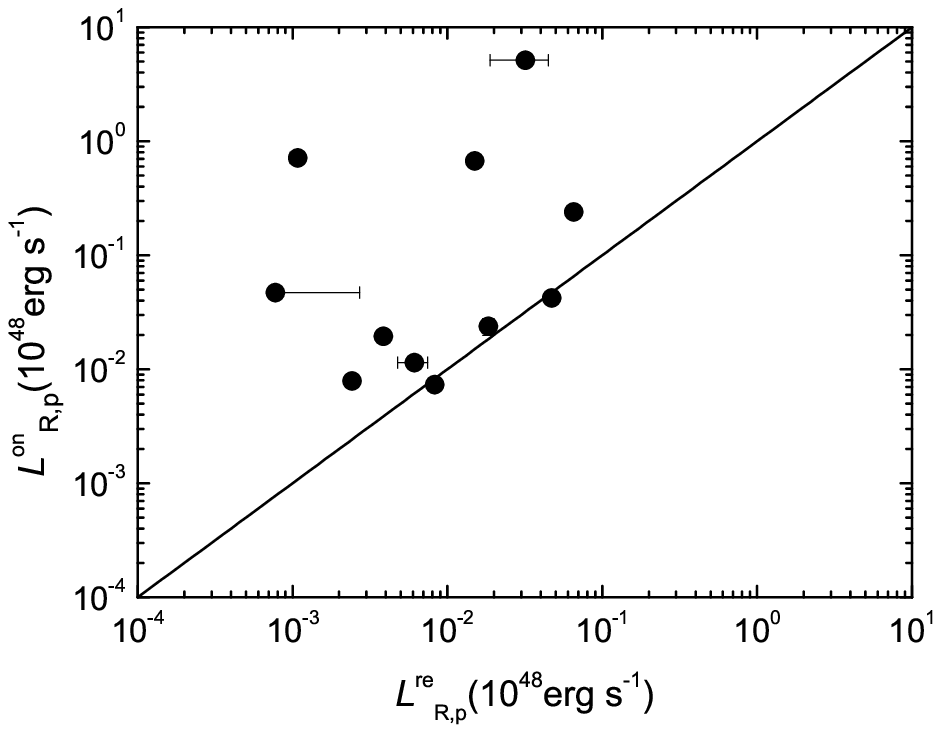}
\includegraphics[angle=0,scale=0.5]{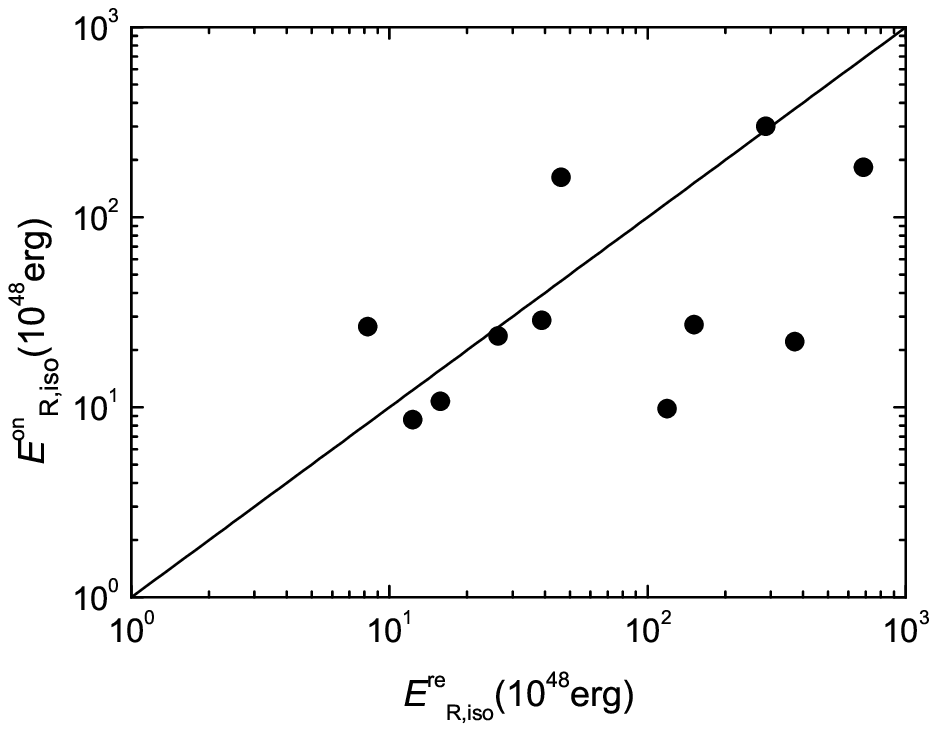}
\caption{Comparisons of the isotropic peak luminosity and energy release in the R-band between the onset and re-brightening bumps for the GRBs with detection of both the onset and re-brightening bumps in their optical lightcurves. The solid line is the equality line.} \label{LR_LB_Corr}
\end{figure}

\clearpage

\begin{figure}
\includegraphics[angle=0,scale=0.5]{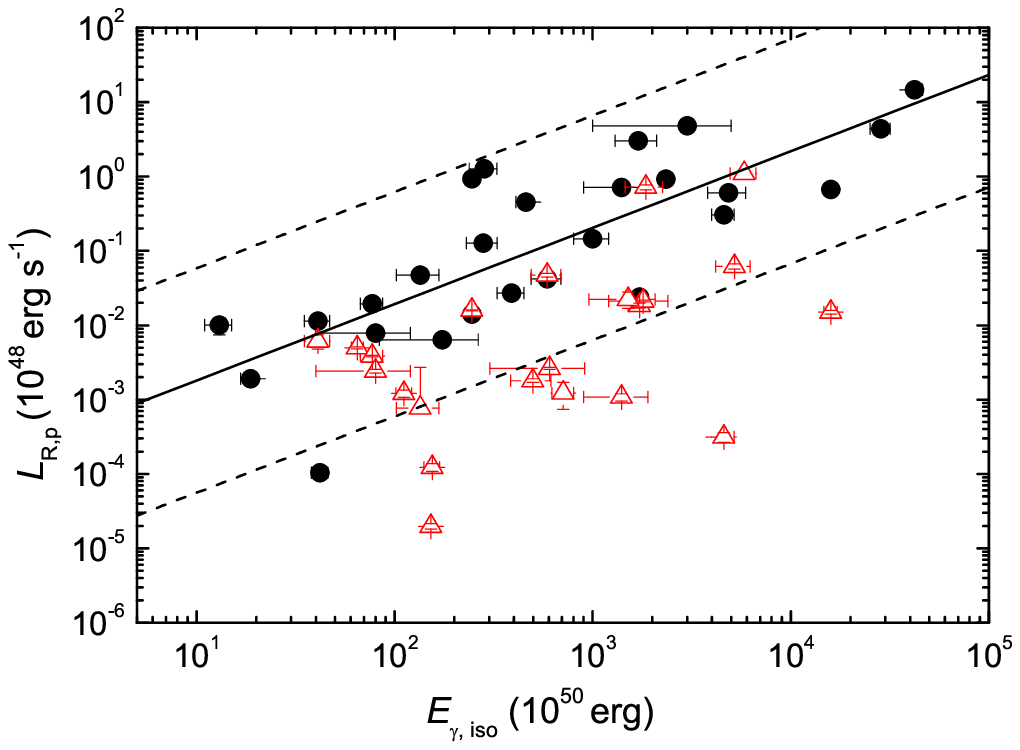}
\includegraphics[angle=0,scale=0.5]{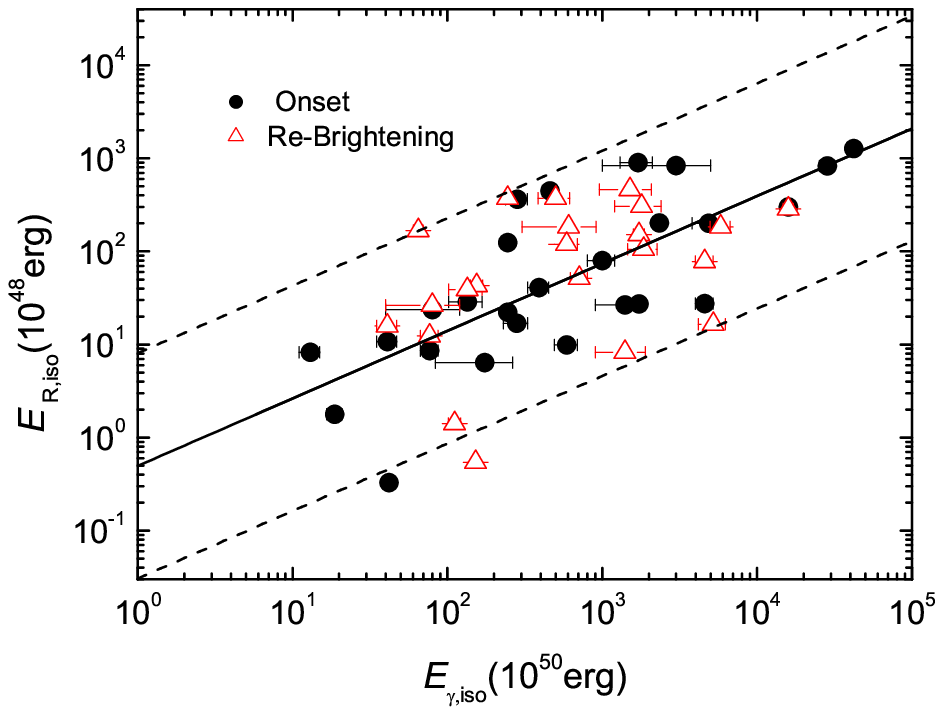}
\caption{Peak luminosity ({\em left panel}) and the isotropic energy release ({\em right panel}) in the R band of the afterglow onset (black solid dots) and re-brightening (open triangles) humps as a function of the isotropic gamma-ray energy release. The solid line is the best linear fit to the data of the afterglow onset bumps and the dashed lines mark the scatter of the data in a 2$\sigma$ region.}
\label{Eiso_Onset_RB_Corr}
\end{figure}

\clearpage

\begin{figure}[htc]
\centering
\includegraphics[angle=0,scale=0.35]{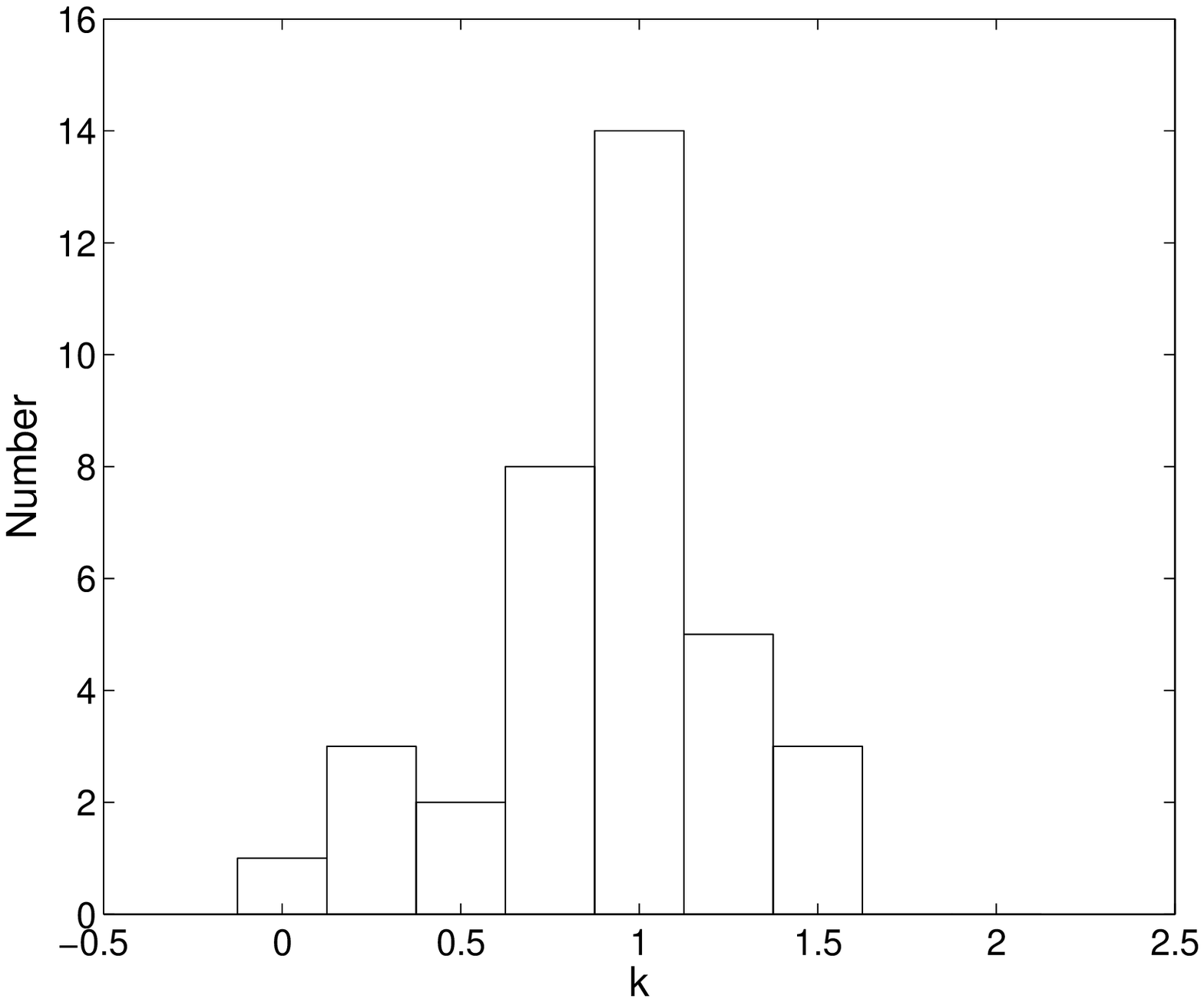}
\includegraphics[angle=0,scale=0.48]{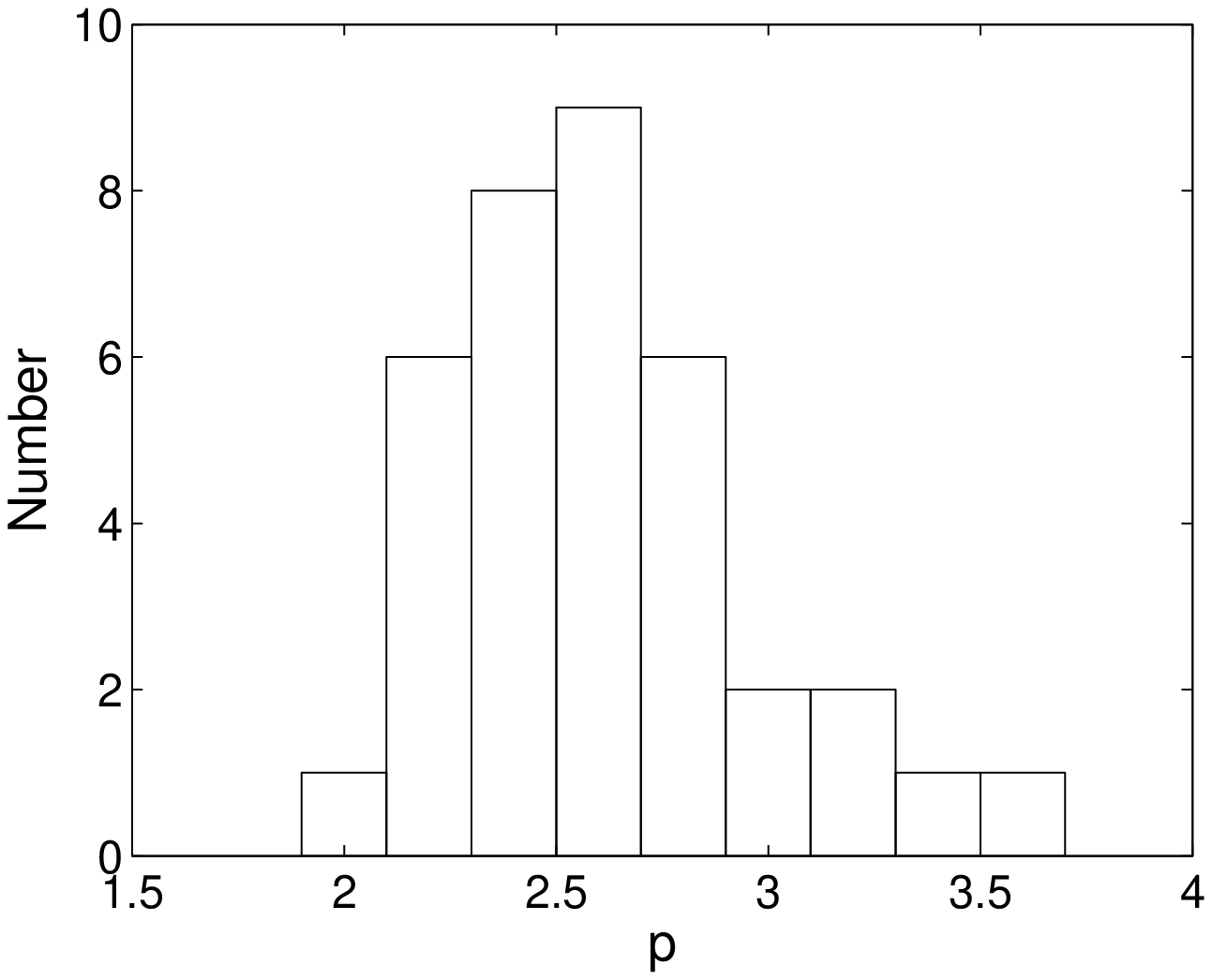}
       \caption{Distributions of electron distribution index \textbf{$k$} ({\em left}) and the ambient density profile index \textbf{$p$ }({\em right}) in our sample.}
           \label{pkonset}
            \end{figure}

\clearpage
\begin{figure}[htc]
\centering
\includegraphics[angle=0,scale=0.5]{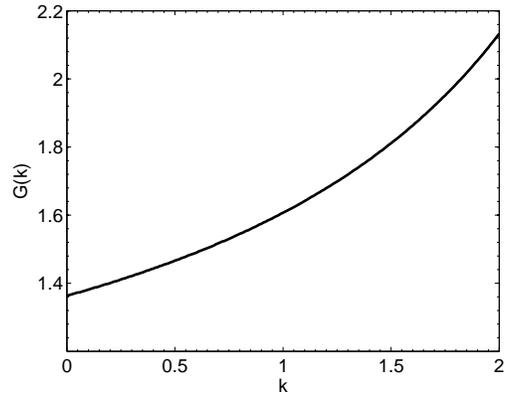}
   \caption{Dimensionless parameter G(k) as a function of the power-law index of the medium density profile $k$.}
   \label{Gk}
      \end{figure}

\clearpage

\begin{figure}[htc]
\centering
\includegraphics[angle=0,scale=0.5]{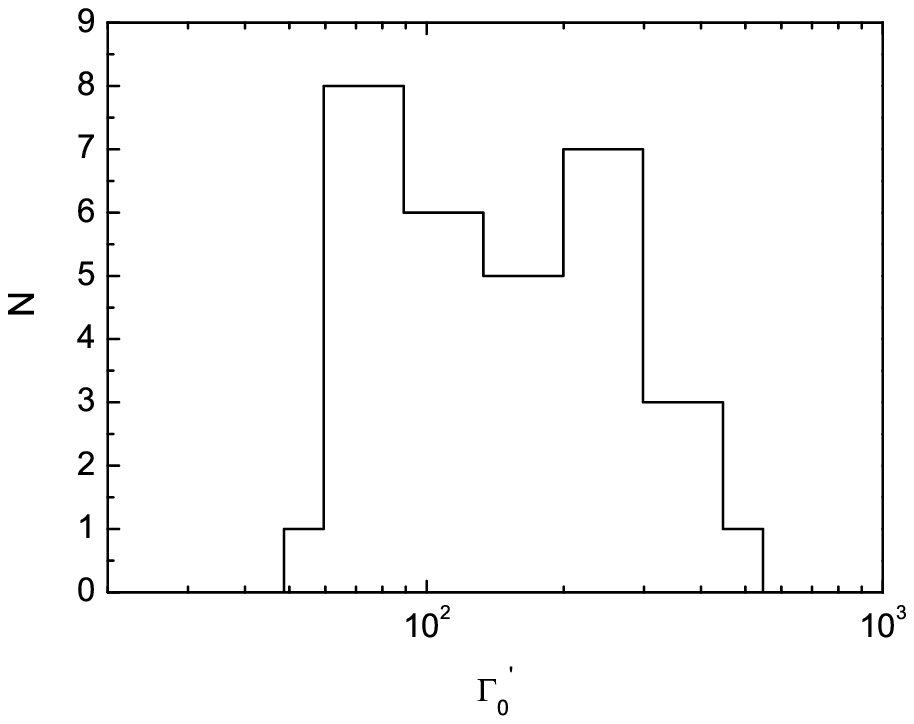}
\includegraphics[angle=0,scale=0.5]{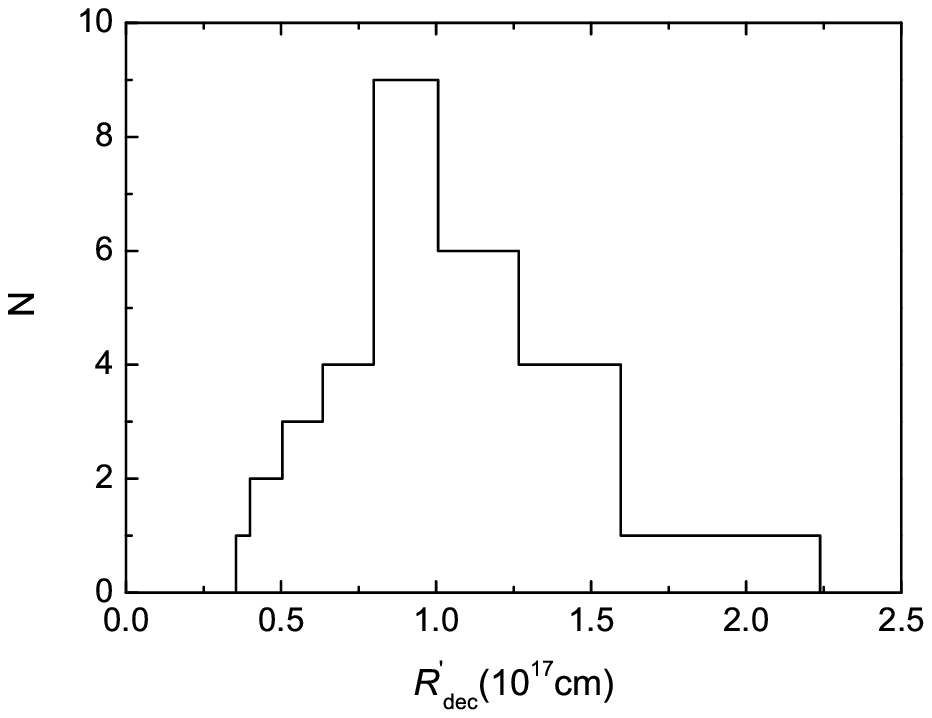}
\caption{Distributions of initial Lorentz factor $\Gamma^{'}_0$ and deceleration radius $R^{'}_{\rm dec}$ for the medium density profile derived from the afterglow onset bumps in our sample.}
\label{Gamma_0}
\end{figure}

\clearpage

\begin{figure}[htc]
\centering
\includegraphics[angle=0,scale=0.75]{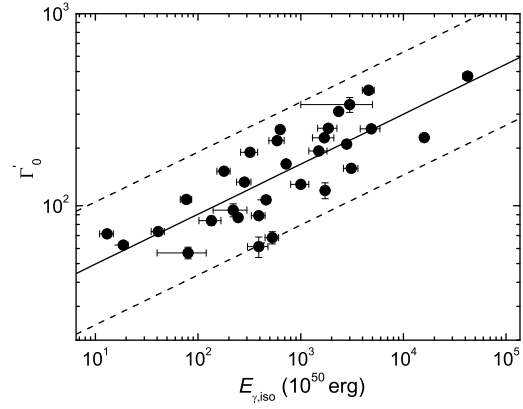}
\caption{Relation between initial Lorentz factor $\Gamma^{'}_{0}$ and isotropic  gamma-ray energy $E_{\rm \gamma, iso}$. The solid line is the best fit to the data and the dashed lines mark the scatter of the data in the 2$\sigma$ region.}
\label{Gamma_0-Eiso}
\end{figure}

\begin{thebibliography}{}
\bibitem[Achterberg et al.(2001)]{2001MNRAS.328..393A} Achterberg, A., Gallant, Y.~A., Kirk, J.~G., \& Guthmann, A.~W.\ 2001, \mnras, 328, 393
\bibitem[Antonelli et al.(2006)]{2006A&A...456..509A} Antonelli, L.~A., Testa, V., Romano, P., et al.\ 2006, \aap, 456, 509
\bibitem[Bednarz \& Ostrowski(1998)]{1998PhRvL..80.3911B} Bednarz, J., \& Ostrowski, M.\ 1998, Physical Review Letters, 80, 3911
\bibitem[Berger et al.(2003)]{2003Natur.426..154B} Berger, E., Kulkarni, S.~R., Pooley, G., et al.\ 2003, \nat, 426, 154
\bibitem[Bj{\"o}rnsson et al.(2004)]{2004ApJ...615L..77B} Bj{\"o}rnsson, G., Gudmundsson, E.~H., \& J{\'o}hannesson, G.\ 2004, \apjl, 615, L77
\bibitem[Burrows et al.(2005)]{2005Sci...309.1833B} Burrows, D.~N., Romano, P., Falcone, A., et al.\ 2005, Science, 309, 1833
\bibitem[Cenko et al.(2006)]{2006ApJ...652..490C} Cenko, S.~B., Kasliwal, M., Harrison, F.~A., et al.\ 2006, \apj, 652, 490
\bibitem[Cenko et al.(2009)]{2009ApJ...693.1484C} Cenko, S.~B., Kelemen, J., Harrison, F.~A., et al.\ 2009, \apj, 693, 1484
\bibitem[Cenko et al.(2011)]{2011ApJ...732...29C} Cenko, S.~B., Frail, D.~A., Harrison, F.~A., et al.\ 2011, \apj, 732, 29
\bibitem[Chester et al.(2008)]{2008AIPC.1000..421C} Chester, M.~M., Wang, X.~Y., Cummings, J.~R., et al.\ 2008, American Institute of Physics Conference Series, 1000, 421
\bibitem[Covino et al.(2008)]{2008MNRAS.388..347C} Covino, S., D'Avanzo, P., Klotz, A., et al.\ 2008, \mnras, 388, 347
\bibitem[Cucchiara et al.(2011)]{2011ApJ...743..154C} Cucchiara, A., Cenko, S.~B., Bloom, J.~S., et al.\ 2011, \apj, 743, 154
\bibitem[Curran et al.(2010)]{2010ApJ...716L.135C} Curran, P.~A., Evans,P.~A., de Pasquale, M., Page, M.~J.,\& van der Horst, A.~J.\ 2010, \apjl, 716, L135
\bibitem[Dai \& Lu(2002)]{2002ApJ...565L..87D} Dai, Z.~G., \& Lu, T.\ 2002, \apjl, 565, L87
\bibitem[Dai \& Wu(2003)]{2003ApJ...591L..21D} Dai, Z.~G., \& Wu, X.~F.\ 2003, \apjl, 591, L21
\bibitem[Dai et al.(2006)]{2006Sci...311.1127D} Dai, Z.~G., Wang, X.~Y., Wu, X.~F., \& Zhang, B.\ 2006, Science, 311, 1127
\bibitem[Daigne \& Mochkovitch(1998)]{1998MNRAS.296..275D} Daigne, F., \& Mochkovitch, R.\ 1998, \mnras, 296, 275
\bibitem[Drenkhahn \& Spruit(2004)]{2004ASPC..312..357D} Drenkhahn, G., \& Spruit, H.~C.\ 2004, Astronomical Society of the Pacific Conference Series, 312, 357
\bibitem[Fan \& Wei(2005)]{2005MNRAS.364L..42F} Fan, Y.~Z., \& Wei, D.~M.\ 2005, \mnras, 364, L42
\bibitem[Ferrero et al.(2008)]{2008AIPC.1000..257F} Ferrero, P., Kann, D.~A., Klose, S., et al.\ 2008, American Institute of Physics Conference Series, 1000, 257
\bibitem[Ferrero et al.(2009)]{2009A&A...497..729F} Ferrero, P., Klose, S., Kann, D.~A., et al.\ 2009, \aap, 497, 729
\bibitem[Fynbo et al.(2009)]{2009ApJS..185..526F} Fynbo, J.~P.~U., Jakobsson, P., Prochaska, J.~X., et al.\ 2009, \apjs, 185, 526
\bibitem[Gehrels et al.(2004)]{2004ApJ...611.1005G} Gehrels, N., Chincarini, G., Giommi, P., et al.\ 2004, \apj, 611, 1005
\bibitem[Gendre et al.(2010)]{2010MNRAS.405.2372G} Gendre, B., Klotz, A., Palazzi, E., et al.\ 2010, \mnras, 405, 2372
\bibitem[Giannios \& Spruit(2006)]{2006A&A...450..887G} Giannios, D., \& Spruit, H.~C.\ 2006, \aap, 450, 887
\bibitem[Golenetskii et al.(2006)]{2006GCN..5837....1G} Golenetskii, S., Aptekar, R., Mazets, E., et al.\ 2006, GRB Coordinates Network, 5837, 1
\bibitem[Golenetskii et al.(2007)]{2007GCN..6879....1G} Golenetskii, S., Aptekar, R., Mazets, E., et al.\ 2007, GRB Coordinates Network, 6879, 1
\bibitem[Gorbovskoy et al.(2012)]{2012MNRAS.421.1874G} Gorbovskoy, E.~S., Lipunova, G.~V., Lipunov, V.~M., et al.\ 2012, \mnras, 421, 1874
\bibitem[Granot et al.(2002)]{2002ApJ...570L..61G} Granot, J., Panaitescu, A., Kumar, P., \& Woosley, S.~E.\ 2002, \apjl, 570, L61
\bibitem[Granot et al.(2003)]{2003Natur.426..138G} Granot, J., Nakar, E., \& Piran, T.\ 2003, \nat, 426, 138
\bibitem[Granot(2005)]{2005ApJ...631.1022G} Granot, J.\ 2005, \apj, 631, 1022
\bibitem[Greiner et al.(2009)]{2009ApJ...693.1610G} Greiner, J., Kr{\"u}hler, T., Fynbo, J.~P.~U., et al.\ 2009, \apj, 693, 1610
\bibitem[Grupe et al.(2007)]{2007ApJ...662..443G} Grupe, D., Gronwall, C., Wang, X.-Y., et al.\ 2007, \apj, 662, 443
\bibitem[Guidorzi et al.(2009)]{2009A&A...499..439G} Guidorzi, C., Clemens, C., Kobayashi, S., et al.\ 2009, \aap, 499, 439
\bibitem[Guidorzi et al.(2011)]{2011MNRAS.417.2124G} Guidorzi, C., Kobayashi, S., Perley, D.~A., et al.\ 2011, \mnras, 417, 2124
\bibitem[He et al.(2011)]{2011ApJ...733...22H} He, H.-N., Wu, X.-F., Toma, K., Wang, X.-Y., \& M{\'e}sz{\'a}ros, P.\ 2011, \apj, 733, 22
\bibitem[Huang et al.(2000)]{2000ApJ...543...90H} Huang, Y.~F., Gou, L.~J., Dai, Z.~G., \& Lu, T.\ 2000, \apj, 543, 90
\bibitem[Huang et al.(2004)]{2004ApJ...605..300H} Huang, Y.~F., Wu, X.~F., Dai, Z.~G., Ma, H.~T., \& Lu, T.\ 2004, \apj, 605, 300
\bibitem[Huang et al.(2009)]{2009AIPC.1133..212H} Huang, K.~Y., Wang, S.~Y., \& Urata, Y.\ 2009, American Institute of Physics Conference Series, 1133, 212
\bibitem[Jia et al.(2012)]{2012RAA....12..411J} Jia, L.-W., Wu, X.-F., L{\"u}, H.-J., Hou, S.-J., \& Liang, E.-W.\ 2012, Research in Astronomy and Astrophysics, 12, 411
\bibitem[Jin et al.(2009)]{2009MNRAS.400.1829J} Jin, Z.~P., Xu, D., Covino, S., et al.\ 2009, \mnras, 400, 1829
\bibitem[Kann et al.(2006)]{2006ApJ...641..993K} Kann, D.~A., Klose, S.,\& Zeh, A.\ 2006, \apj, 641, 993
\bibitem[Kann et al.(2010)]{2010ApJ...720.1513K} Kann, D.~A., Klose, S., Zhang, B., et al.\ 2010, \apj, 720, 1513
\bibitem[Kann et al.(2011)]{2011ApJ...734...96K} Kann, D.~A., Klose, S., Zhang, B., et al.\ 2011, \apj, 734, 96
\bibitem[Kirk et al.(2000)]{2000ApJ...542..235K} Kirk, J.~G., Guthmann, A.~W., Gallant, Y.~A., \& Achterberg, A.\ 2000, \apj, 542, 235
\bibitem[Klotz et al.(2008)]{2008A&A...483..847K} Klotz, A., Gendre, B., Stratta, G., et al.\ 2008, \aap, 483, 847
\bibitem[Klotz et al.(2009)]{2009ApJ...697L..18K} Klotz, A., Gendre, B., Atteia, J.~L., et al.\ 2009, \apjl, 697, L18
\bibitem[Kobayashi \& Zhang(2007)]{2007ApJ...655..973K} Kobayashi, S., \& Zhang, B.\ 2007, \apj, 655, 973
\bibitem[Kobayashi et al.(1997)]{1997ApJ...490...92K} Kobayashi, S., Piran, T., \& Sari, R.\ 1997, \apj, 490, 92
\bibitem[Kr{\"u}hler et al.(2009)]{2009A&A...508..593K} Kr{\"u}hler, T., Greiner, J., Afonso, P., et al.\ 2009, \aap, 508, 593
\bibitem[Kr{\"u}hler et al.(2009)]{2009ApJ...697..758K} Kr{\"u}hler, T., Greiner, J., McBreen, S., et al.\ 2009, \apj, 697, 758
\bibitem[Krimm et al.(2009)]{2009ApJ...704.1405K} Krimm, H.~A., Yamaoka, K., Sugita, S., et al.\ 2009, \apj, 704, 1405
\bibitem[Kuin et al.(2009)]{2009MNRAS.395L..21K} Kuin, N.~P.~M., Landsman, W., Page, M.~J., et al.\ 2009, \mnras, 395, L21
\bibitem[Kumar \& Granot(2003)]{2003ApJ...591.1075K} Kumar, P. \& Granot, J. \ 2003, \apj, 591, 1075
\bibitem[Laas-Bourez et al.(2010)]{2010GCN..11382...1L} Laas-Bourez, M., Klotz, A., Coward, D., et al.\ 2010, GRB Coordinates Network, 11382, 1
\bibitem[Lazzati et al.(2002)]{2002A&A...396L...5L} Lazzati, D., Rossi, E., Covino, S., Ghisellini, G., \& Malesani, D.\ 2002, \aap, 396, L5
\bibitem[Lazzati et al.(2003)]{2003A&A...410..823L} Lazzati, D., Covino, S., di Serego Alighieri, S., et al.\ 2003, \aap, 410, 823
\bibitem[Li \& Filippenko(2008)]{2008GCN..7475....1L} Li, W., \& Filippenko, A.~V.\ 2008, GRB Coordinates Network, 7475, 1
\bibitem[Li et al.(2003)]{2003ApJ...586L...9L} Li, W., Filippenko, A.~V., Chornock, R., \& Jha, S.\ 2003, \apjl, 586, L9
\bibitem[Li et al.(2012)]{2012ApJ...758...27L} Li, L., Liang, E.-W., Tang, Q.-W., et al.\ 2012, \apj, 758, 27 (paper I)
\bibitem[Liang \& Zhang(2006)]{2006ApJ...638L..67L} Liang, E., \& Zhang, B.\ 2006, \apjl, 638, L67
\bibitem[Liang et al.(2008)]{2008AIPC.1000..204L} Liang, E., Racusin, J.~L., Zhang, B., Zhang, B.-B., \& Burrows, D.~N.\ 2008, American Institute of Physics Conference Series, 1000, 204
\bibitem[Liang et al.(2010)]{2010ApJ...725.2209L} Liang, E.-W., Yi, S.-X., Zhang, J., et al.\ 2010, \apj, 725, 2209
\bibitem[Liang\& Zhang(2006)]{2006ApJ...638L..67L} Liang, E., \& Zhang, B.\ 2006,\apjl, 638, L67
\bibitem[Littlejohns et al.(2012)]{2012MNRAS.421.2692L} Littlejohns, O.~M., Willingale, R., O'Brien, P.~T., et al.\ 2012, \mnras, 421, 2692
\bibitem[Liu et al.(2008)]{2008A&A...487..503L} Liu, X.~W., Wu, X.~F., \& Lu, T.\ 2008, \aap, 487, 503
\bibitem[L{\"u} et al.(2012)]{2012ApJ...751...49L} L{\"u}, J., Zou, Y.-C.,Lei, W.-H., et al.\ 2012, \apj, 751, 49
\bibitem[Malesani et al.(2008)]{2008GCN..7436....1M} Malesani, D., Fynbo, J.~P.~U., Vreeswijk, P.~M., \& Villforth, C.\ 2008, GRB Coordinates Network, 7436, 1
\bibitem[Mao et al.(2012)]{2012A&A...538A...1M} Mao, J., Malesani, D., D'Avanzo, P., et al.\ 2012, \aap, 538, A1
\bibitem[Margutti et al.(2010)]{2010MNRAS.406.2149M} Margutti, R., Guidorzi, C., Chincarini, G., et al.\ 2010, \mnras, 406, 2149
\bibitem[Melandri et al.(2009)]{2009MNRAS.395.1941M} Melandri, A., Guidorzi, C., Kobayashi, S., et al.\ 2009, \mnras, 395, 1941
\bibitem[Melandri et al.(2010)]{2010ApJ...723.1331M} Melandri, A., Kobayashi, S., Mundell, C.~G., et al.\ 2010, \apj, 723, 1331
\bibitem[M{\'e}sz{\'a}ros et al.(2006)]{2006AIPC..842.1007M} M{\'e}sz{\'a}ros, P., Razzaque, S., \& Wang, X.~Y.\ 2006, Particles and Nuclei, 842, 1007
\bibitem[M{\'e}sz{\'a}ros \& Rees(1997)]{1997ApJ...476..232M} M{\'e}sz{\'a}ros, P., \& Rees, M.~J.\ 1997, \apj, 476, 232
\bibitem[M{\'e}sz{\'a}ros \& Rees(2003)]{2003ApJ...591L..91M} M{\'e}sz{\'a}ros, P., \& Rees, M.~J.\ 2003, \apjl, 591, L91
\bibitem[Meszaros et al.(1998)]{1998ApJ...499..301M} M\'esz\'aros, P., Rees, M.~J., \& Wijers, R.~A.~M.~J.\ 1998, \apj, 499, 301
\bibitem[Molinari et al.(2007)]{2007A&A...469L..13M} Molinari, E., Vergani, S.~D., Malesani, D., et al.\ 2007, \aap, 469, L13
\bibitem[Nakar \& Granot(2007)]{2007MNRAS.380.1744N} Nakar, E., \& Granot, J.\ 2007, \mnras, 380, 1744
\bibitem[Nardini et al.(2006)]{2006A&A...451..821N} Nardini, M., Ghisellini, G., Ghirlanda, G., et al.\ 2006, \aap, 451, 821
\bibitem[Nardini et al.(2011)]{2011A&A...531A..39N} Nardini, M., Greiner, J., Kr{\"u}hler, T., et al.\ 2011, \aap, 531, A39
\bibitem[Oates et al.(2009)]{2009MNRAS.395..490O} Oates, S.~R., Page,M.~J., Schady, P., et al.\ 2009, \mnras, 395, 490
\bibitem[Page et al.(2009)]{2009MNRAS.400..134P} Page, K.~L., Willingale, R., Bissaldi, E., et al.\ 2009, \mnras, 400, 134
\bibitem[Panaitescu \& Kumar(2001)]{2001ApJ...554..667P} Panaitescu, A., \& Kumar, P.\ 2001, \apj, 554, 667
\bibitem[Panaitescu\& Kumar(2001)]{2001ApJ...560L..49P} Panaitescu, A., \& Kumar, P.\ 2001, \apjl, 560, L49
\bibitem[Panaitescu \& Vestrand(2008)]{2008MNRAS.387..497P} Panaitescu, A., \& Vestrand, W.~T.\ 2008, \mnras, 387, 497
\bibitem[Panaitescu \& Vestrand(2011)]{2011MNRAS.414.3537P} Panaitescu, A., \& Vestrand, W.~T.\ 2011, \mnras, 414, 3537
\bibitem[Panaitescu et al.(1998)]{1998ApJ...503..314P} Panaitescu, A., M{\'e}sz{\'a}ros, P., \& Rees, M.~J.\ 1998, \apj, 503, 314
\bibitem[Pandey et al.(2010)]{2010ApJ...714..799P} Pandey, S.~B., Swenson, C.~A., Perley, D.~A., et al.\ 2010, \apj, 714, 799
\bibitem[Pe'er et al.(2006)]{2006ApJ...642..995P} Pe'er, A., M{\'e}sz{\'a}ros, P., \& Rees, M.~J.\ 2006, \apj, 642, 995
\bibitem[Pelassa \& Ohno(2010)]{2010arXiv1002.2863P} Pelassa, V., \& Ohno, M.\ 2010, arXiv:1002.2863
\bibitem[Perley et al.(2008)]{2008ApJ...688..470P} Perley, D.~A., Li, W., Chornock, R., et al.\ 2008, \apj, 688, 470
\bibitem[Perley et al.(2010)]{2010MNRAS.406.2473P} Perley, D.~A., Bloom, J.~S., Klein, C.~R., et al.\ 2010, \mnras, 406, 2473
\bibitem[Perna et al.(2006)]{2006ApJ...636L..29P} Perna, R., Armitage, P.~J., \& Zhang, B.\ 2006, \apjl, 636, L29
\bibitem[Prochaska et al.(2007)]{2007ApJS..168..231P} Prochaska, J.~X., Chen, H.-W., Bloom, J.~S., et al.\ 2007, \apjs, 168, 231
\bibitem[Proga \& Zhang(2006)]{2006MNRAS.370L..61P} Proga, D., \& Zhang, B.\ 2006, \mnras, 370, L61
\bibitem[Racusin et al.(2008)]{2008Natur.455..183R} Racusin, J.~L., Karpov, S.~V., Sokolowski, M., et al.\ 2008, \nat, 455, 183
\bibitem[Rana et al.(2009)]{2009AAS...21361003R} Rana, V., Cenko, B., Harrison, F., Fox, D., \& Kelemen, J.\ 2009, American Astronomical Society Meeting Abstracts \#213, 213, \#610.03
\bibitem[Rees \& M{\'e}sz{\'a}ros(1994)]{1994ApJ...430L..93R} Rees, M.~J., \& M{\'e}sz{\'a}ros, P.\ 1994, \apjl, 430, L93
\bibitem[Rees \& M{\'e}sz{\'a}ros(2005)]{2005ApJ...628..847R} Rees, M.~J., \& M{\'e}sz{\'a}ros, P.\ 2005, \apj, 628, 847
\bibitem[Rees \& Meszaros(1992)]{1992MNRAS.258P..41R} Rees, M.~J., \& Meszaros, P.\ 1992, \mnras, 258, 41P
\bibitem[Resmi et al.(2005)]{2005A&A...440..477R} Resmi, L., Ishwara-Chandra, C.~H., Castro-Tirado, A.~J., et al.\ 2005, \aap, 440, 477
\bibitem[Robertson \& Ellis(2012)]{2012ApJ...744...95R} Robertson, B.~E., \& Ellis, R.~S.\ 2012, \apj, 744, 95
\bibitem[Rossi et al.(2002)]{2002MNRAS.332..945R} Rossi, E., Lazzati, D., \& Rees, M.~J.\ 2002, \mnras, 332, 945
\bibitem[Rossi et al.(2011)]{2011A&A...529A.142R} Rossi, A., Schulze, S., Klose, S., et al.\ 2011, \aap, 529, A142
\bibitem[Ruiz-Velasco et al.(2007)]{2007ApJ...669....1R} Ruiz-Velasco, A.~E., Swan, H., Troja, E., et al.\ 2007, \apj, 669, 1
\bibitem[Rykoff et al.(2004)]{2004ApJ...601.1013R} Rykoff, E.~S., Smith, D.~A., Price, P.~A., et al.\ 2004, \apj, 601, 1013
\bibitem[Rykoff et al.(2009)]{2009ApJ...702..489R} Rykoff, E.~S.,Aharonian, F., Akerlof, C.~W., et al.\ 2009, \apj, 702, 489
\bibitem[Sakamoto et al.(2010)]{2010GCN..11169...1S} Sakamoto, T., Barthelmy, S.~D., Baumgartner, W.~H., et al.\ 2010, GRB Coordinates Network, 11169, 1
\bibitem[Sari \& Piran(1999)]{1999ApJ...520..641S} Sari, R., \& Piran, T.\ 1999, \apj, 520, 641
\bibitem[Sari et al.(1998)]{1998ApJ...497L..17S} Sari, R., Piran, T., \& Narayan, R.\ 1998, \apjl, 497, L17
\bibitem[Schady et al.(2008)]{2008AIPC.1000..200S} Schady, P., de Pasquale, M., Page, M.~J., et al.\ 2008, American Institute of Physics Conference Series, 1000, 200
\bibitem[Schlegel et al.(1998)]{1998ApJ...500..525S} Schlegel, D.~J., Finkbeiner, D.~P., \& Davis, M.\ 1998, \apj, 500, 525
\bibitem[Shen et al.(2006)]{2006MNRAS.371.1441S} Shen, R., Kumar, P., \& Robinson, E.~L.\ 2006, \mnras, 371, 1441
\bibitem[Th{\"o}ne et al.(2010)]{2010A&A...523A..70T} Th{\"o}ne, C.~C., Kann, D.~A., J{\'o}hannesson, G., et al.\ 2010, \aap, 523, A70
\bibitem[Thompson(1994)]{1994MNRAS.270..480T} Thompson, C.\ 1994, \mnras, 270, 480
\bibitem[Uehara et al.(2011)]{2011A&A...526A..92U} Uehara, T., Uemura, M., Arai, A., et al.\ 2011, \aap, 526, A92
\bibitem[Uhm et al.(2012)]{2012arXiv1208.2347U} Uhm, Z.~L., Zhang, B., Hascoet, R., et al.\ 2012, arXiv:1208.2347
\bibitem[Usov(1992)]{1992Natur.357..472U} Usov, V.~V.\ 1992, \nat, 357, 472
\bibitem[Wren et al.(2009)]{2009GCN..9778....1W} Wren, J., Vestrand, W.~T., Wozniak, P.~R., Davis, H., \& Norman, B.\ 2009, GRB Coordinates Network, 9778, 1
\bibitem[Wu et al.(2005)]{2005MNRAS.357.1197W} Wu, X.~F., Dai, Z.~G., Huang, Y.~F., \& Lu, T.\ 2005, \mnras, 357, 1197
\bibitem[Xue et al.(2009)]{2009A&A...498..671X} Xue, R.-R., Fan, Y.-Z., \& Wei, D.-M.\ 2009, \aap, 498, 671
\bibitem[Yost et al.(2003)]{2003ApJ...597..459Y} Yost, S.~A., Harrison,F.~A., Sari, R., \& Frail, D.~A.\ 2003, \apj, 597, 459
\bibitem[Yost et al.(2006)]{2006ApJ...636..959Y} Yost, S.~A., Alatalo, K., Rykoff, E.~S., et al.\ 2006, \apj, 636, 959
\bibitem[Yuan et al.(2008)]{2008AIPC.1065..103Y} Yuan, F., Rykoff, E.~S., Schaefer, B.~E., et al.\ 2008, American Institute of Physics Conference Series, 1065, 103
\bibitem[Yuan et al.(2010)]{2010ApJ...711..870Y} Yuan, F., Schady, P., Racusin, J.~L., et al.\ 2010, \apj, 711, 870
\bibitem[Zafar et al.(2011)]{2011A&A...532A.143Z} Zafar, T., Watson, D., Fynbo, J.~P.~U., et al.\ 2011, \aap, 532, A143
\bibitem[Zhang \& M{\'e}sz{\'a}ros(2002a)]{2002ApJ...566..712Z} Zhang, B., \& M{\'e}sz{\'a}ros, P.\ 2002a, \apj, 566, 712
\bibitem[Zhang \& M{\'e}sz{\'a}ros(2002b)]{2002ApJ...571..876Z} Zhang, B., \& M{\'e}sz{\'a}ros, P.\ 2002b, \apj, 571, 876
\bibitem[Zhang \& M{\'e}sz{\'a}ros(2004)]{2004IJMPA..19.2385Z} Zhang, B., \& M{\'e}sz{\'a}ros, P.\ 2004, International Journal of Modern Physics A, 19, 2385
\bibitem[Zhang \& Yan(2011)]{2011ApJ...726...90Z} Zhang, B., \& Yan, H.\ 2011, \apj, 726, 90
\bibitem[Zhang et al.(2003)]{2003ApJ...595..950Z} Zhang, B., Kobayashi, S., \& M{\'e}sz{\'a}ros, P.\ 2003, \apj, 595, 950
\bibitem[Zhang et al.(2004)]{2004ApJ...601L.119Z} Zhang, B., Dai, X., Lloyd-Ronning, N.~M., \& M{\'e}sz{\'a}ros, P.\ 2004, \apjl, 601, L119
\bibitem[Zhang et al.(2006)]{2006ApJ...642..354Z} Zhang, B., Fan, Y.~Z., Dyks, J., et al.\ 2006, \apj, 642, 354
\bibitem[Zhang et al.(2007)]{2007ApJ...655L..25Z} Zhang, B., Zhang, B.-B., Liang, E.-W., et al.\ 2007, \apjl, 655, L25
\bibitem[Zhang(2007)]{2007ChJAA...7....1Z} Zhang, B.\ 2007, \cjaa, 7, 1
\bibitem[Zhang(2011)]{2011CRPhy..12..206Z} Zhang, B.\ 2011, Comptes Rendus Physique, 12, 206

\end{thebibliography}
\end{document}